\documentclass[11pt]{article}

\usepackage{amssymb,amsmath}
\usepackage[dvips]{graphicx}
\usepackage{subfigure}
\usepackage{latexsym}
\usepackage[dvips]{graphicx}

\def\abstract#1{\vskip 7mm 
        \begin{center}{\large Abstract}\par \smallskip
                \begin{minipage}[c]{15.5cm}
                       #1
                \end{minipage}
        \end{center}
}
\def\title#1{\begin{center}{\Large\bf #1}\end{center}}
\def\author#1{\vskip 5mm \begin{center}{#1}\end{center}}
\def\address#1{\begin{center}{\it #1}\end{center}}

\numberwithin{equation}{section}

\newcommand{\eqb}{\begin{equation}}
\newcommand{\eqe}{\end{equation}}
\newcommand{\eqbnon}{\begin{equation*}}
\newcommand{\eqenon}{\end{equation*}}

\newcommand{\eqab}{\begin{eqnarray}}
\newcommand{\eqae}{\end{eqnarray}}
\newcommand{\eqabnon}{\begin{eqnarray*}}
\newcommand{\eqaenon}{\end{eqnarray*}}

\newcommand{\seqb}{\begin{subequations}}
\newcommand{\seqe}{\end{subequations}}

\newcommand{\eref}[1]{\eqref{#1}}

\newcommand{\defeq}{:=}
\newcommand{\defeqr}{=:}

\newcommand{\pd}[2]{\dfrac{\partial #1}{\partial #2}}

\newcommand{\Fds}{F_{\rm ds}}
\newcommand{\Sds}{S_{\rm ds}}
\newcommand{\Tds}{T_{\rm ds}}
\newcommand{\Ads}{A_{\rm ds}}
\newcommand{\sigmads}{\sigma_{\rm ds}}
\newcommand{\Eds}{E_{\rm ds}}
\newcommand{\fds}{f_{\rm ds}}
\newcommand{\rds}{r_{\rm ds}}

\newtheorem{ass-ds}{Assumption dS --}
\newtheorem{ass-sds}{Assumption SdS --}
\newtheorem{hypo-ds}{Working Hypothesis dS --}
\newtheorem{hypo-sds}{Working Hypothesis SdS --}
\newtheorem{key}{Key Point}

\begin{document}

\title{
Limit of Universality of Entropy-Area Law for Multi-Horizon Spacetimes
\footnote{
Based on an invited contribution as a chapter in a book, "Classical and Quantum Gravity: Theory, Analysis and Application", 2011, Nova Science Publ.
}}

\author{Hiromi Saida~\footnote{Email : saida@daido-it.ac.jp}}
\address{Department of Physics, Daido University, 10-3 Takiharu Minami-ku, Nagoya 457-8530, Japan}

\abstract{
It may be a common understanding at present that, once event horizons are in thermal equilibrium, the entropy-area law holds inevitably. 
However, no rigorous verification is given to such a very strong universality of the law in multi-horizon spacetimes. 
In this article, based on thermodynamically consistent and rigorous discussion, we investigate thermodynamics of Schwarzschild-de~Sitter spacetime in which the temperatures of two horizons are different. 
We recognize that \emph{three} independent state variables exist in thermodynamics of the horizons. 
One of the three variables represents the effect of ``external gravity'' acting on one horizon due to another one. 
Then we find that thermodynamic formalism with three independent variables suggests the breakdown of entropy-area law, and clarifies the necessary and sufficient condition for the entropy-area law. 
As a by-product, the special role of cosmological constant in thermodynamics of horizons is also revealed. 
Finally we propose two discussions; one of them is on the quantum statistics of underlying quantum gravity, and another is on the Schwarzschild-de~Sitter black hole evaporation from the point of view of non-equilibrium thermodynamics.
}

\section{Introduction}
\label{sec:intro}

In searching for quantum theory of gravity, many plausible ideas have been proposed. 
However, non of them is completed. 
Each of the present incomplete theory includes conceptual and/or technical difficult problems. 
Now, it may be useful and meaningful to refine reliable and rigorous basis of the search for complete quantum gravity. 
The black hole thermodynamics, especially the so-called entropy-area law, seems to be one of the reliable and rigorous basis. 
This article inspects the range of validity of the entropy-area law, and reveal a rather unexpected limit of the law for multi-horizon spacetimes.
Discussions in this article are based on two papers~\cite{ref:euclidean.ds,ref:euclidean.sds}.

The entropy-area law, which is regarded as an equation of state for an event horizon, claims the equilibrium entropy of event horizon is equal to one-quarter of its spatial area in Planck units~\cite{ref:bht.1,ref:bht.2,ref:hr,ref:gsl}. 
This law is already verified for spacetimes possessing a \emph{single} event horizon~\cite{ref:euclidean.sch,ref:euclidean.kn,ref:euclidean.sads,ref:euclidean.ds}. 
Then we may naively expect that the entropy-area law holds also for multi-horizon spacetimes, once every horizon is individually in thermal equilibrium. 
This expectation is equivalent to consider that the thermal equilibrium of each horizon is the necessary and sufficient condition to ensure the entropy-area law for each horizon. 
However this expectation has not been rigorously verified in multi-horizon spacetimes. 
(Comments on existing researches on Schwarzschild-de~Sitter spacetime will be given later in this section.) 
At present, there remains the possibility that the thermal equilibrium may be simply the necessary condition of entropy-area law. 
If we find an example that some event horizon does not satisfy the entropy-area law even when it is in thermal equilibrium, then we recognize the thermal equilibrium as simply the necessary condition of the entropy-area law.

We can consider a situation in which the entropy-area law may break down in multi-horizon spacetime~\footnote{
All discussions in this article are based on the ordinary general relativity. 
The other modified theories of gravity are not considered. 
Even if there is a breakdown of entropy-area law due to exotic fields of modified theory, such a breakdown in modified theory is out of the scope of this article.
}. 
To explain it, it is necessary to \emph{distinguish} thermodynamic state of each horizon and that of the total system (multi-horizon spacetime) composed of several horizons. 
Even when every horizon in a multi-horizon spacetime is in an equilibrium state \emph{individually}, the total system composed of several horizons is never in any equilibrium state if the equilibrium state of one horizon is different from that of the other horizon. 
For example, if the temperatures of horizons in a multi-horizon spacetime are different from each other, then a net energy flow arises from a high temperature horizon to a low temperature one. 
Such multi-horizon spacetime can not be understood to be in any equilibrium state, since, exactly speaking, no energy flow arises in thermal equilibrium states. 
The thermal equilibrium of total system (multi-horizon spacetime) is realized if and only if the temperatures of all constituent horizons are equal. 
Therefore, when the temperatures of horizons are not equal, the multi-horizon spacetime should be understood as it is in a \emph{non-equilibrium} state. 
Here it should be noticed that, generally in non-equilibrium physics, once the system under consideration becomes non-equilibrium, the equation of state for non-equilibrium case takes different form in comparison with that for equilibrium case. 
Especially the non-equilibrium entropy deviates from the equilibrium entropy (when a non-equilibrium entropy is well defined). 
Indeed, although a quite general formulation of non-equilibrium thermodynamics remains unknown at present, the difference of non-equilibrium entropy from equilibrium one is already revealed for some restricted class of non-equilibrium systems~\cite{ref:noneq,ref:noneq.rad}.
Hence, for multi-horizon spacetimes composed of horizons of different temperatures, it seems to be reasonable to expect the breakdown of entropy-area law. 
However, since a ``non-equilibrium thermodynamics'' applicable to multi-horizon spacetime has not been constructed at present, we need to make use of ``equilibrium thermodynamics'' to investigate thermodynamic properties of multi-horizon spacetimes.

Motivated by the above consideration, this article treats Schwarzschild-de~Sitter (SdS) spacetime as the representative of multi-horizon spacetimes~\cite{ref:euclidean.sds}. 
We construct \emph{two thermal equilibrium systems} in SdS spacetime; one of them is for black hole event horizon (BEH) and another is for cosmological event horizon (CEH). 
Note that, since the temperature of BEH is always higher than that of CEH in SdS spacetime~\cite{ref:temperature} (see Sec.\ref{sec:limit}), we need a good way to obtain thermal equilibrium systems of BEH and CEH. 
As will be explained in detail in Sec.\ref{sec:limit}, we will adopt the same way of constructing \emph{two} thermal equilibrium systems as Gibbons and Hawking have used in calculating the Hawking temperatures of BEH and CEH~\cite{ref:temperature}; it is to introduce, between BEH and CEH, a thin wall which reflects perfectly the Hawking radiation coming from the horizons. 
The region enclosed by the wall and BEH (CEH) settles down to a thermal equilibrium state, and we obtain two thermal equilibrium systems separated by the perfectly reflecting wall. 
Then we will examine the entropy-area law for the two thermal equilibrium systems \emph{individually}. 
(Although we are motivated by a non-equilibrium thermodynamic consideration in previous paragraph, the whole analysis in this article is based on equilibrium thermodynamics and we discuss the \emph{two} equilibrium thermodynamics for BEH and CEH individually.) 
As will be explained in Sec.\ref{sec:limit}, our two thermal systems are treated in the canonical ensemble to obtain the free energies of BEH and CEH. 
Hence we will make use of the Euclidean action method which is regarded as one technique to obtain the partition function of canonical ensemble of quantum gravity~\cite{ref:euclidean} 
(see Sec.\ref{sec:pre} or Appendix~A of papers~\cite{ref:euclidean.ds,ref:euclidean.sds}). 
Then we will find that the free energies are functions of three independent state variables. 
The existence of three independent state variables for SdS spacetime was not recognized in existing works on multi-horizon spacetimes~\cite{ref:temperature,ref:sds.special.1,ref:sds.special.2,ref:sds.existing.1,ref:sds.existing.2,ref:sds.existing.3,ref:sds.effective,ref:ss}. 
But in this article, thermodynamically rigorous analysis with three independent state variables will suggest a reasonable evidence of breakdown of entropy-area law for CEH. 
The validity of the law for BEH will not be judged, but we will clarify the key issue for BEH's entropy.

These results imply that the thermal equilibrium of each horizon may not be the necessary and sufficient condition but simply the necessary condition of entropy-area law. 
The necessary and sufficient condition of the law may be implied via some existing works as follows:
Let us note that some proposals for thermodynamics of BEH and CEH in SdS spacetime are already given to a case with some special matter fields and for an extreme case with magnetic/electric charge~\cite{ref:sds.special.1,ref:sds.special.2}. 
These examples are artificial to vanish the temperature difference of BEH and CEH, and show that the entropy-area law holds for SdS spacetime if the temperatures of BEH and CEH are equal. 
However, in this article, we consider a more general case which is not extremal and does not include artificial matter fields. 
In all analyses in this article, the temperatures of BEH and CEH remain different and the discussions in those examples~\cite{ref:sds.special.1,ref:sds.special.2} can not be applied. 
If we find the breakdown of entropy-area law for the case that horizons have different temperatures, then it is suggested that \emph{the necessary and sufficient condition of entropy-area law is the thermal equilibrium of the total system composed of several horizons in which the net energy flow among horizons disappears.}

Here we should make comments on the case that horizons have different temperatures. 
The construction of SdS thermodynamics with leaving horizon temperatures different has already been tried in some existing works~\cite{ref:temperature,ref:sds.existing.1,ref:sds.existing.2,ref:sds.existing.3,ref:sds.effective,ref:ss}. 
Some of those works~\cite{ref:temperature,ref:sds.existing.1,ref:sds.existing.2,ref:sds.existing.3,ref:sds.effective} assume some geometrical conserved quantities to be state variables of SdS spacetime, and derive the so-called mass formula which is simply a geometrical relation and looks similar to the first law of black hole thermodynamics. 
Remaining of those works~\cite{ref:ss} discusses the entropy of SdS spacetime with assuming a naive formula for the horizon entropy which uses the ``effective surface gravity'' obtained via the so-called tunneling method. 
However, the \emph{thermodynamic consistency} has not been confirmed in all of those works. 
Here, ``thermodynamic consistency'' means that the state variables satisfy not only the four laws of thermodynamics but also the appropriate differential relations; for example, the differential of free energy $F_o$ with respect to temperature $T_o$ is equivalent to the minus of entropy, $S_0 \equiv - \partial F_o/\partial T_o$~\footnote{
There are many other similar differential relations in thermodynamics. 
Those relations are the ones required in the ``thermodynamic consistency'', and necessary to understand thermodynamic properties of the system under consideration; e.g. phase transition, thermal and mechanical stabilities, and equations of states.
}. 
It is obvious that thermodynamic entropy should be given in the theory satisfying thermodynamic consistency. 
In other words, if one asserts some theoretical framework to be a ``thermodynamics'', that framework must satisfy the thermodynamic consistency. 
Therefore, exactly speaking, it remains unclear whether those existing works~\cite{ref:temperature,ref:sds.existing.1,ref:sds.existing.2,ref:sds.existing.3,ref:sds.effective,ref:ss} are appropriate as ``thermodynamics''~\footnote{
Some of those existing works~\cite{ref:temperature,ref:sds.existing.1,ref:sds.existing.2,ref:sds.existing.3,ref:sds.effective} preserve/assume the entropy-area law without confirming thermodynamic consistency. Hence, if the breakdown of entropy-area law is concluded via the thermodynamic consistency, those existing works can not be regarded as ``thermodynamic'' theory. 
}. 
On the other hand, it seems to be preferable that the number of assumptions for thermodynamic formulations of BEH and CEH is as small as possible. 
In order to introduce the minimal set of assumptions which preserves thermodynamic consistency, we will refer to Schwarzschild canonical ensemble~\cite{ref:euclidean.sch} for BEH. 
Also we will refer to de~Sitter canonical ensemble for CEH. 
Our conceptual essence is the thermodynamic consistency which gives thermodynamically rigorous formulation to SdS spacetime.

Here let us make a comment that de~Sitter canonical ensemble has not been formulated, while its micro-canonical ensemble has been established. 
Therefore, this article includes the formulation of de~Sitter canonical ensemble~\cite{ref:euclidean.ds} before proceeding to SdS canonical ensemble.

On the other hand, some existing works~\cite{ref:sds.existing.3}, motivated by the so-called dS/CFT correspondence conjecture~\cite{ref:ds/cft}, focus their attention on the future and past null infinities in SdS spacetime (see Fig.\ref{fig:limit-1} shown in Sec.\ref{sec:limit}, $I^{\pm}$ is the null infinities). 
Those infinities may be appropriate to discuss some geometrical quantities. 
However, as implied by the causal structure of SdS spacetime, the future null infinity seems to be inappropriate to discuss thermodynamic properties of BEH and CEH, because any observer near future null infinity (not near the future temporal infinity $i^+$) can not ``access'' BEH~\footnote{
The observer going towards the future temporal infinity $i^+$ in Fig.\ref{fig:limit-1} can ``access'' BEH, since the BEH becomes a boundary of the causally connected region of that observer (the region~I in Fig.\ref{fig:limit-1}).
}.

Hence, contrary to the existing works, the analysis in this article is based on the following two points:
\begin{itemize}
\item 
As will be explained precisely in Sec.\ref{sec:limit}, we focus our attention on the region enclosed by BEH and CEH (not on null infinity) in SdS spacetime as the object of thermodynamic interests. 
\item 
We have a high regard to the ``thermodynamic consistency'' preserved by the minimal set of assumptions without referring to some geometrical conserved quantities and dS/CFT correspondence. 
\end{itemize}
Then, as the result of these two points, the evidence of breakdown of entropy-area law will be obtained.

Here let us emphasize that, in the following sections, we will exhibit explicitly the assumptions on which our discussion is based. 
We think readers can judge the approval or disapproval to every part of our assumptions and analyses. 
Therefore, even if some part of our discussion and analysis is not acceptable for some reader, we hope this article can propose one possible issue about the universality of entropy-area law.

This article is organized as follows: 
In Sec.\ref{sec:pre}, we clarify the conceptual foundation of our discussion and summarize the important tools of our analysis. 
Sec.\ref{sec:york} reviews the Schwarzschild thermodynamics formulated by York~\cite{ref:euclidean.sch}, which is the first thermodynamically rigorous formulation of black hole thermodynamics. 
Our discussion is based on the York's formulation. 
In Sec.\ref{sec:ds} we formulate de~Sitter thermodynamics in the canonical ensemble with referring to York's Schwarzschild thermodynamics. 
The special role of cosmological constant in horizon thermodynamics is also revealed in that section. 
Sec.\ref{sec:limit} is devoted to the evidence of the breakdown of entropy-area law for a multi-horizon spacetime. 
Sec.\ref{sec:conc} is for the conclusion.

Throughout this article we use the Planck units, $c = G = \hbar = k_B = 1$.

\section{Preliminary: Canonical Ensemble and Euclidean Action}
\label{sec:pre}

As mentioned in fourth paragraph of Sec.\ref{sec:intro}, we use the Euclidean action method which is a technique to obtain the partition function of canonical ensemble of quantum gravity~\cite{ref:euclidean}. 
For the first, this section summarizes an important meaning of partition function which forms the conceptual basis of this article. 
Then, a review of Euclidean action method and a summary of useful differential formulas follow.

\subsection{Important Meaning of Partition Function}
\label{sec:pre.partition}

For the aim of this article that is the inspection of entropy-area law in the framework of black hole thermodynamics, it is important to recognize clearly the relation between ``thermodynamics'' and ``statistical mechanics''. 
In statistical mechanics, the partition function can not be expressed as a ``function of state variables'' unless the appropriate state variables, on which the partition function depends, are specified \emph{a priori}~\cite{ref:sm,ref:ll}. 
To understand this, consider for example an ordinary gas in a spherical container of radius $R$, in which the number of constituent particles is $N$, the mass of one particle is $m$ and the average speed of particles is $v$. 
The ordinary statistical mechanics, without the help of thermodynamics, yields the partition function $Z_{\rm gas} = Z_{\rm gas}(R, N, m, v)$ as simply a function of ``parameters'', $R$, $N$, $m$ and $v$. 
\emph{Statistical mechanics, solely, can not determine what combinations of those parameters behave as state variables. 
To determine it, the first law of thermodynamics is necessary~\cite{ref:sm,ref:ll}.} 
(Note that the notion of \emph{heat} in the first law is established by purely the argument in thermodynamics, not in statistical mechanics.) 
Comparing the differential of partition function with the first law results in the identification of partition function with the free energy divided by temperature. 
Then, since the free energy of ordinary gases is a function of the temperature and volume due to the ``thermodynamic'' argument, the partition function $Z_{\rm gas}(R, N, m, v)$ should be rearranged to be a function of temperature and volume $Z_{\rm gas}(V, T)$, where $V = (4 \pi/3) R^3$ and $T = m v^2$ for ideal gases due to the law of equipartition of energy~\footnote{
When the number of particles $N$ changes by, for example, a chemical reaction and an exchange of particles with environment, $N$ is also the state variable on which the free energy depends.
}. (The dependence on $N$ is, for example, $Z_{\rm gas} \propto N$ for ideal gases.)

The reason why the temperature and volume are regarded as the state variables of the gas is that they are consistent with the four laws of ``thermodynamics'' and have the appropriate properties as state variable. 
The appropriate properties are that the state variables are macroscopically measurable, the state variables are classified into two categories, \emph{intensive} variables and \emph{extensive} variables, and the extensive variables are additive. 
Those properties of state variables are specified by purely the argument in thermodynamics, not in statistical mechanics. 
Therefore, from the above, it is recognized that statistical mechanics can not yield the partition function as a ``function of appropriate state variables'' without the help of thermodynamics which specifies the appropriate state variables for the partition function.

Turn our discussion to the Euclidean action method for curved spacetimes. 
Since the Euclidean action method is the technique to obtain the ``partition function'' of the spacetime under consideration (see next subsection), it is necessary to specify the state variables before calculating the Euclidean action. 
In this article, we formulate the canonical ensemble for de~Sitter thermodynamics in Sec.\ref{sec:ds} with referring to Schwarzschild canonical ensembles~\cite{ref:euclidean.sch} summarized in Sec.\ref{sec:york}, and then introduce the minimal set of assumptions for SdS thermodynamics in Sec.\ref{sec:limit}, which specify the appropriate state variables for the partition function. 
The special role of cosmological constant~\cite{ref:euclidean.ds,ref:euclidean.sds,ref:sds.effective} is also clarified in those discussion.

\subsection{Euclidean Action Method}
\label{sec:pre.euclidean}

The Euclidean action method for systems including gravity is originally introduced by Gibbons and Hawking~\cite{ref:euclidean} in an analogy with the thermal field theory of matter fields in flat spacetime~\cite{ref:tft}. 
For the first, we summarize thermal fields in flat spacetime, and then review its generalization by Gibbons and Hawking.

\subsubsection{Thermal fields in flat spacetime}

Thermal field theory is the statistical mechanics of quantum fields in thermal equilibrium~\cite{ref:tft}. 
The partition function for the canonical ensemble of a field $\phi$ in Minkowski spacetime is defined by the path integral,
\eqb
 Z_{\rm flat} \defeq \int {\mathcal D}\phi\,e^{I_E[\phi]} \, ,
\label{eq:pre.Zflat}
\eqe
where ${\mathcal D}\phi$ is a normalized measure of path integral and $I_E[\phi]$ is the Euclidean action of $\phi$ defined by
\eqb
 I_E[\phi] \defeq i\times \mbox{Lorentzian action with replacing $t$ by $-i\,\tau$} \, ,
\label{eq:pre.IE.flat}
\eqe
where the Lorentzian metric signature is $(- + + +)$, the time coordinate $t$ in the Minkowski spacetime is of ordinary Cartesian coordinates (the time-time component of metric is $-1$), and the replacement of real time $t$ by imaginary time $\tau$ is called the \emph{Wick rotation}. 
By the Wick rotation $t \to - i\, \tau$, the metric in evaluating $I_E[\phi]$ becomes that of flat Euclidean space with signature $(+ + + +)$. 
The ``direction'' of Wick rotation on complexified time plane is ``clockwise'' $t \to -i\,\tau$ (not ``counterclockwise'' $t \to + i\,\tau$) in order to make $Z_{\rm flat}$ correspond to the partition function of (grand-)canonical ensemble in quantum statistics~\cite{ref:tft}. 
In the path integral in Eq.\eref{eq:pre.Zflat}, an appropriate boundary condition is also given to $\phi$ in order to realize a thermal equilibrium state. 
At least, because thermal equilibrium state is static, a periodic boundary condition in the imaginary time direction is required,
\eqb
 \phi(\tau) = \phi(\tau + \beta) \, ,
\eqe
where $\beta$ is the imaginary time period~\footnote{
When the periodic boundary condition in imaginary time is not required, the path integral in Eq.\eref{eq:pre.Zflat} describes an ordinary transition amplitude of $\phi$ in ordinary quantum field theory of zero temperature.
}. 
With this condition, it has already been known~\cite{ref:tft} that $Z_{\rm flat}$ corresponds to the partition function of canonical ensemble of equilibrium temperature $T_{\rm flat}$ defined by
\eqb
 T_{\rm flat} \defeq \dfrac{1}{\beta} \, .
\eqe
$Z_{\rm flat}$ describes thermal equilibrium state of $\phi$ of equilibrium temperature $T_{\rm flat}$ in Minkowski spacetime, and the free energy $F_{\rm flat}$ of the equilibrium state is obtained,
\eqb
 F_{\rm flat} = -T_{\rm flat}\,\ln Z_{\rm flat} \, .
\eqe

\subsubsection{Curved spacetime and thermal fields on it}

In curved spacetime, we consider a thermal equilibrium state of the combined system of spacetime and matter field. 
For the canonical ensemble of our combined system, it is usually assumed that the partition function $Z$ is obtained by replacing flat metric in Eq.\eref{eq:pre.Zflat} with curved one~\cite{ref:euclidean},
\eqb
 Z \defeq \int {\mathcal D}g_E \cdot {\mathcal D}\phi\,e^{I_E[g_E,\phi]} \, ,
\label{eq:pre.Z}
\eqe
where
\eqb
 I_E[g_E,\phi] \defeq
 i\times \mbox{$I_{\rm tot}[g,\phi]$ with Wick rotation $t \to -i\,\tau$} \, ,
\label{eq:pre.IE.curved}
\eqe
where $I_{\rm tot}[g,\phi]$ is the Lorentzian action explained below, and $g_E$ is the Euclidean metric of signature $(+ + + +)$ obtained from the Lorentzian metric $g$ by the Wick rotation $t \to - i \tau$. 
Note that, since the spacetime metric $g$ is also assumed to be quantum metric, $g_E$ appears as an integral variable in the path integral~\eref{eq:pre.Z}. 
The action is given as
\eqb
\label{eq:pre.Itot}
 I_{\rm tot}[g,\phi] \defeq I_{\rm matter}[g,\phi] + I_L[g] \,,
\eqe
where $I_{\rm matter}[g,\phi]$ is the Lorentzian matter action, and $I_L[g]$ is the Lorentzian Einstein-Hilbert action defined as
\eqb
\label{eq:pre.IL}
 I_L \defeq
  \dfrac{1}{16 \pi}\int_{\mathcal{M}} dx^4\,\sqrt{-\det g}\,\left( \mathcal{R} - 2\,\Lambda \right)
 + \dfrac{1}{8\,\pi}\int_{\partial \mathcal{M}} dx^3\,\sqrt{\det h}\,K
 + I_{\rm sub} \, ,
\eqe
where $\mathcal{M}$ is the spacetime region under consideration, $\mathcal{R}$ is the Ricci scalar, $\Lambda$ is the cosmological constant, $h$ and $K$ in the second term are respectively the first fundamental form (induced metric) and the trace of second fundamental form (extrinsic curvature) of the boundary ${\partial \mathcal{M}}$, and $I_{\rm sub}$ is the integration constant of $I_L$ which is sometimes called the subtraction term. 
The second term $\int_{\partial M}$ in Eq.\eref{eq:pre.IL} is required to eliminate the second derivatives of metric from the action~\cite{ref:action}. 
The third term $I_{\rm sub}$ does not contribute to the Einstein equation obtained by $\delta_g I_{\rm tot} = 0$.

In order to consider equilibrium states of spacetime with matter field, the periodic boundary condition in imaginary time is required for not only $\phi$ but also $g_E$,
\eqb
g_{E\,\mu \nu}(\tau) = g_{E\,\mu \nu}(\tau + \beta) \, .
\eqe
Here the equilibrium temperature of $g$ and $\phi$ is not defined simply by $\beta^{-1}$, because the spacetime is curved. 
Instead of the simple inverse $\beta^{-1}$, the temperature $T$ should be defined by the proper length in the Euclidean space of $g_E$ in the imaginary time direction,
\eqb
 T \defeq \left[\, \int_0^{\beta}\,\sqrt{g_{E\,\tau \tau}}\,d\tau \,\right]^{-1} \, .
\label{eq:pre.T}
\eqe
Since the metric component $g_{E\,\tau \tau}$ is a function of spacetime coordinates, the integral in Eq.\eref{eq:pre.T} becomes a function of spatial coordinates. 
Therefore it is important to specify where the temperature is defined. 
Here note that the Euclidean action method is for the canonical ensemble. 
This implies the existence of a heat bath whose temperature coincides with the temperature of the system under consideration, since the system is in a thermal equilibrium with the heat bath. 
Therefore it is reasonable to evaluate $g_{E\,\tau \tau}$ in Eq.\eref{eq:pre.T} at the contact surface of the system with the heat bath. 
The contact surface is the boundary of the spacetime region. 
Hence the temperature $T$ should be evaluated at the spacetime boundary.

In the path integral in Eq.\eref{eq:pre.Z}, the metric and matter field are not necessarily solutions of classical Einstein equation and field equations. 
However when the field $\phi$ is weak enough, the dominant contribution would come from the classical solutions, $g_{cl}$ and $\phi_{cl}$, and we can expand as
\eqb
\label{eq:pre.expand.g}
 g_{\mu\nu} = g_{cl\,\mu\nu} + \delta g_{\mu\nu} \quad,\quad
 \phi = \phi_{cl} + \delta \phi \, ,
\eqe
where $\delta g$ and $\delta \phi$ describe quantum/statistical fluctuations of metric and matter. 
In spacetimes with event horizon, this expansion seems reasonable since the Hawking temperature is usually very low and the matter field $\phi$ of Hawking radiation is weak. 
Then the Euclidean action becomes
\eqb
 I_E[g_E,\phi] = I_E[g_{E\,cl},\phi_{cl}] + I_E[\delta g_E] + I_E[\delta \phi]
             + \mbox{higher order terms} \, ,
\label{eq:pre.expand.IE}
\eqe
where $g_{E\,cl}$ is the Euclidean metric obtained from $g_{cl}$, and the second and third terms are quadratic in fluctuations by definition of classical field equations. 
The partition function becomes
\eqb
 \ln Z = I_E[g_{E\,cl},\phi_{cl}] + \ln\int \mathcal{D}(\delta g_E)\,e^{I_E[\delta g_E]}
         + \ln\int \mathcal{D}(\delta \phi)\,e^{I_E[\delta \phi]} + \cdots \, .
\eqe
The leading term $I_E[g_{E\,cl},\phi_{cl}]$ includes only the classical solutions. 
Hence the partition function $Z_{cl}$ of the thermal equilibrium state of background classical spacetime and matter is defined by
\eqb
 \ln Z_{cl} \defeq I_E[g_{E\,cl},\phi_{cl}] \, .
\label{eq:pre.Zcl}
\eqe
The state variables obtained from $Z_{cl}$ describe thermal equilibrium states of background spacetime and matter. 
For spacetimes with event horizon, $Z_{cl}$ is interpreted as the partition function of the event horizon. 
For empty background spacetimes ($\phi_{cl} = 0$) like Schwarzschild and de~Sitter spacetimes, $Z_{cl}$ is determined by only classical metric, $\ln Z_{cl} = I_E[g_{E\,cl}]$. 
This $I_E[g_{E\,cl}]$ describes the canonical ensemble of the thermal equilibrium states of the background classical spacetime, where the thermal equilibrium is achieved by the interaction with the quantum fluctuations of metric. 
Then the free energy of those classical background spacetimes are determined by
\eqb
 F = -T\,\ln Z_{cl} = - T\,I_E[g_{E\,cl}] \, ,
\label{eq:pre.F}
\eqe
where $T$ is defined by Eq.\eref{eq:pre.T} with replacing $g_E$ by $g_{E\,cl}$. 
This $T$ is the equilibrium temperature of the event horizon.

Finally let us make a comment: 
In this article, we take the standpoint that the Euclidean action method is simply one (promising) formalism of obtaining the partition function of spacetimes. 
At present, since we do not know a complete quantum gravity theory, the use of Euclidean action is to be understood as one assumption.

\subsection{Useful Differential Relations}
\label{sec:pre.relations}

Let us exhibit useful differential formulas for calculations of thermodynamic state variables. 
The reader can skip this subsection and return here when the formulas displayed below are referred in the analysis in following sections.

\subsubsection{First case}

Let $f$ be a function of $\alpha_1$, $\alpha_2$ and $\alpha_3$, $f = f(\alpha_1,\alpha_2,\alpha_3)$.
And consider the case that $\alpha_i$ ($i = 1$,~$2$,~$3$) are also functions of $y_1$, $y_2$ and $y_3$, $\alpha_i = \alpha_i(y_j)$ ($j = 1$,~$2$,~$3$). 
Then define $f(y_1,y_2,y_3) \defeq f(\,\alpha_i(y_j)\,)$. 
Let us aim to express the partial derivatives $\partial f(\alpha_1,\alpha_2,\alpha_3)/\partial \alpha_i$ by the derivatives with respect to $y_j$. 
Standard differential calculus gives, $\partial_{y_j}f = \sum_{i=1}^3 (\partial_{\alpha_i}f) \, (\partial_{y_j}\alpha_i)$, which is expressed in vector form as
\eqb
 \begin{pmatrix}
   \partial_{y_1}f \cr
   \partial_{y_2}f \cr
   \partial_{y_3}f
 \end{pmatrix}
 =
 P\, \begin{pmatrix}
       \partial_{\alpha_1}f \cr
       \partial_{\alpha_2}f \cr
       \partial_{\alpha_3}f
      \end{pmatrix}
 \quad,\quad
 P \defeq
 \begin{pmatrix}
   \partial_{y_1}\alpha_1 & \partial_{y_1}\alpha_2 & \partial_{y_1}\alpha_3 \cr
   \partial_{y_2}\alpha_1 & \partial_{y_2}\alpha_2 & \partial_{y_2}\alpha_3 \cr
   \partial_{y_3}\alpha_1 & \partial_{y_3}\alpha_2 & \partial_{y_3}\alpha_3
 \end{pmatrix}
 \, .
\label{eq:pre.vec}
\eqe
Then, when $\det P \neq 0$, we obtain
\eqb
 \begin{pmatrix}
   \partial_{\alpha_1}f \cr
   \partial_{\alpha_2}f \cr
   \partial_{\alpha_3}f
 \end{pmatrix}
 =
 P^{-1}\,
 \begin{pmatrix}
   \partial_{y_1}f \cr
   \partial_{y_2}f \cr
   \partial_{y_3}f
 \end{pmatrix}    \, .
\label{eq:pre.3parameters}
\eqe

\subsubsection{Second case}

Use the same definitions with previous subsection. 
If $f$ has no $\alpha_3$-dependence ($f = f(\alpha_1,\alpha_2)$\,) and $\alpha_i$ has no $y_3$-dependence ($\alpha_i = \alpha_i(y_1,y_2)$, $i = 1$, $2$), then Eq.\eref{eq:pre.3parameters} reduces to
\eqb
 \pd{f(\alpha_1,\alpha_2)}{\alpha_1}
 = \dfrac{(\partial_{y_1} f)\,(\partial_{y_2} \alpha_2)
         - (\partial_{y_2} f)\,(\partial_{y_1} \alpha_2)}
        {(\partial_{y_1} \alpha_1)\,(\partial_{y_2} \alpha_2)
         - (\partial_{y_2} \alpha_1)\,(\partial_{y_1} \alpha_2)}
\label{eq:pre.2parameters} \, ,
\eqe
and a similar formula given by exchanging $\alpha_1$ and $\alpha_2$.

\subsubsection{Third case}

Use the same definitions with previous subsection. 
If $f$ has no $\alpha_3$-dependence ($f = f(\alpha_1,\alpha_2)$\,) and $\alpha_2$ has no $y_2$- and $y_3$-dependence ($\alpha_2 = \alpha_2(y_1)$\,) while $\alpha_1$ depends on all of $y_j$ ($\alpha_1 = \alpha_1(y_1,y_2,y_3)$\,), then Eq.\eref{eq:pre.vec} reduces to
\eqab
 \begin{pmatrix}
   \partial_{y_1}f \cr
   \partial_{y_2}f \cr
   \partial_{y_3}f
 \end{pmatrix}
 =
 \begin{pmatrix}
   \partial_{y_1}\alpha_1 & \partial_{y_1}\alpha_2 \cr
   \partial_{y_2}\alpha_1 & 0 \cr
   \partial_{y_3}\alpha_1 & 0
 \end{pmatrix}\,
 \begin{pmatrix}
   \partial_{\alpha_1}f \cr
   \partial_{\alpha_2}f
 \end{pmatrix} \, .
\eqae
This gives
\eqab
 \pd{f(\alpha_1,\alpha_2)}{\alpha_1}
 &=& \dfrac{\partial_{y_2}f}{\partial_{y_2}\alpha_1}
     = \dfrac{\partial_{y_3}f}{\partial_{y_3}\alpha_1}
\label{eq:pre.3rd} \\
 \pd{f(\alpha_1,\alpha_2)}{\alpha_2}
 &=& \dfrac{(\partial_{y_1} f)\,(\partial_{y_2} \alpha_1)
            - (\partial_{y_2} f)\,(\partial_{y_1} \alpha_1)}
          {(\partial_{y_1} \alpha_2)\,(\partial_{y_2} \alpha_1)} \\
 &=&  \dfrac{(\partial_{y_1} f)\,(\partial_{y_3} \alpha_1)
            - (\partial_{y_3} f)\,(\partial_{y_1} \alpha_1)}
          {(\partial_{y_1} \alpha_2)\,(\partial_{y_3} \alpha_1)}
\nonumber \, .
\eqae
Furthermore, if $f$ depends only on $\alpha_1$ ($f = f(\alpha_1)$\,), then Eq.\eref{eq:pre.3rd} reduces to
\eqb
 \pd{f(\alpha_1)}{\alpha_1}
 = \dfrac{\partial_{y_1}f}{\partial_{y_1}\alpha_1}
 = \dfrac{\partial_{y_2}f}{\partial_{y_2}\alpha_1}
 = \dfrac{\partial_{y_3}f}{\partial_{y_3}\alpha_1}
\label{eq:pre.3rd.2} \, .
\eqe

\section{York's Formulation of Schwarzschild Thermodynamics}
\label{sec:york}

The first thermodynamically rigorous formulation of black hole thermodynamics was given by York for Schwarzschild black hole~\cite{ref:euclidean.sch}. 
As discussed in following sections, by referring to the York's Schwarzschild thermodynamics, we can learn the minimal set of assumptions of de~Sitter and SdS themodynamics which provide the appropriate \emph{state variables} of partition function (Euclidean action). 
This section summarizes the essence of York's theory~\cite{ref:euclidean.sch}.

\begin{figure}[t]
 \begin{center}
 \includegraphics[height=30mm]{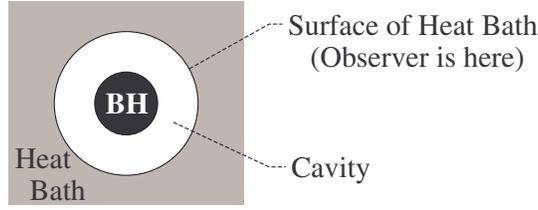}
 \end{center}
\caption{Schematic image of thermal equilibrium of black hole with heat bath. This is described in the canonical ensemble. State variables of black hole are defined at the surface of heat bath. With those state variables, the consistent thermodynamic formulation is realized using the Euclidean action method~\cite{ref:euclidean.sch}.}
\label{fig:york-1}
\end{figure}

There are three key points in the canonical ensemble for Schwarzschild thermodynamics. 
The first one is the zeroth law which describes the existence and construction of thermal equilibrium states: 
\begin{key}[Zeroth law of black hole] 
Place a black hole in a spherical cavity as shown in Fig.\ref{fig:york-1} and also the observer at the surface of the heat bath. 
Through the Hawking radiation by black hole and the black body radiation by heat bath, the black hole interacts with the heat bath. 
Then, by appropriately adjusting the temperature of heat bath, the combined system of black hole and heat bath settles down to a thermal equilibrium state.
\end{key}

The equilibrium state of black hole under the contact with heat bath is described in the canonical ensemble. 
And the equilibrium state variables of black hole are defined by the quantities measured at the surface of heat bath where the observer is. 
Then the ``thermodynamically consistent'' Schwarzschild canonical ensemble is constructed as follows.

The second key point is the difference of black hole thermodynamics from thermodynamics of ordinary laboratory systems: 
\begin{key}[Peculiar scaling law of black hole] 
Extensive and intensive state variables of black hole show a peculiar scaling law: 
When a length size $L$ (e.g. event horizon radius) is scaled as $L \to \lambda\,L$ with an arbitrary scaling rate $\lambda \, (>0)$, then the extensive variables $X$ (e.g. entropy) and intensive variables $Y$ (e.g. temperature) are scaled as $X \to \lambda^2\,X$ and $Y \to \lambda^{-1}\,Y$, while the thermodynamic functions $\Phi$ (e.g. free energy) are scaled as $\Phi \to \lambda\,\Phi$. 
This implies that, because the system size is one of the extensive variables, the thermodynamic system size of equilibrium system constructed in the key point~1 should have the areal dimension. 
Indeed the surface area of heat bath, $4 \pi r_w^2$, behaves as the consistent extensive variable of the system size, where $r_w$ is the radius of the surface of heat bath. 
\end{key}

Here recall that, in thermodynamics of ordinary laboratory systems, the intensive variables remain un-scaled under the scaling of system size, the extensive variables scales as the volume, and the thermodynamic functions are the members of extensive variables. 
However, as noted in the key point~2, the black hole thermodynamics has the peculiar scaling law of state variables. 
Although the scaling law differs from that in thermodynamics of ordinary laboratory systems, the peculiar scaling law in black hole thermodynamics retains the thermodynamic consistency as noted in the next key point.

The third key point is the similarity of black hole thermodynamics with thermodynamics of ordinary laboratory systems:
\begin{key}[Euclidean action method and thermodynamic consistency] 
The free energy $F_{\rm BH}$ of black hole is yielded by the Euclidean action method, where the integration constant (the so-called subtraction term) of the action integral is determined with referring to flat spacetime. 
The action integral is evaluated in the region, $2 M < r < r_w$, which is in thermal equilibrium as noted in the key point~1. 
Here $M$ is the mass parameter.
For the equilibrium system of Schwarzschild black hole constructed in the key point~1, the Euclidean action in Eq.\eref{eq:pre.Zcl} becomes
\eqb
\label{eq:york.IE}
 I_{E {\rm (BH)}} =
 4 \pi M\,\left[\, M - 2\,r_w\,\left(\, 1-2M/r_w - \sqrt{1-2M/r_w} \,\right) \,\right] \,.
\eqe
The free energy is given by Eq.\eref{eq:pre.F}. 
Then, as for the ordinary thermodynamics, this free energy is expressed as a function of two independent state variables, temperature and system size;
\eqb
 F_{\rm BH}(T_{\rm BH} , 4 \pi r_w^2) \,,
\eqe
where the intensive variable $T_{\rm BH} \defeq \left(8 \pi M \sqrt{1-2 M/r_w}\right)^{-1}$ is the Hawking temperature measured by the observer at $r_w$, and the factor $\sqrt{1-2 M/r_w}$ is the so-called Tolman factor~\cite{ref:tolman} which expresses the gravitational redshift affecting the Hawking radiation propagating from the black hole horizon to the observer. 
This Hawking temperature is obtained by Eq.\eref{eq:pre.T}. 
In order to let $T_{\rm BH}$ and $4 \pi r_w^2$ be independent variables in $F_{\rm BH}$, the mass parameter $M$ and the heat bath radius $r_w$ are regarded as two independent variables. 
Then the thermodynamic consistency holds as follows: 
The entropy $S_{\rm BH}$ and the ``surface pressure'' $\sigma_{\rm BH}$ are defined by
\eqb
\label{eq:york.S.sigma_BH}
 S_{\rm BH} \defeq -\pd{F_{\rm BH}(T_{\rm BH} , 4 \pi r_w^2)}{T_{\rm BH}}
                     = \dfrac{A_{\rm BH}}{4} \quad,\quad
 \sigma_{\rm BH} \defeq -\pd{F_{\rm BH}(T_{\rm BH} , 4 \pi r_w^2)}{ (4 \pi r_w^2)} \,,
\eqe
where $\sigma_{\rm BH}$ has the dimension of force par unit area because the system size $4 \pi r_w^2$ has the dimension of area. 
See Appendix~B in reference~\cite{ref:euclidean.ds} for a detail explanation of thermodynamic meaning of $\sigma_{\rm BH}$. 
(The temperature of heat bath should be adjusted to be $T_{\rm BH}$ in the key point~1.) 
These differential relations among the free energy, entropy and surface pressure are the same with those obtained in thermodynamics of ordinary laboratory systems. 
Furthermore, as for the ordinary thermodynamics, the internal energy and the other thermodynamic functions are defined by the Legendre transformation of the free energy; for example the internal energy $E_{\rm BH}$ is
\eqb
 E_{\rm BH}(S_{\rm BH},4 \pi r_w^2) \defeq F_{\rm BH} + T_{\rm BH}\,S_{\rm BH} \,.
\eqe
The enthalpy, Gibbs energy and so on are also defined by the Legendre transformation. 
Then the differential relations among those thermodynamic functions and the other state variables also hold, for example $T_{\rm BH} \equiv \partial E_{\rm BH}/\partial S_{\rm BH}$. 
Furthermore, with the state variables obtained above, we can check that the first, second and third laws of thermodynamics hold for black holes. 
\end{key}

The above three key points hold also for the other single-horizon black hole spacetimes, and those black hole thermodynamics has already been established~\cite{ref:euclidean.sch,ref:euclidean.kn,ref:euclidean.sads}.

Here let us remark about the heat bath introduced in the key point~1. 
In York's consistent black hole thermodynamics~\cite{ref:euclidean.sch}, the heat bath is essential to establish the thermodynamic consistency in the canonical ensemble as explained below: 
Generally in thermodynamics, as noted in the key point~3, thermodynamic functions are defined as a function of two independent state variables. 
Especially the free energy should be expressed as a function of the temperature and the extensive state variable which represents the system size. 
This thermodynamic requirement is satisfied by introducing the heat bath, which gives us two independent variables; the mass parameter $M$ and the radius of heat bath $r_w$. 
These two independent variables make it possible to define the temperature $T_{\rm BH}$ and the surface area $4 \pi r_w^2$ as the two independent state variables in free energy $F_{\rm BH}(T_{\rm BH},4\pi r_w^2)$. 
Therefore the heat bath is necessary to establish manifestly the thermodynamic consistency.

Concerning the heat bath, let us make another comment here. 
It is possible to take the limit $r_w \to \infty$ after constructing the consistent black hole thermodynamics with the heat bath of finite $r_w$.
Here one may think that the limit $r_w \to \infty$ corresponds to the micro-canonical ensemble, since state variables are expressed as functions of only one parameter $M$ and the heat bath seems to disappear (run away to infinitely distant region). 
However it should be emphasized that, generally in statistical mechanics, the micro-canonical ensemble is not some limiting case of the canonical ensemble. 
Therefore the limit $r_w\to\infty$ does not mean to consider the micro-canonical ensemble (the system without heat bath), but it is just the large limit of the cavity size in the canonical ensemble. 
Hence the black hole thermodynamics have been established in the canonical ensemble.

\section{de~Sitter Thermodynamics in the Canonical Ensemble}
\label{sec:ds}

Recall that the aim of this article is to inspect the entropy-area law for SdS spacetime~\cite{ref:euclidean.sds}, in which we will refer to the canonical ensembles for Schwarzschild and de~Sitter spacetimes. 
However, the established formalism of cosmological event horizon (CEH) in de~Sitter spacetime is based on the micro-canonical ensemble~\cite{ref:sds.special.2,ref:micro}. 
Here note that, generally in the ordinary thermodynamics and statistical mechanics, both of the micro-canonical and canonical ensembles yield the same equation of state for any thermodynamic system. 
This implies the existence of de~Sitter canonical ensemble. 
Hence, in this section, we re-formulate the thermodynamics of de~Sitter CEH in the canonical ensemble~\cite{ref:euclidean.ds}.

\subsection{Micro-Canonical Ensemble}
\label{sec:ds.micro}

For the first in this section, let us summarize the micro-canonical ensemble of de~Sitter horizon. 
Thermodynamics of CEH in de~Sitter spacetime can be formulated for the region bounded by the CEH solely (the region I in Fig.\ref{fig:ds-1}), which is filled with the Hawking radiation of CEH and settles down to a thermal equilibrium state without introducing a heat bath (see Sec.\ref{sec:ds.action} for a summary of de~Sitter geometry). 
This region can be regarded as an isolated thermodynamic system without the contact with heat bath. 
This implies that thermodynamics of CEH can be established in the micro-canonical ensemble. 
Indeed, Hawking and Ross~\cite{ref:sds.special.2} have proposed that, when the thermal equilibrium system of a horizon is isolated and has no contact with heat bath, the Euclidean action $I_E^{\rm (micro)}$ of such system can be interpreted as the number of micro-states $W$ or the density of micro-states~\cite{ref:micro} of underlying quantum gravity, which satisfies $W = e^{I_E^{\rm (micro)}}$. 
Then, because $I_E^{\rm (micro)} = \pi \rds^2$ is obtained (in Sec.\ref{sec:ds.action}) for the isolated system of de~Sitter CEH of radius $\rds$, the entropy-area law can be obtained by the Boltzmann's relation, 
\eqb
\label{eq:ds.micro.Sds}
 \Sds \defeq \ln W = \pi \rds^2 \,,
\eqe
where $\Sds$ is the entropy of CEH. 
The entropy-area law for CEH has already been verified in the micro-canonical ensemble. 
This relation~\eref{eq:ds.micro.Sds} should be obtained also in the canonical ensemble of de~Sitter CEH.

\subsection{Basic Assumptions of de~Sitter Canonical Ensemble}
\label{sec:ds.assumptions}

Let us introduce the basic assumptions of de~Sitter canonical ensemble. 
Those assumptions give us the way to construct thermal equilibrium state under the contact with heat bath, and specifies the appropriate state variables for the partition function (see Sec.\ref{sec:pre.partition}). 
Furthermore, to make our discussion exact logically, the use of Euclidean action method will also be listed as one assumption (see the comment at the end of Sec.\ref{sec:pre.euclidean}).

\begin{figure}[t]
 \begin{center}
 \includegraphics[height=30mm]{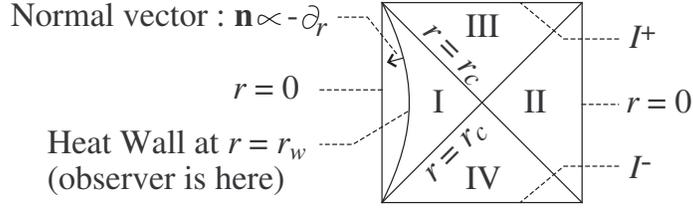}
 \end{center}
\caption{Penrose diagram of de~Sitter spacetime for $r \ge 0$. $I^{\pm}$ is the future/past null infinity. $\rds$ is the CEH radius. The heat wall of radius $r_w$ is placed at the center of the region surrounded by CEH. Our observer is at the wall. This wall reflects perfectly the Hawking radiation, and the observer sees that the region enclosed by the wall and CEH is in a thermal equilibrium state.}
\label{fig:ds-1}
\end{figure}

The CEH in de~Sitter spacetime is a spherically symmetric null hyper-surface defined as the boundary of causal past of the observer's world line~\cite{ref:temperature}. 
The observer detects the Hawking radiation of thermal spectrum emitted by the CEH~\cite{ref:temperature}. 
This implies that the observer can regard the CEH as an object in thermal equilibrium. 
It is expected that a thermodynamically consistent canonical ensemble of CEH can be constructed in the way similar to the Schwarzschild canonical ensemble. 
Then, referring to the key point~1 of Schwarzschild thermodynamics, the first basic assumption of de~Sitter canonical ensemble is the zeroth law:
\begin{ass-ds}[Zeroth law of CEH]
Place a spherically symmetric thin wall of radius $r_w$ at the center of the region surrounded by the CEH as shown in Fig.\ref{fig:ds-1}. 
The wall is smaller than CEH, $r_w < \rds$, where $\rds$ is the CEH radius. 
The mass energy of this wall is negligible, and the geometry of the region $r_w < r < \rds$ is of de~Sitter spacetime. 
Let the wall reflect perfectly the Hawking radiation, and we call this perfectly reflecting wall the ``heat wall'' hereafter. 
We put our observer at the heat wall. 
Then, this observer sees that the region enclosed by the heat wall and CEH settles down to a thermal equilibrium state. 
The equilibrium state variables of CEH are defined by the quantity measured at the heat wall where the observer is.
\end{ass-ds}

Since this equilibrium state of CEH has the contact with the heat wall, we expect that de~Sitter thermodynamics is obtained in the canonical ensemble. 
However before calculating the partition function, as explained in Sec.\ref{sec:pre.partition}, we have to specify the appropriate state variables for the partition function, the temperature and extensive variable of system size. 
To do so, it is necessary to clarify the notion of extensivity and intensivity, which is the significant scaling property of state variables under the scaling of system size. 
The scaling behavior provides a guideline to specify the state variable of system size, and also provides the basis to adopt the temperature defined in Eq.\eref{eq:pre.T} as an intensive state variable.

Concerning the extensivity and intensivity, note that, in general, thermodynamics seems to be a very universal formalism which can be applied to any system if it is in thermal equilibrium. 
This implies that the scaling behavior of state variables shown in the key point~2 of Schwarzschild thermodynamics is common to any horizon system. 
Then, the second assumption of de~Sitter canonical ensemble is as follows:
\begin{ass-ds}[Scaling law and system size of CEH]
State variables of CEH are classified into three categories, extensive variables, intensive variables and thermodynamic functions, and those state variables satisfy the same scaling law as explained in the key point~2: 
When a length size $L$ (e.g. CEH radius) is scaled as $L \to \lambda\,L$ with an arbitrary scaling rate $\lambda \, (>0)$, then the extensive variables $X$ (e.g. system size) and intensive variables $Y$ (e.g. temperature) are scaled as $X \to \lambda^2\,X$ and $Y \to \lambda^{-1}\,Y$, while the thermodynamic functions $\Phi$ (e.g. free energy) are scaled as $\Phi \to \lambda\,\Phi$. 
This implies that the thermodynamic system size of the equilibrium system constructed in assumption~dS-1 should have the areal dimension, and the surface area of heat wall, $\Ads \defeq 4 \pi r_w^2$, behaves as the extensive variable of system size. 
(Indeed, it will be shown later that thermodynamically consistent formulation of de~Sitter canonical ensemble can be established with using $\Ads$.) 
\end{ass-ds}

The concrete functional form of free energy can not be determined by only the assumption~dS-2. 
But, as implied by the key point~3 of Schwarzschild thermodynamics, the Euclidean action method can yield the concrete form of free energy. 
Hence, the third assumption of de~Sitter canonical ensemble is the declaration of using the Euclidean action method:
\begin{ass-ds}[Euclidean action and State variables of CEH] 
Euclidean action $I_E$ of a thermal equilibrium state of CEH yields the partition function of  canonical ensemble via Eq.\eref{eq:pre.Zcl}, where the integration in $I_E$ is calculated over the region enclosed by the heat wall and CEH ($r_w < r < \rds$). 
And the free energy $\Fds(\Tds,\Ads)$ of CEH is defined by Eq.\eref{eq:pre.F}, where $\Tds$ is the temperature of CEH determined by Eq.\eref{eq:pre.T} and $\Ads$ is the area of heat wall discussed in assumption~dS-2. 
Once $\Fds$ is determined, any state variable of CEH is defined from $\Fds$ in the same way with the state variable in thermodynamics of ordinary laboratory systems. 
For example, CEH entropy $\Sds$ is defined by $\Sds \defeq -\partial \Fds/\partial \Tds$.
\end{ass-ds}

Let us emphasize that, exactly and logically speaking, once these assumptions are adopted, we also have to adopt supplemental working hypotheses as explained below.

\subsection{Supplemental Working Hypotheses}
\label{sec:ds.hypotheses}

When we require the thermodynamic consistency, the differential relations like Eq.\eref{eq:york.S.sigma_BH} must be explicitly satisfied. 
And note that, as required in assumption~dS-3, the free energy $\Fds$ should be a function of two independent variables $\Tds$ and $\Ads$. 
That is, two independent state variables must exist. 
On the other hand, we have only two parameters $\Lambda$ and $r_w$ in our equilibrium system constructed in assumption~dS-1. 
Hence we have to adopt the following working hypothesis:
\begin{hypo-ds}[Two independent variables]
To construct de~Sitter canonical ensemble in a thermodynamically consistent way, we require that the cosmological constant $\Lambda$ is an independent ``working variable''. 
Then, we have two independent variables $\Lambda$ and $r_w$ which can ensure the free energy to be a function of two independent state variables. 
When we regard non-variable $\Lambda$ as the physical situation, it is obtained by the ``constant $\Lambda$ process'' in the ``generalized'' de~Sitter thermodynamics in which $\Lambda$ is regarded as a working variable to ensure the thermodynamic consistency.
\end{hypo-ds}

A verification of this working hypothesis will be discussed later in Sec.\ref{sec:ds.lambda}. 
This working hypothesis can be combined with the assumption~dS-3 to obtain a ``complete free energy'' as a function of two independent state variables. 
Here, in order to emphasize the role of $\Lambda$ as a working variable, we divide the requirement of ``complete free energy'' into the assumption~dS-3 and the working hypothesis~dS-1. 
The assumption~dS-3 and working hypothesis~dS-1 may be combined to form one assumption.

Here let us comment on this working hypothesis. 
This article is not the first which requires the variable $\Lambda$. 
For example, in order to consider a thermodynamic formalism for Schwarzschild-de~Sitter black hole, it has been reported that $\Lambda$ has to be regarded as an variable, otherwise thermodynamic consistency is lost~\cite{ref:sds.existing.3,ref:sds.effective}. 
The variable $\Lambda$ seems to be a practical idea/assumption to treat CEH in thermodynamic framework.

Under the working hypothesis~dS-1, the Euclidean action is to be expressed as a function of two ``independent'' state variables; temperature and surface area of heat wall (see assumption~dS-2). 
In addition, the integration constant $I_{\rm sub}$ in Eq.\eref{eq:pre.IL} should be determined. 
Here recall that, as mentioned at the end of Sec.\ref{sec:ds.micro}, the entropy-area law should be reproduced from our Euclidean action of de~Sitter canonical ensemble, since this law is the equation of state verified already in de~Sitter micro-canonical ensemble~\cite{ref:sds.special.2,ref:micro}. 
This requirement gives us the guiding principle to determine $I_{\rm sub}$, which we summarize in the following working hypothesis:
\begin{hypo-ds}[Consistency with the micro-canonical ensemble] 
The entropy-area law, which is the equation of state verified already in the micro-canonical ensemble, should be reproduced in our de~Sitter canonical ensemble. 
In other words, the integration constant $I_{\rm sub}$ in the action $I_L$ should be determined so as to reproduce the entropy-area law. 
To do so, in calculating the Euclidean action of de~Sitter spacetime, we set for the time being,
\eqb
\label{eq:ds.IEsub}
 I_{E {\rm sub}} = \alpha_w \,I^{\rm (flat)}_E \, ,
\eqe
where $I_{E {\rm sub}}$ is the integration constant in de~Sitter Euclidean action, $I^{\rm (flat)}_E$ is the Euclidean action of flat spacetime ($\Lambda = 0$) whose concrete form will be shown in Sec.\ref{sec:ds.ds}, and $\alpha_w$ is a dimensionless factor. 
In order not to let $I_{E {\rm sub}}$ affect the variational principle in obtaining Einstein equation, the factors $\alpha_w$ and $I^{\rm (flat)}_E$ should be expressed by only the quantity determined at the boundary of Euclidean de~Sitter space (at the heat wall).~\footnote{
Indeed, as will be shown in Sec.\ref{sec:ds.ds}, $I^{\rm (flat)}$ is expressed by only such quantity. 
The concrete form of $\alpha_w$ will be obtained in Sec.\ref{sec:ds.ds}.
}
\end{hypo-ds}

Let us emphasize that Eq.\eref{eq:ds.IEsub} is just a working hypothesis including the unknown factor $\alpha_w$. 
It seems that, even if one starts calculations of de~Sitter Euclidean action with $I_{E {\rm sub}}$ which is proportional (not to $I^{\rm (flat)}_E$ but) to some other action integral determined by only the boundary of Euclidean space, then the requirement of preserving the entropy-area law results in the same form with our $I_{E {\rm sub}}$ obtained in Sec.\ref{sec:ds.ds}. 
The essence of the working hypothesis~dS-2 is not Eq.\eref{eq:ds.IEsub} but the preservation of entropy-area law which is the equation of state verified already in the micro-canonical ensemble. 
This working hypothesis~dS-2 is based on the statistical mechanical requirement that both the micro-canonical and canonical ensembles yield the same equation of state for any thermodynamic system~\cite{ref:micro,ref:sm}.

\subsection{Euclidean Action of de~Sitter canonical ensemble}
\label{sec:ds.action}

\subsubsection{Euclidean de~Sitter space}

The Lorentzian de~Sitter metric in static chart is
\eqb
 ds^2 = -\fds(r)\,dt^2 + \dfrac{dr^2}{\fds(r)} + r^2\,d\Omega^2 \, ,
\eqe
where $d\Omega^2 = d\theta^2 + \sin^2\theta\,d\varphi^2$ is the line element on a unit two-sphere, and
\eqb
 \fds(r) \defeq 1 - H^2\,r^2 \quad,\quad 3\,H^2 = \Lambda \,.
\eqe
The Penrose diagram for $r \ge 0$ is shown in Fig.\ref{fig:ds-1}. 
The static chart with $0 \le r < H^{-1}$ covers the region I or II shown in Fig.\ref{fig:ds-1}. 
The notion of CEH is observer dependent~\cite{ref:temperature}. 
For the observer whose world line is confined in the region I, the radius $\rds$ of CEH is given by $\fds(\rds) = 0$ ;
\eqb
 \rds = \dfrac{1}{H} \, .
\eqe
A Killing vector $\xi \propto \partial_t$ becomes null at the CEH. 
This means that the CEH is the Killing horizon of $\xi$. 
Surface gravity $\kappa$ of the Killing horizon, CEH, is defined by the relation, $\nabla_\xi\,\xi = \kappa\,\xi$ at the CEH. 
The value of $\kappa$ depends on the normalization of $\xi$ by definition~\cite{ref:sg}. 
When the Killing vector is normalized as $\xi = \partial_t$, we get
\eqb
 \kappa = H \, .
\label{eq:ds.sg}
\eqe
This $\kappa$ is the surface gravity of CEH measured by the observer at the central world line $r = 0$, because $t$ is the proper time of the central observer and the norm $\xi^2 = - \fds(r)$ is $-1$ at $r = 0$.

The global chart of Lorentzian de~Sitter spacetime is introduced by the coordinate transformation from $(t,r,\theta,\varphi)$ to $(W,X,\theta,\varphi)$,
\eqb
 W-X \defeq e^{H\,(t-r^{\ast})} \quad,\quad W+X \defeq - e^{-H\,(t+r^{\ast})} \, ,
\label{eq:ds.dS.trans}
\eqe
where $dr^{\ast} \defeq dr/\fds(r)$ which means $r^{\ast} = (1/2 H)\,\ln\left|(1 + H\,r)/(1 - H\,r)\right|$. 
This transformation yields
\eqb
\label{eq:ds.dS.global}
 ds^2 = \dfrac{1}{H^2 \left( W^2 - X^2 -1 \right)^2}
        \left[ -dW^2 + dX^2 + \dfrac{\left( W^2 - X^2 + 1 \right)^2}{4}\,d\Omega^2 \right] \, .
\eqe
The coordinate transformation~\eref{eq:ds.dS.trans} implies the range of coordinates, $X < W < -X$ and $ X < 0$. 
By extending it to the range $-\infty < W < \infty$ and $-\infty < X < \infty$, the global chart covers the whole region of maximally extended de~Sitter spacetime, I, II, III and IV shown in Fig.\ref{fig:ds-1}.

Here concerning the relation of the two charts, note that the Penrose diagram shown in Fig.\ref{fig:ds-1} is depicted according to the global chart~\eref{eq:ds.dS.global}. 
Under the coordinate transformation~\eref{eq:ds.dS.trans}, the direction of timelike Killing vector $\xi \defeq \partial_t$ is future pointing in the region I, and past pointing in the region II. 
Therefore, throughout this section, we consider the region I in using the static chart~\footnote{If the signature of exponent in Eq.\eref{eq:ds.dS.trans} is opposite, $W - X = \exp[-H (t-r^{\ast})]$ and $W + X = -\exp[H (t+r^{\ast})]$, then the direction of $\partial_t$ becomes past pointing in the region I, and future pointing in the region II.}.

Next we proceed to the construction of Euclidean de~Sitter space. 
As explained in Sec.\ref{sec:pre.euclidean}, we apply the Wick rotations $t \to - i \tau$ in the static chart and $W \to - i w$ in the global chart to obtain the Euclidean de~Sitter space. 
These Wick rotations are equivalent, because the transformation~\eref{eq:ds.dS.trans}, $W = e^{-H r^{\ast}}\sinh\left(H t\right)$, implies that the imaginary time $w$ in global chart is defined by $w \defeq e^{-H r^{\ast}}\sin\left(H \tau\right)$, where $\tau$ is the imaginary time in the static chart. 
The Euclidean metric in the static chart is
\eqb
 ds_E^2 = \fds(r)\,d\tau^2 + \dfrac{dr^2}{\fds(r)} + r^2\,d\Omega^2 \, .
\label{eq:ds.dSE.static} \\
\eqe
The Euclidean metric in the global chart is
\eqb
 ds_E^2 = \dfrac{1}{H^2 \left( w^2 + X^2 + 1 \right)^2}
          \left[ dw^2 + dX^2 + \dfrac{\left( w^2 + X^2 - 1 \right)^2}{4}\,d\Omega^2 \right] \, .
\label{eq:ds.dSE.global}
\eqe
About the global chart, we get from the coordinate transformation~\eref{eq:ds.dS.trans},
\eqb
 w^2 + X^2 = \dfrac{1 - H\,r}{1 + H\,r} \, .
\eqe
Because of $w^2 + X^2 \ge 0$, the Euclidean de~Sitter space corresponds to the region I (or~II), $0<r<\rds$, in the Lorentzian de~Sitter spacetime. 
Because the Lorentzian de~Sitter spacetime is regular at $r = 0$ and $r = \rds$, the Euclidean de~Sitter space is also regular at those points. 
To examine the regularity of Euclidean space, we make use of the metric in static chart~\eref{eq:ds.dSE.static}. 
The regularity at $r = 0$ is rather obvious, since the metric near $r = 0$ is flat and regular, $ds_E^2 \simeq d\tau^2 + dr^2 + r^2\,d\Omega^2$.

To examine the regularity of Euclidean space at $r = \rds$, let us define a coordinate $y$
 and a function $b(y)$ by
\eqb
 y^2 \defeq \rds - r \quad,\quad
 b(y) \defeq \sqrt{\fds(\rds - y^2)} \, .
\eqe
We get $b'(y) \defeq d b(y)/dy = 2\,H^2\,r\,y/b(y)$ which denotes
\eqb
 \lim_{y \to 0}b'(y) = 2\,H\,\lim_{y\to 0}\dfrac{y}{b(y)}
                     = 2\,H\,\dfrac{1}{\displaystyle \lim_{y \to 0}b'(y)} \, .
\eqe
This means $b'(0) = \sqrt{2\,H}$, and near the CEH, $\fds \simeq \left[\, b(0) + b'(0)\,y \,\right]^2 = 2\,H\,y^2$. 
Therefore the Euclidean metric near the CEH is
\eqb
 ds_E^2 \simeq
 \dfrac{2}{H}\,\left[\, y^2\,d(H \tau)^2 + dy^2 \,\right] + \dfrac{1}{H^2}\,d\Omega^2 \, .
\eqe
It is obvious that the Euclidean de~Sitter space is regular at CEH if the imaginary time has the period $\beta$ defined by
\eqb
 0 \le \tau < \beta \defeq \dfrac{2\,\pi}{H} \, .
\label{eq:ds.period}
\eqe
Throughout our discussion, the imaginary time $\tau$ has the period $\beta$.

\subsubsection{Euclidean action}

Let us calculate the de~Sitter's Euclidean action $I_E$ defined by Eq.\eref{eq:pre.IE.curved}. 
To obtain $I_E$, we should specify the Lorentzian action $I_L$ given in Eq.\eref{eq:pre.IL}. 
The integral region ${\mathcal M}$ in $I_L$ is the Lorentzian region which forms the thermal equilibrium state constructed in the assumption~dS-1. 
Therefore ${\mathcal M}$ is given by $r_w < r < \rds$, and its boundary $\partial{\mathcal M}$ is at the heat wall, $r = r_w$. 
There is another boundary at $r = \rds$ in the Lorentzian region ${\mathcal M}$. 
However we do not need to consider it, because, as shown above, the points at $r = \rds$ in \emph{Euclidean} space do not form a boundary but are the regular points when the imaginary time $\tau$ has the period~\eref{eq:ds.period}. 
Then the first fundamental form $h_{i j}$ ($i, j = 0, 2, 3$) of $\partial{\mathcal M}$ in the static chart (in Lorentzian de~Sitter spacetime) is
\eqb
 \left. ds^2 \right|_{r = r_w} = h_{i j}\, dx^i \, dx^j
 = - \fds(r_w)\,dt^2 + r_w^2\,d\Omega^2 \, ,
\eqe
where $\fds(r_w) = 1 - H^2\,r_w^2$~. 
Here, since ${\mathcal M}$ is the region enclosed by the heat wall and CEH, the direction of unit normal vector $\mathbf{n}$ to $\partial{\mathcal M}$ is pointing towards the ``center'' $r=0$, which means $\mathbf{n} \propto - \partial_r$ (see Fig.\ref{fig:ds-1}). 
Then the second fundamental form of $\partial{\mathcal M}$ in the static chart is
\eqb
 K_{i j} =
 - \sqrt{\fds(r_w)}\, {\rm diag.}\left[\, H^2\,r_w \,,\, r_w \,,\, r_w\,\sin^2\theta \, \right] \, ,
\label{eq:ds.Kij}
\eqe
where diag. means the diagonal matrix form.

On the other hand, the second fundamental form $K_{i j}^{\rm (flat)}$ of a spherically symmetric timelike hyper-surface $r = r_w$ in Minkowski spacetime is obtained by setting $H=0$ in Eq.\eref{eq:ds.Kij},
\eqb
 K_{i j}^{\rm (flat)} =
 - {\rm diag.}\left[\, 0 \,,\, r_w \,,\, r_w\,\sin^2\theta \, \right] \, ,
\eqe
where the normal vector $\mathbf{n}^{\rm (flat)}$ to this surface is set to be $\mathbf{n}^{\rm (flat)} \propto - \partial_r$ as that in $K_{i j}$. 
For Minkowski spacetime $\mathcal{R} = 0$ and $\Lambda=0$, and its Lorentzian action $I^{\rm (flat)}$ is expressed by only the surface term in Eq.\eref{eq:pre.IL},
\eqb
 I^{\rm (flat)} =
 \dfrac{1}{8 \pi}\int_{\partial \mathcal{M}} dx^3\,\sqrt{\det h\,}\,\,K^{\rm (flat)} \,.
\eqe
In applying this $I^{\rm (flat)}$ to Eq.\eref{eq:ds.IEsub} of working hypothesis~dS-2, the integral element $\sqrt{\det h}$ in $I^{\rm (flat)}$ should be that of de~Sitter spacetime, because the background spacetime on which the integral is calculated is the de~Sitter spacetime.

From the above, applying the Wick rotation $t \to -i \tau$ to the Lorentzian action $I_L$ of de~Sitter spacetime, we obtain the Euclidean action $I_E$ via Eqs.\eref{eq:pre.IE.curved} and~\eref{eq:ds.IEsub},
\eqab
I_E
 &=& \dfrac{3\,H^2}{8\,\pi} \int_{\mathcal M_E}dx_E^4\,\sqrt{g_E}
   + \dfrac{1}{8\,\pi}\int_{\partial\mathcal M} dx_E^3 \,\sqrt{h_E}\,
     \left(\,K_E + \alpha_w\,K_E^{\rm (flat)}\,\right) \nonumber \\
 &=& \dfrac{\pi}{H^2}\,\left(\, 1 - 2\,H\,r_w\,\fds(r_w) - 2\,\alpha_w\,H\,r_w\,\sqrt{\fds(r_w)}\,\right) \, ,
\label{eq:ds.IE_alphaw}
\eqae
where the relation for de~Sitter spacetime $\mathcal{R} = 4\,\Lambda = 12\,H^2$ is used in the first equality, $Q_E$ is the quantity $Q$ evaluated on the Euclidean de~Sitter space, and ${\mathcal M_E}$ is the Euclidean region expressed by $0 \le \tau < \beta$ , $r_w \le r \le \rds$ , $0 \le \theta \le \pi$ and $0 \le \varphi < 2\,\pi$. 
This $I_E$ corresponds to $I_E[g_{E\,cl}]$ in Eq.\eref{eq:pre.F}, which yields the partition function for the thermal equilibrium of spacetime with quantum fluctuations (whose effects are neglected in Eq.\eref{eq:pre.Zcl}). 
A remark on the limit $r_w \to 0$ of this $I_E$ will be given in next subsetion after specifying the form of $\alpha_w$.

Let us make a comment for the micro-canonical ensemble. 
The action $I_E^{\rm (micro)}$ used in Eq.\eref{eq:ds.micro.Sds} is given by only the bulk term (first term) of $I_L$ in Eq.\eref{eq:pre.IL}. 
It yields for Euclidean de~Sitter space,
\eqb
\label{eq:ds.micro.IE}
 I_E^{\rm (micro)} = \pi\,\rds^2 \,,
\eqe
This is used in Eq.\eref{eq:ds.micro.Sds}.

\subsection{de~Sitter Thermodynamics in the Canonical Ensemble}
\label{sec:ds.ds}

In this subsection, principal state variables of CEH are calculated successively, and the ``thermodynamic consistency'' is also shown explicitly.

\subsubsection{Temperature}

As explained in Eq.\eref{eq:pre.T}, the temperature $\Tds$ of CEH is defined by the integral in imaginary time direction at the heat wall,
\eqb
 \Tds \defeq \left[\, \int_0^{\beta}\,\sqrt{\fds(r_w)}\,d\tau \,\right]^{-1}
      = \dfrac{H}{2 \pi \sqrt{\fds(r_w)}} \, ,
\label{eq:ds.T}
\eqe
where $\beta$ is defined in Eq.\eref{eq:ds.period} and $\fds(r_w) = 1 - H^2\,\rds^2$. 
Note that, as implied by Eq.\eref{eq:ds.sg}, $H/2\pi$ coincides with the Hawking temperature measured at $r=0$~\cite{ref:temperature}, and the factor $\sqrt{\fds(r_w)}$ in Eq.\eref{eq:ds.T} is the Tolman factor which expresses the gravitational redshift affecting the Hawking radiation propagating from CEH to the observer~\cite{ref:tolman}. 
This $\Tds$ is the temperature measured by the observer at heat wall.

\subsubsection{Surface area and homothetic variation of the system}

In the ordinary laboratory systems, the size of the system under consideration is its volume. 
For de~Sitter spacetime, the volume can not be defined uniquely, since the choice of spatial slice in the region ${\mathcal M}$ of $r_w < r < \rds$ is not determined in a natural way. 
However, an area can be uniquely and naturally assigned to our system. 
It is the surface area $\Ads$ of the heat wall.

The above discussion supports the assumption~dS-2 in which $\Ads$ is regarded as the state variable. 
Therefore, we adopt $\Ads$ as the extensive state variable of system size,
\eqb
 \Ads \defeq 4\,\pi\,r_w^2 \, .
\label{eq:ds.A}
\eqe
Here we have to note that, as explained in Appendix~B in reference~\cite{ref:euclidean.ds}, the variation of system size is restricted to homothetic variations. 
For our system, the homothetic variation is the spherical variation due to the spherical symmetry.

\subsubsection{Choice of $\alpha_w$ of working hypothesis~dS-2}

This subsection determines the form of $\alpha_w$ which appears in Eq.\eref{eq:ds.IE_alphaw}. 
As required in the working hypothesis~dS-2, $\alpha_w$ is a dimensionless factor composed of the quantity determined at the heat wall. 
This implies that $\alpha_w$ is a function of the parameter,
\eqb
\label{eq:ds.x}
 x \defeq H\,r_w \, ,
\eqe
which means $\fds(r_w) = 1 - x^2$. 
This $x$ is regarded as the dimensionless quantity determined at the heat wall.
Then our Euclidean action $I_E$ is expressed as
\eqb
 I_E = \dfrac{\pi}{H^2}\,\left(\, 1 - 2\,x\,\fds(r_w) - 2\,x\,\alpha_w(x)\,\sqrt{\fds(r_w)} \,\right) \, .
\eqe
In order to determine $\alpha_w(x)$ so as to preserve the entropy-area law which is the equation of state verified already in the micro-canonical ensemble~\cite{ref:sds.special.2,ref:micro}, we need the free energy. 
The free energy $\Fds$ of CEH is obtained via Eq.\eref{eq:pre.F},
\eqb
\label{eq:ds.F_def}
 \Fds(\Tds,\Ads) \defeq 
 - \dfrac{1}{2 H \sqrt{\fds(r_w)}}\,
   \left(\, 1 - 2\,x\,\fds(r_w) - 2\,x\,\alpha_w(x)\,\sqrt{\fds(r_w)} \,\right) \, ,
\eqe
where $H$ and $x$ are regarded as functions of $\Tds$ and $\Ads$. 
Then, by the assumption~dS-3, the entropy $\Sds$ of CEH is defined by,
\eqb
\label{eq:ds.S_def}
 \Sds \defeq - \pd{\Fds(\Tds,\Ads)}{\Tds} = - \dfrac{\partial_H \Fds}{\partial_H \Tds}
 = \dfrac{\pi}{H^2}\,D\left[\,\alpha_w(x)\,\right] \, ,
\eqe
where
\eqb
 D\left[\,\alpha_w(x)\,\right] \defeq
 - 2\,x^2\,\fds(r_w)^{3/2}\,\dfrac{{\rm d}\alpha_w}{{\rm d}x}
 - \left(\, 1 - 2\,x^3 \,\right)\,\fds(r_w) + x^2 \, .
\eqe
Therefore, to preserve the entropy-area law, $\alpha_w(x)$ has to be a solution of the first-order differential equation, $D\left[\,\alpha_w(x)\,\right] = 1$. 
Then we obtain
\eqb
\label{eq:ds.alphaw}
 \alpha_w(x) = \left(\, \dfrac{1}{x} - 1 \,\right)\,\sqrt{\fds(r_w)} + k_w \, ,
\eqe
where $k_w$ is the integration constant. 
The value of $k_w$ can not be determined at present. 
However, as will be shown at Eq.\eref{eq:ds.sigma_kw}, by the existence of surface pressure at a limit $r_w \to 0$, we find $k_w$ is zero,
\eqb
\label{eq:ds.kw}
 k_w = 0 \, .
\eqe
Although the verification of this value is shown later, we proceed our calculations with setting $k_w = 0$ for simplicity of discussion. 
Then, adopting Eq.\eref{eq:ds.kw}, the Euclidean action~\eref{eq:ds.IE_alphaw} is determined,
\eqb
\label{eq:ds.IE}
 I_E = - \dfrac{\pi}{H^2}\,\left( 1 - 2\, x^2 \right) \, .
\eqe

It is conceptually important to consider the ``small heat wall limit'', $r_w \to 0$. 
The Eucliean action $I_E$ given in Eq.\eref{eq:ds.IE} takes the limit value $I_E \to - \pi \rds^2$ as $r_w \to 0$. 
This value differs from $I_E^{\rm (micro)}$ in Eq.\eref{eq:ds.micro.IE} by the negative signature. 
Here one may naively think that our $I_E$ should coincide with $I_E^{\rm (micro)}$ at this limit. 
This naive requirement seems reasonable from the point of view of spacetime geometry, but is not necessarily reasonable from the point of view of statistical mechanics, because the coincidence of $I_E$ with $I_E^{\rm (micro)}$ at the limit $r_w \to 0$ means that the micro-canonical ensemble is some limiting case of the canonical ensemble. 
In statistical mechanics, the micro-canonical ensemble is not some limiting case of the canonical ensemble. 
In de~Sitter thermodynamics, the limit $r_w \to 0$ is just the case of an arbitrarily small heat wall and not the case without heat wall. 
Therefore, in statistical mechanical sense, there seems to be no reason to require that $I_E$ coincides with $I_E^{\rm (micro)}$ at the limit $r_w \to 0$. 
(Concerning this discussion, see also the end of Sec.\ref{sec:york} in which the limiting case of large heat bath for black hole thermodynamics is summarized.)

\subsubsection{Free energy, entropy and second law}

Previous subsection has given us the free energy $\Fds$ and the entropy $\Sds$.
The free energy is
\eqb
\label{eq:ds.F}
 \Fds(\Tds,\Ads) = \dfrac{1 - 2\,x^2}{2\,H\,\sqrt{\fds(r_w)}} \, ,
\eqe
where $x$ is defined in Eq.\eref{eq:ds.x}. 
As noted at the working hypothesis~dS-1, $\Fds$ should be regarded as a function of $\Tds$ and $\Ads$. 
And the entropy is
\eqb
\label{eq:ds.S}
 \Sds = \dfrac{\pi}{H^2} \, .
\eqe

Concerning the entropy, it should be noted that, exactly speaking, the second law is not a ``theorem'' proven by some other assumptions, but the basic assumption which can not be proven in the framework of thermodynamics. 
The best way to believe the second law is to ``check'' the statement of the law for as many processes as we can. 
For the black hole thermodynamics, the generalized second law is checked for some representative processes~\cite{ref:bht.2,ref:gsl}. 
For de~Sitter thermodynamics, we need to check the generalized second law for as many processes as we can. 
However let us expect that the generalized second law holds also for the CEH in de~Sitter spacetime, since our basic assumptions~dS-1,~2, and~3 are the natural extension of consistent black hole thermodynamics.

\subsubsection{Internal energy}

The internal energy $\Eds$ of CEH is defined by the argument of ordinary statistical mechanics,
\eqb
 \Eds \defeq
  - \left.\pd{\ln Z_{cl}}{(1/\Tds)}\right|_{\Ads=\mbox{const.}}
  = \pd{(\Fds/\Tds)}{(1/\Tds)}
  = \Fds + \Tds \, \Sds
\label{eq:ds.E.2}
\eqe
where Eq.\eref{eq:pre.F} is used at the second equality and the definition of $\Sds$ in Eq.\eref{eq:ds.S_def} is used at the third equality. 
The third equality, $\Eds = \Fds + \Tds \, \Sds$, is the Legendre transformation between $\Fds(\Tds,\Ads)$ and $\Eds$. 
This implies that $\Eds$ is a function of $\Sds$ and $\Ads$, which is consistent with the ordinary thermodynamic argument that the internal energy is a function of extensive state variables. 
Then we get
\eqb
\label{eq:ds.E}
 \Eds(\Sds,\Ads) \defeq \dfrac{1}{H}\,\sqrt{\fds(r_w)} \, ,
\eqe
where $H$ and $x$ are regarded as functions of $\Sds$ and $\Ads$ via Eqs.\eref{eq:ds.A},~\eref{eq:ds.x} and~\eref{eq:ds.S}. 
The origin of internal energy $\Eds$ will be discussed later in this subsection.

\subsubsection{Surface pressure}

As explained in Appendix~B of reference~\cite{ref:euclidean.ds}, the conjugate state variable to $\Ads$ is the surface pressure $\sigmads$ defined by
\eqb
\label{eq:ds.sigma_def}
 \sigmads \defeq - \pd{\Fds(\Tds,\Ads)}{\Ads} =
  - \dfrac{(\partial_H \Fds)\,(\partial_{r_w} \Tds) - (\partial_{r_w} \Fds)\,(\partial_H \Tds)}
         {(\partial_H \Ads)\,(\partial_{r_w} \Tds) - (\partial_{r_w} \Ads)\,(\partial_H \Tds)} \, ,
\eqe
where Eq.\eref{eq:pre.2parameters} is used at the second equality.

As mentioned at eq.\eref{eq:ds.kw}, we determine the integration constant $k_w$ here.
To do so, we calculate $\sigmads$ without setting $k_w$ zero. 
From Eqs.\eref{eq:ds.F_def} and~\eref{eq:ds.alphaw}, the free energy $\tilde{F}$ with non-zero $k_w$ is $\tilde{F} = \Fds + k_w\,r_w$, where $\Fds$ is shown in Eq.\eref{eq:ds.F}. 
Then we obtain from Eq.\eref{eq:ds.sigma_def},
\eqb
\label{eq:ds.sigma_kw}
 \sigmads = \dfrac{H}{8\,\pi\,\sqrt{\fds(r_w)}} - \dfrac{k_w}{8\,\pi\,r_w} \, .
\eqe
It is obvious that, if $k_w \neq 0$, then the surface pressure diverges in the limit $r_w \to 0$. 
Hence, when we require the existence of finite $\sigmads$ for thermal equilibrium states of CEH with arbitrarily small heat wall, it is natural to set $k_w = 0$. 
Then, with adopting this choice $k_w = 0$, we obtain
\eqb
\label{eq:ds.sigma}
 \sigmads = \dfrac{H}{8\,\pi\,\sqrt{\fds(r_w)}} = \dfrac{1}{4}\,\Tds \, .
\eqe

\subsubsection{First and third laws}

By definitions of $\Sds$ and $\sigmads$, 
\eqb
 d\Fds(\Tds,\Ads) = - \Sds \,d\Tds - \sigmads\,d\Ads \, .
\eqe
Then the first law holds automatically via the Legendre transformation in Eq.\eref{eq:ds.E.2},
\eqb
 d\Eds(\Sds,\Ads) = d(\Fds + \Tds \,\Sds) = \Tds \,d\Sds - \sigmads\,d\Ads \, .
\eqe

Next, to discuss the third law, note that $\Tds$ is monotonically increasing as a function of $r_w$ as shown by $\partial_{r_w} \Tds(H,r_w) = H^3\,r_w/(2\,\pi\,\fds(r_w)^{3/2}) > 0$. 
The minimum value of $\Tds$ as a function of $r_w$ is given at $r_w = 0$, $\left.\Tds\right|_{r_w\to 0} = H/2\,\pi$. 
This denotes the zero-temperature state is achieved only by the process $H \to 0$ and $r_w \to 0$. 
However it is obvious from Eq.\eref{eq:ds.E} that the infinite energy is required to realize the process $H \to 0$. 
The infinite energy supply is unphysical. 
Hence the third law holds in the sense that the zero-temperature state can not be achieved by any physical process.

\subsubsection{Scaling law, Euler relation and $\Lambda$ as a ``hidden'' variable}

This subsection demonstrates the consistency of the scaling law, which is introduced in the assumption~dS-2, with the other assumptions~dS-1 and~3.

Let us consider the scaling of length size,
\eqb
 r_w \to \lambda\,r_w \quad,\quad
 \rds \to \lambda\,\rds \, .
\eqe
The scaling of CEH radius denotes $H \to \lambda^{-1}\,H$. 
Then from Eqs.\eref{eq:ds.T} and~\eref{eq:ds.sigma}, we get the scaling of intensive variables,
\eqb
 \Tds \to \dfrac{1}{\lambda}\,\Tds \quad,\quad \sigmads \to \dfrac{1}{\lambda}\,\sigmads \, .
\eqe
From Eqs.\eref{eq:ds.A} and~\eref{eq:ds.S}, the scaling of extensive variables is
\eqb
 \Ads \to \lambda^2\,\Ads \quad,\quad \Sds \to \lambda^2\,\Sds \, .
\eqe
From Eqs.\eref{eq:ds.F} and~\eref{eq:ds.E}, we get the scaling of thermodynamic functions,
\eqb
 \Fds \to \lambda\,\Fds \quad,\quad \Eds \to \lambda\,\Eds \, .
\eqe
Therefore we find that the scaling law of assumption~dS-2 is consistent with the assumptions~dS-1 and~3. 
Concerning this consistency, the definition of temperature in Eq.\eref{eq:pre.T} should be emphasized. 
It is obvious that the temperature has the dimension of the inverse of length size by definition. 
Hence, the scaling law of assumption~dS-2 is necessary to adopt Eq.\eref{eq:pre.T} as the temperature which should be intensive.

Furthermore, to show a more robust consistency of the scaling law, recall that the internal energy is a function of $\Sds$ and $\Ads$. 
Then the above scaling law implies
\eqb
\label{eq:ds.euler.1}
 \lambda\,\Eds(\Sds,\Ads) = \Eds(\lambda^2 \Sds,\lambda^2 \Ads) \, .
\eqe
This denotes that $\Eds(\Sds,\Ads)$ is the homogeneous expression of degree $1/2$. 
By the partial differential of this equation with respect to $\lambda$, we get the Euler relation,
\eqb
 \dfrac{1}{2}\,\Eds(\Sds,\Ads) = \Tds\,\Sds - \sigmads\,\Ads \, .
\label{eq:ds.euler.2}
\eqe
Note that the concrete functional forms of state variables is not used in deriving this Euler relation.
The Euler relation~\eref{eq:ds.euler.2} is obtained from the scaling behavior~\eref{eq:ds.euler.1} and the differential relations implied by the first law, $\Tds \equiv \partial \Eds(\Sds,\Ads)/\partial \Sds$ and $\sigmads \equiv -\partial \Eds(\Sds,\Ads)/\partial \Ads$. 
On the other hand, we can check that the concrete forms of state variables $\Tds$, $\Sds$, $\sigmads$ and $\Ads$ obtained in previous subsections satisfy the relations~\eref{eq:ds.euler.1} and~\eref{eq:ds.euler.2}. 
Hence the state variables obtained so far are completely consistent with the scaling law of assumption~dS-2.

Finally in this subsection, recall that the cosmological constant $\Lambda$ is regarded as an independent variable in the working hypothesis~dS-1. 
It is obvious that the assumption~dS-2, via the relation $\Lambda = 3\,H^2$, excludes the ``bare $\Lambda$'' from state variables, since $\Lambda$ is neither intensive nor extensive. 
Hence the bare $\Lambda$ can not be a ``state variable'' but a ``hidden variable'' in the consistent de~Sitter thermodynamics.

\subsubsection{Heat capacity and thermal stability}

The thermodynamic consistency of our de~Sitter canonical ensemble has been clearly checked so far. 
This subsection researches the thermal stability of CEH. 
The appropriate quantity to consider thermal stability is the heat capacity.

The representative heat capacity may be the heat capacity $C_{\Ads}$ at constant $\Ads$. 
Since $\Ads$ depends only on $r_w$, $C_{\Ads}$ describes the response of temperature to the energy supply into CEH with fixing the position of observer $r_w$. 
$C_{\Ads}$ is defined by
\eqb
\label{eq:ds.CA.def}
 C_{\Ads} \defeq \Tds\,\pd{\Sds(\Tds,\Ads)}{\Tds}
  = \Tds\,\dfrac{\partial_H \Sds}{\partial_H \Tds} \, ,
\eqe
where $\Tds$ and $\Ads$ are regarded as independent variables. 
By Eqs.\eref{eq:ds.T} and~\eref{eq:ds.S}, we get
\eqb
 C_{\Ads} = - \dfrac{2\,\pi}{H^2}\,\fds(r_w) \, .
\eqe
Obviously the heat capacity $C_{\Ads}$ is negative definite. 
When the energy is supplied to (extracted from) the CEH, then the temperature $\Tds$ decreases (increases). 
This denotes the CEH is thermally unstable. 
However, when we consider the constant $\Lambda$ process as the physical one, this heat capacity $C_{\Ads}$ is the capacity for unphysical process, since $C_{\Ads}$ is defined by the derivative with respect to $H$ as seen in Eq.\eref{eq:ds.CA.def}. 
The thermal instability due to negative $C_{\Ads}$ seems to be unphysical.

When we consider the constant $\Lambda$ process as the physical one, the heat capacity $C_{\Lambda}$ at constant $\Lambda$ is of interest. 
It is defined by
\eqb
 C_{\Lambda} \defeq \Tds\,\pd{\Sds(\Tds,H)}{\Tds}
  = \Tds\,\dfrac{\partial_{r_w} \Sds}{\partial_{r_w} \Tds} \, ,
\eqe
where $\Tds$ and $\Lambda$ are regarded as independent variables. 
Then, since $\Sds$ is independent of $r_w$, we find
\eqb
\label{eq:ds.CLambda}
 C_{\Lambda} = 0 \, .
\eqe
Since $C_{\Lambda}$ is not negative but \emph{zero}, thermal equilibrium of CEH is not thermally unstable but thermally \emph{marginal} stable for constant $\Lambda$ process.

Since the constant $\Lambda$ process means the variation of only the position of observer $r_w$ with fixing the CEH radius $\rds = \sqrt{3/\Lambda}$, the vanishing heat capacity~\eref{eq:ds.CLambda} means that the observer's position $r_w$ can change without a heat supply to the CEH. 
Here note that, Eq.\eref{eq:ds.CLambda} does not imply that changing $r_w$ has no thermodynamic effect on the CEH. 
For example, changing $r_w$ in constant $\Lambda$ process gives rise to the change of surface pressure, $\partial \sigmads/\partial r_w \neq 0$. 
The vanishing heat capacity at constant $\Lambda$ process~\eref{eq:ds.CLambda} means simply the disappearance of heat supply in changing $r_w$ with fixing $\Lambda$. 
This is a peculiar thermodynamic property of the CEH.

\subsubsection{Surface compressibility and mechanical stability}

Let us research the mechanical stability of our thermal equilibrium system of CEH. 
As explained in Appendix~B of reference~\cite{ref:euclidean.ds}, the appropriate quantity to consider the mechanical stability may be the isothermal surface compressibility $\kappa_{\Tds}$ defined by $\kappa_{\Tds} \defeq \Ads^{-1}\,\partial \Ads(\Tds,\sigmads)/\partial \sigmads$. 
However, since $\sigmads$ is proportional to $\Tds$ as shown in Eq.\eref{eq:ds.sigma}, the definition of $\kappa_{\Tds}$ becomes meaningless. 
Then, instead of $\kappa_{\Tds}$, let us consider the ``isentropic'' surface compressibility $\kappa_{\Sds}$ defined by
\eqb
 \kappa_{\Sds} \defeq \dfrac{1}{\Ads}\,\pd{\Ads(\Sds,\sigmads)}{\sigmads} \, ,
\eqe
where $\Sds$ and $\sigmads$ are regarded as independent variables.
Since $\Sds$ depends only on $\Lambda$ as shown in Eq.\eref{eq:ds.S}, $\kappa_{\Sds}$ is equivalent to the surface compressibility at constant $\Lambda$, and therefore it seems to be the physical quantity. 
By Eqs.\eref{eq:ds.A},~\eref{eq:ds.S} and~\eref{eq:ds.sigma}, we get
\eqb
 \kappa_{\Sds} = \dfrac{1}{\Ads}\,\dfrac{\partial_{r_w}\Ads}{\partial_{r_w}\sigmads}
  = \dfrac{16\,\pi\,\fds(r_w)^{3/2}}{H\,x^2} \, .
\eqe
Obviously $\kappa_{\Sds}$ is positive definite. 
When the surface $\Ads$ increases, the surface pressure $\sigmads$ also increases. 
If we take the same criterion of mechanical stability as York~\cite{ref:euclidean.sch} (see the end of Appendix~B of reference~\cite{ref:euclidean.ds}), then the positivity of $\kappa_{\Sds}$ implies that our thermal equilibrium system is mechanically stable.

\subsection{The Role of Cosmological Constant in de~Sitter Thermodynamics}
\label{sec:ds.lambda}

As mentioned in the working hypothesis~dS-1, we regard $\Lambda$ as a working variable to obtain thermodynamically consistent de~Sitter canonical ensemble. 
The validity of working hypothesis~dS-1 can be recognized simply by the following fact: 
The entropy $\Sds = 3 \pi/\Lambda$ given in Eq.\eref{eq:ds.S} depends only on $\Lambda$ as already verified in the micro-canonical ensemble~\cite{ref:sds.special.2,ref:micro}, and consequently the definition of $\Sds$ in Eq.\eref{eq:ds.S_def} is expressed by using the derivatives of $\Fds$ and $\Tds$ with respect to $\Lambda$. 
The derivative with respect to $\Lambda$ requires implicitly the variable $\Lambda$. 
Hence, in order to calculate the entropy in the canonical ensemble, it is necessary to adopt the working hypothesis~dS-1. 
The following role of $\Lambda$ is worth emphasizing;
\begin{itemize}
\item \emph{The canonical ensemble of de~Sitter spacetime constructs the ``generalized'' thermodynamics in which $\Lambda$ behaves as a working variable, and the physical process is described by the constant $\Lambda$ process.}
\end{itemize}

Finally in this section, let us discuss the origin of internal energy $\Eds$, which is related to $\Lambda$ as explained below: 
For the first, recall the Schwarzschild thermodynamics formulated by York~\cite{ref:euclidean.sch}. 
The internal energy $E_{\rm BH}$ in Schwarzschild thermodynamics is related to its mass parameter $M$ by
\eqb
 M = E_{\rm BH} - \dfrac{E_{\rm BH}^{\quad 2}}{2\,r_w} \,,
\eqe
where $r_w$ is the outer-most radius of cavity shown in Fig.\ref{fig:york-1}. 
The second term $E_{\rm BH}^{\quad 2}/(2 r_w)$ can be interpreted as the self-gravitational potential energy of black hole. 
Then $E_{\rm BH}$ is interpreted as the ``bare'' mass energy of the \emph{black hole in cavity}, while $M$ is the ``net'' mass energy including the self-gravitational potential. 
It seems reasonable to consider that the origin of internal energy $E_{\rm BH}$ is the mass of black hole. 
The mass $M$ as the origin of energy $E_{\rm BH}$ can be clearly exhibited in the large cavity limit $r_w \to \infty$. 
In this limit we have $E_{\rm BH}\,|_{r_w\to\infty} = M$, which manifestly shows that $E_{\rm BH}$ is originated from $M$.

Then turn our discussion to de~Sitter thermodynamics. 
Let us consider the small heat wall limit $r_w \to 0$, which seems to correspond to the large cavity limit in Schwarzschild thermodynamics, since the heat wall is most distant from CEH. 
In this limit we have
\eqb
 \lim_{r_w \to 0} \Eds = \dfrac{1}{H} \,.
\eqe
This may show that the origin of internal energy $\Eds$ is the cosmological constant $\Lambda\,(= 3 H^2 )$. 
In the framework of classical general relativity, the de~Sitter spacetime is a vacuum spacetime which includes no energy source. 
However, in the de~Sitter thermodynamics which includes essentially the quantum gravitational effects, $\Lambda$ may be interpreted as a kind of energy source which is responsible to the energy $\Eds$. 
Also $\Lambda$ may be responsible to the entropy $\Sds$.

\section{Limit of Entropy-Area Law for Multi-Horizon Spacetimes}
\label{sec:limit}

Let us proceed to the discussion on the universality of entropy-area law for black hole event horizon (BEH) and cosmological event horizon (CEH) in SdS spacetime. 
As discussed in Sec.\ref{sec:intro}, some reasonable evidence of the breakdown of entropy-area law is going to be revealed in this section, which indicate the following:
\begin{itemize}
\item
Thermal equilibrium of individual horizon in multi-horizon spacetime is just a necessary condition of entropy-area low.
\item 
The necessary and sufficient condition of entropy-area law is the thermal equilibrium of the total system composed of several horizons in which the net energy flow among horizons disappears.
\end{itemize}

\subsection{Summary of Schwarzschild-de~Sitter Geometry}
\label{sec:limit.geometry}

\begin{figure}[t]
 \begin{center}
 \includegraphics[height=35mm]{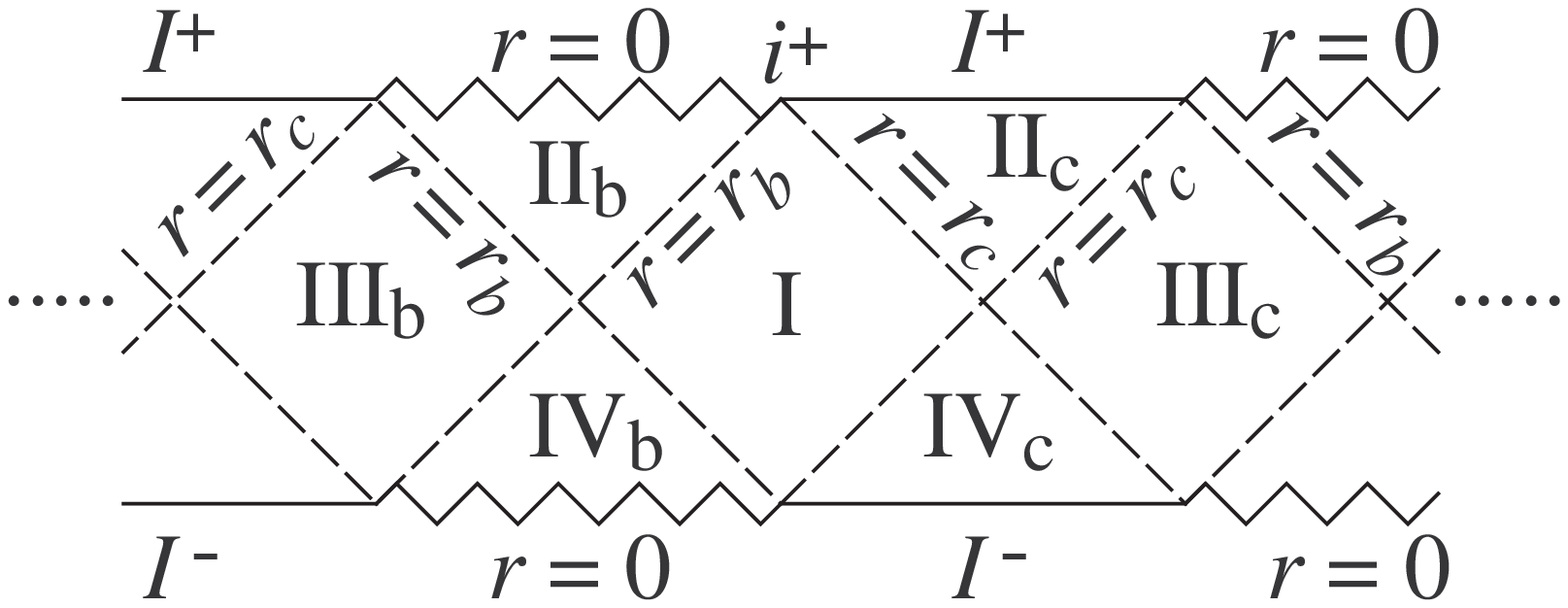}
 \end{center}
\caption{Penrose diagram of SdS spacetime. $I^{\pm}$ are the future/past null infinity, $i^+$ is the future temporal infinity. BEH is at $r=r_b$, and CEH at $r=r_c$. Spacetime singularity is at $r=0$. Static chart covers the region~I. Semi-global black hole chart covers the regions I, II$_b$, III$_b$ and IV$_b$. Semi-global cosmological chart covers the regions I, II$_c$, III$_c$ and IV$_c$. The maximal extension is obtained by connecting the two semi-global charts alternately.}
\label{fig:limit-1}
\end{figure}

In order to prepare some quantities used in following analysis, let us summarize the Lorentzian SdS spacetimes. 
The metric of SdS spacetime in the \emph{static chart} is
\eqb
 ds^2 = -f(r)\,dt^2 + \dfrac{dr^2}{f(r)} + r^2\,d\Omega^2 \, ,
\label{eq:limit.SdS.static}
\eqe
where $d\Omega^2 = d\theta^2 + \sin^2\theta\,d\varphi^2$ is the line element on the unit two-sphere, and
\eqb
\label{eq:limit.SdS.f.H}
 f(r) \defeq 1 - \dfrac{2\,M}{r} - H^2\,r^2 \quad,\quad 3\,H^2 \defeq \Lambda \, ,
\eqe
where $M$ is the mass parameter of black hole and $\Lambda$ is the cosmological constant. 
The Penrose diagram of SdS spacetime is shown in Fig.\ref{fig:limit-1}, and the static chart covers the region~I.

An algebraic equation $f(r)=0$ has one negative root and two positive roots. 
The smaller and larger positive roots are, respectively, the radius of BEH $r_b$ and that of CEH $r_c$. 
The notion of CEH is observer dependent and the CEH at $r_c$ is associated with the observer going towards the temporal future infinity $i^+$ in region~I~\cite{ref:temperature}. 
The equation $f(r) = 0$ is rearranged to $4\,\tilde{r}^3 -3\,\tilde{r} + \sqrt{27}\,M\,H = 0$, where $\tilde{r} \defeq \sqrt{3}\,H\,r/2$. 
Then via a formula, $\sin\theta = -4\,\sin^3(\theta/3) + 3\,\sin(\theta/3)$, we get
\eqb
\label{eq:limit.r}
 r_b = \dfrac{2}{\sqrt{3}\,H}\,\sin\left(\dfrac{\alpha}{3}\right) \quad,\quad
 r_c = \dfrac{2}{\sqrt{3}\,H}\,\sin\left(\dfrac{\alpha + 2 \pi}{3}\right) \, ,
\eqe
where $\alpha$ is defined by, $\sin\alpha \defeq \sqrt{27}\,M\,H$. 
The existence condition of BEH and CEH is $0 < \sqrt{27}\,M\,H < 1$. 
This is equivalent to, $0 < \alpha < \pi/2$, which means
\eqb
 2 M < r_b < 3 M < \dfrac{1}{\sqrt{3} H} < r_c < \dfrac{1}{H} \, .
\label{eq:limit.range.rb.rc}
\eqe
This denotes that $r_b$ is larger than the Schwarzschild radius $2 M$ and $r_c$ is smaller than the de~Sitter's CEH radius $H^{-1}$.

SdS spacetime has a timelike Killing vector $\xi \defeq N\, \partial_t$, where $N$ is a normalization constant~\cite{ref:sg}. 
This $\xi$ becomes null at BEH and CEH. 
This means those horizons are the Killing horizons of $\xi$. 
The surface gravity of BEH $\kappa_b$ and that of CEH $\kappa_c$ are defined by the equations, 
$\nabla_{\xi}\,\xi^{\mu}\,|_{r=r_b}
  = \kappa_b\,\xi^{\mu}\,|_{r=r_b}$ , 
$\nabla_{\xi}\,\xi^{\mu}\,|_{r=r_c}
  = \kappa_c\,\xi^{\mu}\,|_{r=r_c}$. 
The surface gravity depends on $N$. 
Throughout this section we take the normalization $N = 1$ for BEH, and $N = -1$ for CEH to make $\kappa_c$ positive~\footnote{Even if $N = 1$ for CEH, we can keep consistency of our analysis by changing the signature of $\kappa_c$ appropriately.}. 
Then the surface gravities become equal to the absolute value $\left|(1/2)\,df(r)/dr \right|$ at each Killing horizon,
\eqb
\label{eq:limit.kappa}
\begin{array}{rl}
 \kappa_b &= \dfrac{H^2}{2 r_b}\,\left( r_c - r_b \right)\,\left( 2\,r_b + r_c \right)
           = \dfrac{1}{2\,r_b}\,\left( 1 - 3\,H^2\,r_b^2 \right) \, , \\
 \kappa_c &= \dfrac{H^2}{2 r_c}\,\left( r_c - r_b \right)\,\left( r_b + 2\,r_c \right)
           = \dfrac{1}{2\,r_c}\,\left( 3\,H^2\,r_c^2 - 1 \right) \, ,
\end{array}
\eqe
where Eq.\eref{eq:limit.r} is used in the second equality in each equation. 
From the inequality $r_b < r_c$ in Eq.\eref{eq:limit.range.rb.rc}, we get
\eqb
 \kappa_b > \kappa_c \, .
\label{eq:limit.kappa.rel}
\eqe
This implies the Hawking temperature of BEH is higher than that of CEH, which will be verified by Eqs.\eref{eq:limit.Tb} and~\eref{eq:limit.Tc}.

For later use, let us show some differentials,
\seqb
\eqab
 \pd{\,r_b}{M} = \dfrac{2}{1 - 3\,H^2\,r_b^2} \quad&,&\quad
 \pd{\,r_b}{H} = -\dfrac{r_b}{H} + \dfrac{M}{H}\,\pd{\,r_b}{M}
\label{eq:limit.diff.rb} \, , \\
 \pd{\,r_c}{M} = - \dfrac{2}{3\,H^2\,r_c^2 - 1} \quad&,&\quad
 \pd{\,r_c}{H} = -\dfrac{r_c}{H} + \dfrac{M}{H}\,\pd{\,r_c}{M}
\label{eq:limit.diff.rc} \, ,
\eqae
and
\eqab
 \pd{\,\kappa_b}{M} = -\dfrac{1}{r_b^2}\,\dfrac{1 + 3\,H^2\,r_b^2}{1 - 3\,H^2\,r_b^2}
 \quad&,&\quad
 \pd{\,\kappa_b}{H}
   = \dfrac{1 - 3\,H^2\,r_b^2}{2\,H\,r_b}
    + \dfrac{M}{H}\,\pd{\,\kappa_b}{M}
\label{eq:limit.diff.kappab} \, , \\
 \pd{\,\kappa_c}{M} = -\dfrac{1}{r_b^2}\,\dfrac{3\,H^2\,r_c^2 + 1}{3\,H^2\,r_c^2 - 1}
 \quad&,&\quad
 \pd{\,\kappa_c}{H}
   = \dfrac{3\,H^2\,r_c^2 - 1}{2\,H\,r_c}
    + \dfrac{M}{H}\,\pd{\,\kappa_c}{M}
\label{eq:limit.diff.kappac} \, ,
\eqae
\seqe
where we used a formula, $\cos\theta = 4\,\cos^3(\theta/3) - 3\,\cos(\theta/3)$, and the differentials, $\partial_M \alpha = \sqrt{27} H/\cos\alpha$, and $\partial_H \alpha = \sqrt{27} M/\cos\alpha$, obtained from the definition of $\alpha$, $\sin\alpha \defeq \sqrt{27} M H$.

The metric in \emph{semi-global black hole chart} is given by the coordinate transformation from $(t,r,\theta,\varphi)$ to $(\eta_b,\chi_b,\theta,\varphi)$:
\eqb
 \eta_b - \chi_b \defeq - e^{-\kappa_b (t - r^{\ast})} \quad,\quad
 \eta_b + \chi_b \defeq e^{\kappa_b (t + r^{\ast})} \, ,
\label{eq:limit.trans.beh}
\eqe
where $dr^{\ast} \defeq dr/f(r)$ which means
\eqb
 2\,r^{\ast} =  \ln\left| \dfrac{r}{r_b} - 1 \right|^{1/\kappa_b}
              - \ln\left| 1 - \dfrac{r}{r_c} \right|^{1/\kappa_b}
              + \ln\left| \dfrac{r}{r_b + r_c} + 1 \right|^{1/\kappa_c-1/\kappa_b} \, .
\label{eq:limit.rstar}
\eqe
We get by this transformation,
\eqb
 ds^2 = \Upsilon_b(r)\,\left[\, - d\eta_b^2 + d\chi_b^2 \,\right] + r^2\,d\Omega^2 \, ,
\eqe
where
\eqb
 \Upsilon_b(r) \defeq
 \dfrac{2 M}{\kappa_b^2\,r}\,\left( 1 - \dfrac{r}{r_c} \right)^{1+\kappa_b/\kappa_c}\,
 \left(\dfrac{r}{r_b + r_c} + 1 \right)^{2 - \kappa_b/\kappa_c} \, .
\label{eq:limit.upsilon.beh}
\eqe
The transformation~\eref{eq:limit.trans.beh} implies the range of coordinates, $-\chi_b < \eta_b < \chi_b$ and $0 < \chi_b$, which covers the region~I in Fig.\ref{fig:limit-1}. 
By extending to the range, $-\infty < \eta_b < \infty$ and $-\infty < \chi_b < \infty$, the semi-global black hole chart covers the regions I, II$_b$, III$_b$ and IV$_b$ in Fig.\ref{fig:limit-1}. 
In these regions we find $\Upsilon_b > 0$ since $r < r_c$.

The metric in \emph{semi-global cosmological chart} is given by the coordinate transformation from $(t,r,\theta,\varphi)$ to $(\eta_c,\chi_c,\theta,\varphi)$:
\eqb
 \eta_c - \chi_c \defeq e^{\kappa_c (t - r^{\ast})} \quad,\quad
 \eta_c + \chi_c \defeq -e^{-\kappa_c (t + r^{\ast})} \, ,
\label{eq:limit.trans.ceh}
\eqe
where $r^{\ast}$ is given in Eq.\eref{eq:limit.rstar}. 
By this transformation we get
\eqb
 ds^2 = \Upsilon_c(r)\,\left[\, - d\eta_c^2 + d\chi_c^2 \,\right] + r^2\,d\Omega^2 \, ,
\eqe
where
\eqb
 \Upsilon_c(r) \defeq
 \dfrac{2 M}{\kappa_c^2\,r}\,\left( \dfrac{r}{r_b} - 1 \right)^{1+\kappa_c/\kappa_b}\,
 \left(\dfrac{r}{r_b + r_c} + 1 \right)^{2 - \kappa_c/\kappa_b} \, .
\label{eq:limit.upsilon.ceh}
\eqe
The transformation~\eref{eq:limit.trans.ceh} implies the range of coordinates, $\chi_c < \eta_c < -\chi_c$ and $\chi_c < 0$, which covers the region~I in Fig.\ref{fig:limit-1}. 
By extending to the range, $-\infty < \eta_c < \infty$ and $-\infty < \chi_c < \infty$, the semi-global cosmological chart covers the regions I, II$_c$, III$_c$ and IV$_c$ in Fig.\ref{fig:limit-1}.
In these regions we find $\Upsilon_c > 0$ since $r_b < r$.

The maximally extended SdS spacetime is obtained by connecting the two semi-global charts alternately as shown in Fig.\ref{fig:limit-1}.

\subsection{Minimal Assumptions for Schwarzschild-de~Sitter Thermodynamics}
\label{sec:limit.assumption}

As mentioned in Sec.\ref{sec:pre.partition}, we need to specify the appropriate state variables before culculating the Euclidean action. 
To do so, we introduce the minimal set of assumptions with referring to York's Schwarzschild thermodynamics~\cite{ref:euclidean.sch}. 
As reviewed in Sec.\ref{sec:york}, there are three key points in Schwarzschild thermodynamics from which we can learn how to ensure the ``thermodynamic consistency'' in SdS thermodynamics.

Here, before considering SdS spacetime, we must comment on Anti-de~Sitter (AdS) black holes~\cite{ref:euclidean.sads}.
AdS black hole thermodynamics has a conceptual difference from the other black hole thermodynamics.
The difference appears, for example, in the definition of temperature. 
While the temperatures in asymptotic flat black hole and de~Sitter thermodynamics include the \emph{Tolman factor}~\cite{ref:euclidean.sch,ref:euclidean.kn,ref:euclidean.ds,ref:tolman}, the temperature assigned to AdS black hole~\cite{ref:euclidean.sads} does not include the Tolman factor, where the Tolman factor~\cite{ref:tolman} expresses the gravitational redshift affecting the Hawking radiation propagating from horizon to observer (see for example Eq.\eref{eq:limit.Tb} shown later or the key point~3 of Schwarzschild thermodynamics in Sec.\ref{sec:york}). 
The temperature in AdS black hole thermodynamics can not be measured by a thermometer of the physical observer who are outside the black hole. 
This implies that the state variables in AdS black hole thermodynamics are defined not by the observer outside the black hole, but defined just on the black hole event horizon on which no physical observer can rest.
In this article we do not refer to AdS black hole thermodynamics, since it seems to be preferable to expect that state variables are defined according to a physical observer.

\subsubsection{Zeroth law and independent variables}

\begin{figure}[t]
 \begin{center}
 \includegraphics[height=35mm]{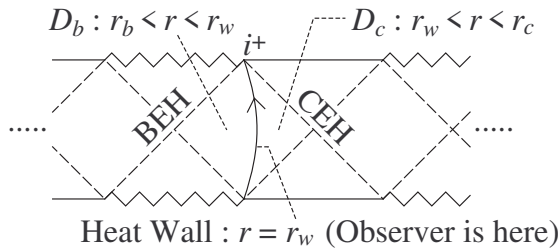}
 \end{center}
\caption{Our thermal equilibrium systems for BEH ($D_b$) and CEH ($D_c$). State variables of them are defined at the heat wall.}
\label{fig:limit-2}
\end{figure}

We will construct two thermal equilibrium systems for BEH and CEH, but place only one observer who can measure the state variables of BEH and CEH. 
Such observer is in the region~I, $r_b < r < r_c$ (see Fig.\ref{fig:limit-1}). 
However, as mentioned at Eq.\eref{eq:limit.kappa.rel}, Hawking temperature of BEH is higher than that of CEH. 
This temperature difference implies that, when the region~I constitutes one thermodynamic system, the thermodynamic state of region~I is in a non-equilibrium state. 
Therefore, by dividing the region~I into two regions, we construct \emph{two} thermal equilibrium systems for BEH and CEH individually which are measured by the same observer. 
To do so, we adopt the following assumption as the zeroth law: 
\begin{ass-sds}[Zeroth law] 
Two thermal ``equilibrium'' systems for BEH and CEH in SdS spacetime are constructed by the following three steps:
\begin{enumerate}
\item
Place a spherically symmetric thin wall at $r = r_w$ in the region~I, $r_b < r_w < r_c$, as shown in Fig.\ref{fig:limit-2}. 
This wall has negligible mass, and reflects perfectly Hawking radiation coming from each horizon. 
We call this wall the ``heat wall'' hereafter. 
The BEH side of heat wall is regarded as a ``heat bath'' of Hawking temperature of BEH due to the reflected Hawking radiation. 
Also the CEH side of heat wall is regarded as a heat bath of Hawking temperature of CEH.
\item
The region $D_b$ enclosed by BEH and heat wall, $r_b < r < r_w$, is filled with Hawking radiation emitted by BEH and reflected by heat wall, and forms a thermal equilibrium system for BEH. 
Similarly the region $D_c$ enclosed by CEH and heat wall, $r_w < r < r_c$, forms a thermal equilibrium system for CEH. 
Hence we have ``two'' thermal equilibrium systems separated by the heat wall. 
In the statistical mechanical sense, these two thermal equilibrium systems are described by the canonical ensemble, since those systems have a contact with the heat wall.
\item
Set our observer at the heat wall. 
When the observer is at the BEH side of heat wall, the observer can measure the state variables of thermal equilibrium system for BEH. 
The same is true of CEH. 
Then the state variables of two thermal equilibrium systems are defined by the quantities measured by the observer at heat wall.
\end{enumerate}
\end{ass-sds}
Note that the two thermal equilibrium systems constructed in this assumption have already been used by Gibbons and Hawking~\cite{ref:temperature} to calculate Hawking temperatures of BEH and CEH. 
Also the above step~3, which gives a criterion of defining state variables, has already been adopted in the consistent thermodynamics of single-horizon spacetimes~\cite{ref:euclidean.sch,ref:euclidean.kn,ref:euclidean.ds}. 
This assumption is a simple extension of the key point~1 of Schwarzschild thermodynamics shown in Sec.\ref{sec:york}.

It is expected that state variables of the thermal equilibrium system for BEH depend on BEH radius $r_b$ and/or BEH surface gravity $\kappa_b$. 
Similarly, state variables of CEH depend on $r_c$ and/or $\kappa_c$. 
These imply that the state variables of BEH and CEH depend on $M$ and $\Lambda$, since the horizon radii and surface gravities depend on $M$ and $\Lambda$ via Eqs.\eref{eq:limit.r} and~\eref{eq:limit.kappa}. 
Furthermore, by the step~3 in assumption~SdS-1, there should be $r_w$-dependence in state variables of BEH and CEH, since the observer is at $r = r_w$. 
Therefore the state variables of BEH and CEH depend on three parameters $M$, $\Lambda$ and $r_w$.

The existence of three parameters $(M,\Lambda,r_w)$ may imply that the CEH is regarded as a source of external gravitational field which affects the thermodynamic state of BEH. 
Also, BEH is a source of external gravitational field affecting the thermodynamic state of CEH. 
Here it is instructive to compare \emph{qualitatively} our two thermal equilibrium systems of horizon with a magnetized gas. 
The gas consists of molecules possessing a magnetic moment, and its thermodynamic state is affected by an external magnetic field. 
The qualitative correspondence between the magnetized gas and our thermal equilibrium systems of horizons is described as follows; 
the gas corresponds to the system $D_b$~($D_c$), and the external magnetic field corresponds to the external gravitational field produced by CEH~(BEH). 
Then, what we should emphasize is the following fact of the magnetized gas: 
When the gas is enclosed in a container of volume $V_{\rm gas}$ and an external magnetic field $\vec{H}_{\rm ex}$ is acting on the gas, the free energy $F_{\rm gas}$ of the gas is expressed as a function of three independent state variables (see for example \S52, 59 and~60 in Landau and Lifshitz~\cite{ref:ll}),
\eqb
\label{eq:limit.Fgas}
 F_{\rm gas} = F_{\rm gas}(T_{\rm gas} , V_{\rm gas} , \vec{H}_{\rm ex}) \, ,
\eqe
where $T_{\rm gas}$ is the temperature of the gas. 
According to this fact of the magnetized gas, it is reasonable for our two thermal equilibrium systems of horizons to require that the free energies are also functions of three independent state variables. 
For the BEH, the free energy $F_b$ is
\eqb
\label{eq:limit.Fb_abst}
 F_b = F_b(T_b , A_b , X_b) \, ,
\eqe
where $T_b$ is the temperature of BEH, $A_b$ is the state variable of system size, and $X_b$ is the state variable which represents the effect of CEH's gravity on the BEH. 
And for the CEH, the free energy $F_c$ is
\eqb
\label{eq:limit.Fc_abst}
 F_c = F_c(T_c , A_c , X_c) \, ,
\eqe
where $T_c$ is the temperature of CEH, $A_c$ is the state variable of system size and $X_c$ is the state variable which represents the effect of BEH's gravity on the CEH. 
Indeed, it will be proven in Secs.~\ref{sec:limit.beh} and~\ref{sec:limit.ceh} that the thermodynamic consistency never hold unless the free energies are functions of three independent variables as shown in Eqs.\eref{eq:limit.Fb_abst} and~\eref{eq:limit.Fc_abst}.

Now we recognize that, because free energies are functions of three independent variables (as will be verified in Secs.~\ref{sec:limit.beh} and~\ref{sec:limit.ceh}), the following working hypothesis is needed: 
\begin{hypo-sds}[Three independent variables] 
To ensure the thermodynamic consistency of our thermal equilibrium systems constructed in assumption~SdS-1, we have to regard the cosmological constant $\Lambda$ as an independent working variable. 
Then the three quantities $(M,\Lambda,r_w)$ are regarded as independent variables, and consequently the free energies $F_b$ and $F_c$ of our thermal equilibrium systems are functions of three independent state variables as shown in Eqs.\eref{eq:limit.Fb_abst} and~\eref{eq:limit.Fc_abst}. 
On the other hand, when we regard the non-variable $\Lambda$ as the physical situation, it is obtained by the ``constant $\Lambda$ process'' in the consistent thermodynamics for BEH and CEH which are constructed with regarding $\Lambda$ as a working variable.
\end{hypo-sds}

This working hypothesis will be verified in Secs.~\ref{sec:limit.beh} and~\ref{sec:limit.ceh}, and we can not preserve thermodynamic consistency without regarding $\Lambda$ as an independent working variable. 
Therefore, as already noticed for de~Sitter canonical ensemble in Sec.\ref{sec:ds.lambda}, the special role of cosmological constant is recognized:
\begin{itemize}
\item
\emph{Thermal equilibrium states of event horizon with positive $\Lambda$ may construct the ``generalized'' thermodynamics in which $\Lambda$ behaves as a working variable and the physical process is described by the constant $\Lambda$ process.} 
\end{itemize}

\subsubsection{Scaling law and system size}

In thermodynamics of ordinary laboratory systems, all state variables are classified into two categories, \emph{extensive} state variables and \emph{intensive} ones. 
The criterion of this classification is the scaling behavior of state variables under the scaling of system size. 
However, as explained by the key point~2 of Schwarzschild thermodynamics in Sec.\ref{sec:york}, the state variables in thermodynamics of single-horizon spacetimes~\cite{ref:euclidean.sch,ref:euclidean.kn,ref:euclidean.ds} have its own peculiar scaling behavior classified into three categories, and the state variable of system size is not a volume but the surface area of heat bath.
Although the scaling behavior differs from that in thermodynamics of ordinary laboratory systems, the peculiar scaling behavior in thermodynamics of single-horizon spacetimes retains the thermodynamic consistency as explained in Secs.\ref{sec:york} and~\ref{sec:ds}. 
Then, we assume that the key point~2 of Schwarzschild thermodynamics is simply extended to our two thermal equilibrium systems constructed in assumption~SdS-1:
\begin{ass-sds}[Scaling law and system size]
All state variables of our thermal equilibrium systems are classified into three categories; extensive variables, intensive variables and thermodynamic functions, and satisfy the following scaling law: 
When the length size $L$ (e.g. horizon radius) is scaled as $L \to \lambda\,L$ with an arbitrary scaling rate $\lambda\,(>0)$, then the extensive variables $X$ (e.g. system size) and intensive variables $Y$ (e.g. temperature) are scaled respectively as $X \to \lambda^2\,X$ and $Y \to \lambda^{-1}\,Y$, while the thermodynamic functions $\Phi$ (e.g. free energy) are scaled as $\Phi \to \lambda\,\Phi$. 
This implies that the thermodynamic system size of our thermal equilibrium systems should have the areal dimension, since the system size is extensive in thermodynamic argument. 
Then we assume that the surface area of heat wall, $A \defeq 4 \pi r_w^2$, behaves as the consistent extensive variable of system size for our thermal equilibrium systems for BEH and CEH. 
This denotes to set $A_b = A_c = A$ in Eqs.\eref{eq:limit.Fb_abst} and~\eref{eq:limit.Fc_abst}.
\end{ass-sds}

Accepting this assumption, the length size scaling in our thermal equilibrium systems for BEH and CEH should be specified. 
Here recall that, due to the working hypothesis~SdS-1, the fundamental independent variables in our thermal equilibrium systems are $M$, $\Lambda$ and $r_w$. 
Therefore the fundamental length size scaling for our thermal equilibrium systems is composed of the following three scalings;
\eqb
\label{eq:limit.scaling}
 M \to \lambda\,M \quad,\quad
 H \to \dfrac{1}{\lambda}\,H \quad,\quad
 r_w \to \lambda\,r_w \, ,
\eqe
where $3 H^2 = \Lambda$, and $\lambda\,(>0)$ is an arbitrary scaling rate. 
The extensivity and intensivity of each state variable of our thermal equilibrium systems should be defined under these fundamental length size scalings as explained in assumption~SdS-2.
\footnote{
As explained in Appendix~B of paper~\cite{ref:euclidean.ds}, when we regard $A \defeq 4\,\pi\,r_w^2$ as a state variable of system size, the scaling of system size should be restricted to the homothetic one, which is the spherical scaling due to the spherical symmetry of SdS spacetime. 
The fundamental length size scaling~\eref{eq:limit.scaling} is consistent with this restriction. 
See Appendix~B of paper~\cite{ref:euclidean.ds} for details of such restriction.
}

\subsubsection{Euclidean action method and how to obtain state variables}

We need to specify how to get the state variables. 
As noted in the step~2 in assumption~SdS-1, thermodynamics of our two thermal equilibrium systems should be constructed in the canonical ensemble. 
Therefore we use the Euclidean action method which is the technique to calculate the partition function of quantum gravity~\cite{ref:euclidean}. 
Indeed, the Euclidean action method has already made successes to obtain the partition function of canonical ensemble for the thermodynamics of single-horizon spacetimes~\cite{ref:euclidean.sch,ref:euclidean.kn,ref:euclidean.sads,ref:euclidean.ds}. 
The key point~3 in Sec.\ref{sec:york} explains how to use the Euclidean action and to define state variables for Schwarzschild thermodynamics. 
Then, referring to the key point~3, we adopt the following assumption:
\begin{ass-sds}[State variables and Euclidean action method] 
Euclidean actions $I_{Eb}$ and $I_{Ec}$ of our two thermal equilibrium systems yield the partition functions of canonical ensembles by Eq.\eref{eq:pre.Zcl}. 
And the free energies $F_b$ and $F_c$ are defined by Eq.\eref{eq:pre.F}, where the temperatures are defined by Eq.\eref{eq:pre.T}. 
Then, once $F_b$ and $F_c$ are determined, all state variables of BEH and CEH are defined from $F_b$ and $F_c$ as for thermodynamics of ordinary laboratory systems. 
For example, BEH entropy $S_b$ is defined by $S_b \defeq -\partial F_b/\partial T_b$, where $T_b$ is the temperature of BEH.
\end{ass-sds}

When we use the Euclidean action method, it is necessary to specify the integration constant $I_{\rm sub}$. 
It is natural to require that our thermal equilibrium systems for BEH and CEH should reproduce, respectively, the Schwarzschild thermodynamics in the limit $\Lambda \to 0$ and the de~Sitter thermodynamics in the limit $M \to 0$. 
Then the following working hypothesis is naturally required:
\begin{hypo-sds}[Integration constants in Euclidean action]
For the thermal equilibrium system for BEH, the integration constant in Euclidean action is determined with referring to Schwarzschild canonical ensemble formulated by York~\cite{ref:euclidean.sch}. 
For the thermal equilibrium system for CEH, the integration constant in Euclidean action is determined with referring to de~Sitter canonical ensemble~\cite{ref:euclidean.ds} formulated in Sec.\ref{sec:ds}.
\end{hypo-sds}

We introduce this working hypothesis as if this is a statement separated from the assumption~SdS-3. 
However the determination of integration constant accompanies necessarily the Euclidean action method. 
The working hypothesis~SdS-2 is regarded as a part of the assumption~SdS-3.

\subsubsection{Effects of external gravitational fields}

By the assumption~SdS-3 together with the working hypothesis~SdS-2, the concrete functional form of free energies $F_b$ and $F_c$ can be determined as functions of three independent working variables $(M,\Lambda,r_w)$. 
However, since the form of the state variables $X_b$ and $X_c$ have not been specified yet, we can not rearrange $F_b$ and $F_c$ to functions of independent \emph{state variables}, $(T_b,A,X_b)$ and $(T_c,A,X_c)$.

As explained at Eqs.\eref{eq:limit.Fb_abst} and~\eref{eq:limit.Fc_abst}, the CEH (BEH) is regarded as the source of external gravitational field which affects the thermodynamic state of BEH (CEH). 
This means that the state variables $X_b$ and $X_c$ represent, respectively, the thermodynamic effect of CEH's gravity on BEH and that of BEH's gravity on CEH. 
Then it is reasonable to expect that $X_b$ depends on the quantity characterizing the CEH's gravity, and $X_c$ depends on the quantity characterizing the BEH's gravity. 
Moreover, due to the step~3 in assumption~SdS-1, $X_b$ and $X_c$ should be measurable for the observer at $r_w$. 
Then, we can offer two candidates for the pair of \emph{dimensionless} characteristic quantities of BEH's and CEH's gravities; 
\begin{itemize}
\item
First candidate pair consists of $\kappa_b r_w$ and $\kappa_c r_w$, where $\kappa_b r_w$ is for BEH's gravity and $\kappa_c r_w$ is for CEH's gravity. This pair means that both of BEH's and CEH's gravities are characterized by three quantities $(M , \Lambda , r_w)$, since $\kappa_b$ and $\kappa_c$ depend on $M$ and $\Lambda$. 
\item
Second candidate pair consists of $M/r_w$ and $H r_w$, where $M/r_w$ is for BEH's gravity and $H r_w$ is for CEH's gravity. This pair means that the BEH's gravity is characterized by $(M , r_w)$, and the CEH's gravity is characterized by $(\Lambda , r_w)$. 
\end{itemize}
Here, purely logically, we can consider the ``inverse'' pair of second one, where $H r_w$ is for BEH's gravity and $M/r_w$ is for CEH's gravity. 
This means that the BEH's gravity is characterized by $(\Lambda , r_w)$, and the CEH's gravity is characterized by $(M , r_w)$. 
However this is physically unacceptable, since we do not expect that the BEH does not depend on $M$ and the CEH does not depend on $\Lambda$. 
Therefore, the reasonable candidates for the pair of characteristic quantities of BEH's and CEH's gravities are the two candidates listed above. 
Then, $X_b$ should be a function of $( \kappa_c  \,,\, r_w )$ or $( H \,,\, r_w )$, and $X_c$ should be a function of $( \kappa_b \,,\, r_w )$ or $( M \,,\, r_w )$.

On the other hand, as will be mathematically verified in Secs.\ref{sec:limit.beh} and~\ref{sec:limit.ceh}, the state variables $X_b$ and $X_c$ are the extensive variables and proportional to $r_w^2$. 
The proportionality to $r_w^2$ is consistent with the scaling law of extensive variables denoted in assumption~SdS-2. 
And, according to the previous paragraph, the factor of proportionality should be a function of the characteristic quantity of BEH's or CEH's gravity. 
Although the verification of the extensivity of $X_b$ and $X_c$ are shown later, we accept it in the following assumption~SdS-4 for the simplicity of our discussion.

From the above, it is reasonable to adopt the following assumption:
\begin{ass-sds}[Extensive variable of ``external field''] 
The state variables $X_b$ and $X_c$ in Eqs.\eref{eq:limit.Fb_abst} and~\eref{eq:limit.Fc_abst} are the extensive variables. 
(This will be verified in Secs.{\ref{sec:limit.beh}} and {\ref{sec:limit.ceh}}). 
Then, there are two candidates for the functional forms of $X_b$ and $X_c$. 
One of them is based on the quantities $(\kappa_b r_w \,,\, \kappa_c r_w)$:
\eqb
\label{eq:limit.X.1}
 X_b = r_w^2\, \Psi_b(\kappa_c r_w) \quad,\quad X_c = r_w^2\,\Psi_c(\kappa_b r_w) \, ,
\eqe
where $\Psi_b$ and $\Psi_c$ are arbitrary functions of single argument, whose functional forms are not specified at present. 
Another candidate of $X_b$ and $X_c$ is based on the quantities $(M/r_w \,,\, H\,r_w)$:
\eqb
\label{eq:limit.X.2}
 X_b = r_w^2\, \Psi_b(H\,r_w) \quad,\quad X_c = r_w^2\,\Psi_c(M/r_w) \, .
\eqe
At least for the present author, no criterion to choose one of these candidates is found, and the way for determining the functional forms of $\Psi_b$ and $\Psi_c$ are also unknown. 
An obvious constraint on $\Psi_b$ and $\Psi_c$ is that they never be constant to make $X_b$ and $X_c$ independent of the state variable of system size $A \defeq 4 \pi r_w^2$.
\end{ass-sds}

In Sec.\ref{sec:conc}, we will make some comments on the issue which of Eqs.\eref{eq:limit.X.1} and~\eref{eq:limit.X.2} is valid. 
Those comments will suggest that Eq.\eref{eq:limit.X.1} may be appropriate, but we do not have mathematical verification to choose Eq.\eref{eq:limit.X.1} as the general form of $X_b$ and $X_c$. 
Therefore, to retain the logical strictness of this article, we list the two possibilities~\eref{eq:limit.X.1} and~\eref{eq:limit.X.2} in the assumption~SdS-4.

From the above, we recognize that the minimal set of assumptions for ``consistent'' thermodynamics of our two thermal equilibrium systems should be composed of four assumptions. 
However, the determination of functions $\Psi_b$ and $\Psi_c$ remains as a future task and we can not find concrete functional forms of them. 
Although the state variables $X_b$ and $X_c$ are not specified in this article, the existence of them enables us to examine the validity of entropy-area law in SdS spacetime as shown in Secs.~\ref{sec:limit.beh} and~\ref{sec:limit.ceh}.

\subsection{Euclidean Actions for Two Horizons}
\label{sec:limit.action}

Referring to assumption~SdS-3 and working hypothesis~SdS-2, we calculate Euclidean actions for the two thermal equilibrium systems for BEH and CEH constructed in assumption~SdS-1. 

\subsubsection{Euclidean action for BEH}

Euclidean space of thermal equilibrium system for BEH is obtained by the Wick rotations $t \to -i\,\tau$ in the static chart and $\eta_b \to -i\,\omega_b$ in the semi-global black hole chart. 
These Wick rotations are equivalent, because the coordinate transformation~\eref{eq:limit.trans.beh}, $\eta_b = e^{\kappa_b\, r^{\ast}}\sinh\left(\kappa_b\, t\right)$, implies that the imaginary time $\omega_b$ in the semi-global chart is defined by $\omega_b \defeq e^{\kappa_b\, r^{\ast}}\sin\left(\kappa_b\, \tau\right)$, where $\tau$ is the imaginary time in the static chart. 
Euclidean metric in the static chart is
\eqb
 ds_E^2 = f(r)\,d\tau^2 + \dfrac{dr^2}{f(r)} + r^2\,d\Omega^2 \, .
\label{eq:limit.SdSE.static} \\
\eqe
Euclidean metric in the semi-global black hole chart is
\eqb
 ds_E^2 = \Upsilon_b(r)\,\left[\,d\omega_b^2 + d\chi_b^2 \right] + r^2\,d\Omega^2 \, ,
\label{eq:limit.SdSE.global.beh}
\eqe
where $\Upsilon_b$ is defined in Eq.\eref{eq:limit.upsilon.beh}. 
About the semi-global chart, we get from the coordinate transformation~\eref{eq:limit.trans.beh},
\eqb
 \omega_b^2 + \chi_b^2 =
  \left( \dfrac{r}{r_b} - 1 \right)\,\left( 1 - \dfrac{r}{r_c} \right)^{-\kappa_b/\kappa_c}
  \left( \dfrac{r}{r_b + r_c} + 1 \right)^{-1+\kappa_b/\kappa_c} \, .
\label{eq:limit.SdSE.radial}
\eqe
Then, because the thermal equilibrium system for BEH is the region $D_b$, $r_b < r < r_w$, in Lorentzian SdS spacetime, we find the topology of Euclidean space of thermal equilibrium system for BEH is $D^2\times S^2$. 
Because $\omega_b^2 + \chi_b^2 = 0$ at $r = r_b$, the center of $D^2$-part is at the BEH $r = r_b$, and the boundary of $D^2$-part is at the heat wall $r = r_w$. 
The topology of heat wall boundary is $S^1 \times S^2$, where $S^1$ is along the $\tau$-direction.

Because the Lorentzian SdS spacetime is regular at $r_b$, the Euclidean space is also regular at $r_b$.
To examine the regularity of Euclidean space at $r_b$, we make use of the static chart~\eref{eq:limit.SdSE.static}. 
Let us define a coordinate $y_b$ and a function $\gamma(y_b)$ by
\eqb
 y_b^2 \defeq r - r_b \quad,\quad
 \gamma(y_b) \defeq \sqrt{f(r_b + y_b^2)} \, .
\eqe
We get $\gamma'(y_b) \defeq d\gamma(y_b)/dy_b = [ y_b/\gamma(y_b) ]\,df(r)/dr$, from which we find
\eqb
 \lim_{y_b \to 0}\gamma'(y_b)
  = \left.\dfrac{d C}{dr}\right|_{r_b}\,\lim_{y_b\to 0}\dfrac{y_b}{\gamma(y_b)}
  = 2\,\kappa_b\,\dfrac{1}{\displaystyle \lim_{y_b \to 0}\gamma'(y_b)} \, .
\eqe
This means $\gamma'(0) = \sqrt{2\,\kappa_b}$, and near the BEH, $f \simeq \left[ \gamma(0) + \gamma'(0)\,x \right]^2 = 2\,\kappa\,y_b^2$. 
Therefore the Euclidean metric near BEH is
\eqb
 ds_E^2 \simeq
 \dfrac{2}{\kappa_b}\,\left[\, y_b^2\,d(\kappa_b\, \tau)^2 + dy_b^2 \,\right] + r^2\,d\Omega^2 \, .
\eqe
It is obvious that the Euclidean space is regular at BEH if the imaginary time has the period $\beta_b$ defined by
\eqb
 0 \le \tau < \beta_b \defeq \dfrac{2\,\pi}{\kappa_b} \, .
\label{eq:limit.period.beh}
\eqe
Throughout our discussion, $\tau$ has the period $\beta_b$ in the Euclidean space of thermal equilibrium system for BEH.

Now let us proceed to the calculation of the Euclidean action $I_{Eb}$ of thermal equilibrium system for BEH via Eq.\eref{eq:pre.IE.curved}, where the definition of Lorentzian action $I_L$ is given in Eq.\eref{eq:pre.IL}. 
Following the working hypothesis~SdS-2, we use the same integration constant $I_{\rm sub}$ as Schwarzschild canonical ensemble~\cite{ref:euclidean,ref:euclidean.sch}, which gives us
\eqb
\label{eq:limit.Isub.beh}
 I_{\rm sub}
  \defeq - I_L^{\rm (flat)}
  = - \dfrac{1}{8 \pi}\,\int_{\partial \mathcal M}\,dx^3 \,
      \sqrt{\det h\,}\,\,K^{\rm (flat)} \, ,
\eqe
where $I_L^{\rm (flat)}$ is the Lorentzian Einstein-Hilbert action for flat spacetime which reduces to only the surface term (the second term in Eq.\eref{eq:pre.IL} ) due to $R = 0$ and $\Lambda = 0$ for flat spacetime, and $K^{\rm (flat)}$ is the trace of second fundamental form of $\partial \mathcal M$ in flat spacetime~\cite{ref:euclidean,ref:euclidean.sch}. 
Here note that, the integral element $\sqrt{\det h}$ in $I_L^{\rm (flat)}$ should be given by that of SdS spacetime when $I_L^{\rm (flat)}$ is used as the integration constant of action integral of SdS spacetime, because the background spacetime on which the integral in $I_L^{\rm (flat)}$ is calculated is SdS spacetime.

For our thermal equilibrium system for BEH in SdS spacetime, the region ${\mathcal M}$ in $I_L$ is $D_b$, $r_b < r < r_w$, and its boundary $\partial D_b$ is at $r_w$. 
There is another boundary at $r_b$ in the Lorentzian region $D_b$. 
However we do not need to consider it, because the points at $r_b$ in the Euclidean space do not form a boundary but are the regular points when $\tau$ has the period~\eref{eq:limit.period.beh}. 
Then the first fundamental form $h_{i j}$ ($i, j = 0, 2, 3$) of $\partial D_b$ in the static chart is
\eqb
 \left. ds^2 \right|_{r = r_w} = h_{i j}\, dx^i \, dx^j
 = - f_w\,dt^2 + r_w^2\,d\Omega^2 \, ,
\eqe
where
\eqb
 f_w \defeq f(r_w) = 1 - \dfrac{2 M}{r_w} - H^2\,r_w^2 \, .
\label{eq:limit.fw}
\eqe
Here, since $D_b$ is the region enclosed by BEH and heat wall, the direction of unit normal vector $n^{\mu}$ to $\partial D_b$ is pointing towards CEH, $n^{\mu} \propto \partial_r$. 
Then the second fundamental form of $\partial D_b$ in the static chart is
\eqb
\label{eq:limit.Kij.beh}
 K_{i j} =
 \sqrt{f_w}\, 
 {\rm diag.}\left[\, - \dfrac{M}{r_w^2} + H^2\,r_w \,,\, r_w \,,\, r_w\,\sin^2\theta \, \right] \, ,
\eqe
where diag. means the diagonal matrix form.

On the other hand, the second fundamental form $K_{ij}^{\rm (flat)}$ of a spherically symmetric timelike hypersurface of radius $r_w$ in flat spacetime is given by setting $M = 0$ and $H = 0$ in Eq.\eref{eq:limit.Kij.beh},
\eqb
 K_{i j}^{\rm (flat)} =
 {\rm diag.}\left[\, 0 \,,\, r_w \,,\, r_w\,\sin^2\theta \, \right] \, .
\eqe
This gives $K^{\rm (flat)} \defeq h^{ij}\,K_{ij}^{\rm (flat)} = 2\,r_w^{-1}$.

From the above, applying the Wick rotation $t \to -i \tau$ to the Lorentzian action $I_L$ in Eq.\eref{eq:pre.IL}, we obtain the Euclidean action $I_{E b}$ of the thermal equilibrium system for BEH via Eq.\eref{eq:pre.IE.curved},
\eqb
\label{eq-setting.IEb}
\begin{array}{rl}
 I_{Eb}\,
 =& \dfrac{3\,H^2}{8\,\pi} \int_{D_{Eb}}dx_E^4\,\sqrt{g_E}
    + \dfrac{1}{8\,\pi}\int_{\partial D_{Eb}} dx_E^3 \,
                      \sqrt{h_E}\,\left(\,K_E - K_E^{\rm (flat)}\,\right) \\
 =& \dfrac{\beta_b}{2}\,
    \left[\, 3\,M - r_b + 2\,r_w \left( f_w - \sqrt{f_w} \right) \,\right] \, ,
\end{array}
\eqe
where the relation for SdS spacetime $R = 4\,\Lambda = 12\,H^2$ is used in the first equality, the relation $M - H^2 r_b^3 = 3 M - r_b$ due to $f(r_b) = 0$ is used in the second equality, $Q_E$ is the quantity $Q$ evaluated on Euclidean space, and $D_{Eb}$ is the Euclidean region denoted by $0 \le \tau < \beta_b$ , $r_b \le r \le r_w$ , $0 \le \theta \le \pi$ and $0 \le \varphi < 2\,\pi$. 
This $I_{Eb}$ corresponds to $I_E[g_{E\,cl}]$ in Eq.\eref{eq:pre.F}, which yields the partition function of our thermal equilibrium system for BEH.

Note that $I_{Eb}$ should reproduce the Euclidean action~\eref{eq:york.IE} of Schwarzschild canonical ensemble as required in the working hypothesis~SdS-2. 
To check if this is satisfied, take the limit $\Lambda \to 0$, 
\eqb
 \lim_{\Lambda \to 0} I_{Eb} = 4 \pi M\,\left[\, M - 2\,r_w \left( f_w - \sqrt{f_w} \right) \,\right]_{\Lambda=0} \, .
\eqe
This coincides with Eq.\eref{eq:york.IE}.

\subsubsection{Euclidean action for CEH}

Euclidean space of thermal equilibrium system for CEH is obtained by the Wick rotations $t \to -i\,\tau$ in the static chart and $\eta_c \to -i\,\omega_c$ in the semi-global cosmological chart. 
These Wick rotations are equivalent, because the coordinate transformation~\eref{eq:limit.trans.ceh}, $\eta_c = e^{-\kappa_c\, r^{\ast}}\sinh\left(\kappa_c\, t\right)$, implies that the imaginary time $\omega_c$ in the semi-global chart is defined by $\omega_c \defeq e^{-\kappa_c\, r^{\ast}}\sin\left(\kappa_c\, \tau\right)$, where $\tau$ is the imaginary time in the static chart. 
Euclidean metric in the static chart is given by Eq.\eref{eq:limit.SdSE.static}. 
Euclidean metric in the semi-global cosmological chart is $ds_E^2 = \Upsilon_c(r)\,\left[\,d\omega_c^2 + d\chi_c^2 \right] + r^2\,d\Omega^2$, where $\Upsilon_c$ is defined in Eq.\eref{eq:limit.upsilon.ceh}. 
About the semi-global chart, we get from the coordinate transformation~\eref{eq:limit.trans.ceh},
\eqb
 \omega_c^2 + \chi_c^2 =
  \left( \dfrac{r}{r_b} - 1 \right)^{-\kappa_c/\kappa_b}\,\left( 1 - \dfrac{r}{r_c} \right)
  \left( \dfrac{r}{r_b + r_c} + 1 \right)^{-1+\kappa_c/\kappa_b} \, .
\eqe
Then, because the thermal equilibrium system for CEH is the region $D_c$, $r_w < r < r_c$, in Lorentzian SdS spacetime, we find the topology of Euclidean space of thermal equilibrium system for CEH is $D^2\times S^2$. 
Because $\omega_c^2 + \chi_c^2 = 0$ at $r = r_c$, the center of $D^2$-part is at the CEH $r = r_c$, and the boundary of $D^2$-part is at the heat wall $r = r_w$. 
The topology of heat wall boundary is $S^1 \times S^2$, where $S^1$ is along the $\tau$-direction.

Because the Lorentzian SdS spacetime is regular at $r_c$, the Euclidean space is also regular at $r_c$.
Using the static chart~\eref{eq:limit.SdSE.static} and defining a coordinate $y_c$ and a function $\gamma(y_c)$ by $y_c^2 \defeq r_c - r$ and $\gamma(y_c) \defeq \sqrt{f(r_c - y_c^2)}$, we obtain the Euclidean metric near CEH,
\eqb
 ds_E^2 \simeq
 \dfrac{2}{\kappa_c}\,\left[\, y_c^2\,d(\kappa_c\, \tau)^2 + dy_c^2 \,\right] + r^2\,d\Omega^2 \, .
\eqe
It is obvious that the Euclidean space is regular at CEH if the imaginary time has the period $\beta_c$ defined by
\eqb
 0 \le \tau < \beta_c \defeq \dfrac{2\,\pi}{\kappa_c} \, .
\label{eq:limit.period.ceh}
\eqe
Throughout our discussion, $\tau$ has the period $\beta_c$ in the Euclidean space of thermal equilibrium system for CEH.

Now we proceed to the calculation of Euclidean action $I_{Ec}$ of the thermal equilibrium system for CEH. 
Lorentzian action is defined in Eq.\eref{eq:pre.IL} for the spacetime region $D_c$, $r_w < r < r_c$.
The calculation of Euclidean action $I_{Ec}$ is parallel to that of $I_{Eb}$ except for the direction of unit normal vector $n^{\mu}$ to $\partial D_c$ and the integration constant $I_{\rm sub}$ in $I_L$. 
Concerning the vector $n^{\mu}$, since $D_c$ is the region enclosed by CEH and heat wall, the direction of $n^{\mu}$ is pointing towards BEH, $n^{\mu} \propto - \partial_r$.

Concerning the integration constant, following the working hypothesis~SdS-2, we determine the integration constant $I_{\rm sub}$ for CEH with referring to the de~Sitter canonical ensemble~\cite{ref:euclidean.ds} formulated in Sec.\ref{sec:ds}. 
At the limit $M \to 0$, the term $I_{\rm sub}$ should reduce to the integration constant $I_{\rm sub}^{\rm (dS)}$ of the de~Sitter canonical ensemble which is read from Eqs.\eref{eq:ds.IEsub},~\eref{eq:ds.alphaw} and~\eref{eq:ds.kw},
\eqb
 I_{\rm sub}(M=0) = I_{\rm sub}^{\rm (dS)}
 \defeq \left(\dfrac{1}{H\,r_w} - 1 \right)\,\sqrt{f_w(M=0)} \, I_L^{\rm (flat)} \, ,
\eqe
where $I_L^{\rm (flat)}$ is the action of flat spacetime used in Eq.\eref{eq:limit.Isub.beh}. 
Note that the CEH radius in de~Sitter spacetime is $H^{-1}$, and the factor $H\,r_w$ is the ratio of heat wall radius $r_w$ to CEH radius. 
Hence we set $I_{\rm sub}$ for the CEH in SdS spacetime,
\eqb
\label{eq:limit.Isub.ceh}
 I_{\rm sub} \defeq \left(\dfrac{r_c}{r_w} - 1 \right)\,\sqrt{f_w}\,I_L^{\rm (flat)} \, ,
\eqe
where it should be emphasized that the signature of $K^{\rm (flat)}$ in the integrand of $I_L^{\rm (flat)}$ (shown in Eq.\eref{eq:limit.Isub.beh}) should be reversed, because the direction of normal vector $n^{\mu}$ is reversed as mentioned in previous paragraph.

Then we obtain the Euclidean action $I_{Ec}$ of our thermal equilibrium system for CEH via Eqs.\eref{eq:pre.IL} and~\eref{eq:pre.IE.curved},
\eqb
\label{eq:limit.IEc}
 I_{Ec}
 = - \dfrac{\beta_c}{2}\,\left[\, 3\,M - r_c + 2\,r_c\,f_w \,\right] \, ,
\eqe
where the relation $M - H^2 r_c^3 = 3 M - r_c$ due to $f(r_c) = 0$ is used, and the overall minus signature comes from the direction of normal vector $n^{\mu}$ to $\partial D_c$. 
This $I_{Ec}$ corresponds to $I_E[g_{E\,cl}]$ in Eq.\eref{eq:pre.F}, which yields the partition function of our thermal equilibrium system for CEH.

Note that $I_{Ec}$ should reproduce the Euclidean action~\eref{eq:ds.IE} of de~Sitter canonical ensemble as required in the working hypothesis~SdS-2. 
To check if this is satisfied, take the limit $M \to 0$, 
\eqb
 \lim_{M \to 0} I_{Ec} = - \dfrac{\pi}{H^2}\,\left[\, 1 - 2\,(H r_w)^2 \,\right] \, .
\eqe
This coincides with Eq.\eref{eq:ds.IE}.

\subsection{Breakdown of Entropy-Area Law for the Black Hole Event Horizon}
\label{sec:limit.beh}

We examine whether the entropy-area law holds for ``consistent'' thermodynamics of our thermal equilibrium system for BEH.

\subsubsection{Temperature and free energy of BEH}

By the assumption~SdS-3, the temperature $T_b$ of BEH is defined by Eq.\eref{eq:pre.T}, which relates $T_b$ to the proper length in the imaginary time direction at the boundary $\partial D_b$ (the direction along $S^1$ part of boundary topology $S^1 \times S^2$ in Euclidean space),
\eqb
 T_b \defeq \left[\, \int_0^{\beta_b}\,\sqrt{f_w}\,d\tau \,\right]^{-1}
     = \frac{\kappa_b}{2\,\pi\,\sqrt{f_w}} \, ,
\label{eq:limit.Tb}
\eqe
where $\beta_b$ is the imaginary time period~\eref{eq:limit.period.beh} and $f_w$ is in Eq.\eref{eq:limit.fw}. 
Under the length size scaling~\eref{eq:limit.scaling}, this temperature is scaled as $T_b \to \lambda^{-1}\,T_b$. 
Therefore, by the assumption~SdS-2, $T_b$ is an intensive state variable of BEH.

Note that this $T_b$ coincides with the Hawking temperature of BEH derived originally by Gibbons and Hawking~\cite{ref:temperature}, and the factor $\sqrt{f_w}$ is the so-called Tolman factor~\cite{ref:tolman} which expresses the gravitational redshift affecting the Hawking radiation propagating from BEH to observer at $r_w$. 
Therefore this $T_b$ is the temperature measured by the observer at heat wall.

By the assumption~SdS-2, the extensive state variable of system size for our thermal equilibrium system is the surface area of heat wall,
\eqb
 A \defeq 4\,\pi\,r_w^2 \, .
\eqe

By the assumption~SdS-3, the free energy $F_b$ of BEH in Eq.\eref{eq:limit.Fb_abst} is defined by Eq.\eref{eq:pre.F},
\eqb
 F_b(T_b,A,X_b) \defeq - T_b\,I_{Eb} = r_w - r_w\,\sqrt{f_w} - \frac{3\,M - r_b}{2\,\sqrt{f_w}} \, .
\label{eq:limit.Fb}
\eqe
Under the length size scaling~\eref{eq:limit.scaling}, this free energy satisfies the scaling law of thermodynamic functions, $F_b \to \lambda\,F_b$. 
As discussed at Eq.\eref{eq:limit.Fb}, $F_b$ is regarded as a function of three independent state variables $T_b$, $A$ and the state variable $X_b$ of CEH's gravitational effect on BEH. 
However, since $X_b$ is not specified as mentioned in the assumption~SdS-4, the form of $F_b$ as a function of $( T_b\,,\,A\,,\,X_b )$ remains unknown. 
Instead, Eq.\eref{eq:limit.Fb} shows $F_b$ as a function of independent parameters $( M\,,\,H\,,\,r_w )$.

Let us verify that the free energy $F_b$ is a function of three independent state variables. 
We use the reductive absurdity: 
Assume that only two (not three) state variables are independent. 
This assumption means that, as for the ordinary non-magnetized gases, $F_b$ is a function of $T_b$ and $A$, $F_b(T_b,A)$. 
Here it is obvious from Eq.\eref{eq:limit.Tb} that $T_b$ depends on three parameters $(M,H,r_w)$, while $A$ depends only on $r_w$. 
Then, via Eq.\eref{eq:pre.3rd}, a mathematical relation $(\partial_M F_b)/(\partial_M T_b) = (\partial_H F_b)/(\partial_H T_b)$ must hold if $F_b$ is a function of $(T_b , A)$. 
However we find this relation does not hold, $(\partial_M F_b)/(\partial_M T_b) \neq (\partial_H F_b)/(\partial_H T_b)$, via Eq.\eref{eq:limit.dFb_dTb} shown below. 
Hence the assumption of two independent state variables is denied by the reductive absurdity. 
Now the working hypothesis~SdS-1, which assumes $F_b$ to be a function of three independent state variables, is verified.

To support the discussion in previous paragraph and for later use, we show some differentials:
\seqb
\eqab
 \pd{T_b}{M} &=&
  \frac{1}{2 \pi r_b^2 f_w^{3/2}}
  \left[ \frac{r_b}{2 r_w} \left( 1 - 3 H^2 r_b^2 \right)
       - \frac{1 + 3 H^2 r_b^2}{1 - 3 H^2 r_b^2}\, f_w \right]
\label{eq:limit.dTb/dM} \, , \\
 \pd{T_b}{H} &=&
  \frac{1}{4 \pi H r_b f_w^{3/2}}
  \left[ \left( 1 - \frac{2 M}{r_w} \right) \left( 1 - 3 H^2 r_b^2 \right)
       - \frac{1 + 3 H^2 r_b^2}{1 - 3 H^2 r_b^2}\,\, \frac{2 M f_w}{r_b} \right]
\label{eq:limit.dTb/dH} ,
\eqae
and
\eqab
\label{eq:limit.dFb/dM}
 \pd{F_b}{M} &=&
  \frac{1}{2 f_w^{3/2}}
  \left[ - \frac{3 M - r_b}{r_w} + \frac{1 + 3 H^2 r_b^2}{1 - 3 H^2 r_b^2}\, f_w \right] \, , \\
\label{eq:limit.dFb/dH}
 \pd{F_b}{H} &=&
  \frac{1}{2 f_w^{3/2}}
  \left[ \left( 2 r_w + r_b - 7 M - 2 H^2 r_w^3 \right) H r_w^2
       + \frac{2 H r_b^3 f_w}{1 - 3 H^2 r_b^2} \right] \, ,
\eqae
\seqe
where the differentials in Eqs.\eref{eq:limit.diff.rb} and~\eref{eq:limit.diff.kappab} are used. 
Then we get
\seqb
\label{eq:limit.dFb_dTb}
\eqab
 \pd{F_b}{M} &=& - \pi\,r_b^2\,\pd{T_b}{M}
\label{eq:limit.dFb/dM.dTb/dM} \, , \\
 \pd{F_b}{H} &\not\propto& \pi\,r_b^2\,\pd{T_b}{H}
\label{eq:limit.dFb/dH.dTb/dH} \, ,
\eqae
\seqe
where the definition of $r_b$, $f(r_b) = 0$, is used in the first relation. 
These relations are used in examining the entropy-area law in next subsection.

\subsubsection{Entropy of BEH}

By the assumption~SdS-3, the entropy $S_b$ of BEH is defined as the thermodynamic conjugate variable to $T_b$,
\eqb
 S_b \defeq - \pd{F_b(T_b,A,X_b)}{T_b} \, .
\label{eq:limit.Sb.def}
\eqe
Under the length size scaling~\eref{eq:limit.scaling}, this entropy satisfies the extensive scaling law, $S_b \to \lambda^2\,S_b$. 
By Eq.\eref{eq:pre.2parameters}, $S_b$ is rearranged to
\eqb
 S_b =
 -\frac{(\partial_M F_b)\,(\partial_H X_b) - (\partial_H F_b)\,(\partial_M X_b)}
       {(\partial_M T_b)\,(\partial_H X_b) - (\partial_H T_b)\,(\partial_M X_b)} \, .
\label{eq:limit.Sb}
\eqe
Then we get by Eq.\eref{eq:limit.dFb/dM.dTb/dM},
\eqb
 S_b =
 \frac{\pi\,r_b^2 \,(\partial_M T_b)\,(\partial_H X_b) + (\partial_H F_b)\,(\partial_M X_b)}
      {(\partial_M T_b)\,(\partial_H X_b) - (\partial_H T_b)\,(\partial_M X_b)} \, .
\label{eq:limit.Sb.2}
\eqe
This denotes the following: If $\partial_M X_b \equiv 0$, then the entropy-area law $S_b \equiv \pi\,r_b^2$ holds. 
However, if $\partial_M X_b \not\equiv 0$, then Eq.\eref{eq:limit.dFb/dH.dTb/dH} together with Eq.\eref{eq:limit.Sb.2} imply that the entropy-area law breaks down $S_b \not\equiv \pi\,r_b^2$. 
Therefore we find that the entropy-area law holds if and only if $\partial_M X_b \equiv 0$.

In summary, although the validity of entropy-area law for BEH can not be judged at present, we can clarify the issue on the entropy-area law for BEH: 
\emph{The necessary and sufficient condition to ensure the entropy-area law for BEH is that the BEH is in thermal equilibrium and the state variable $X_b$ satisfies $\partial_M X_b \equiv 0$. 
If the CEH's gravitational effect on BEH, $X_b$, is characterized by the CEH's quantity $\kappa_c r_w$ as shown in Eq.\eref{eq:limit.X.1}, then $\partial_M X_b \not\equiv 0$ and the entropy-area law breaks down for BEH. 
If the CEH's gravitational effect $X_b$ is characterized by $H r_w$ as shown in Eq.\eref{eq:limit.X.2}, then $\partial_M X_b \equiv 0$ and the entropy-area law holds.
The validity of entropy-area law for BEH will be judged by revealing which of $\kappa_c$ or $H$ is appropriate as the characteristic quantity of CEH's gravity.}

At the end of this subsection, some physical discussions which may support the breakdown of entropy-area law for BEH are given. 
And in next subsection, it is revealed mathematically that the entropy-area law breaks down for CEH.

\subsubsection{Thermodynamic consistency of BEH}

Let us confirm the ``thermodynamic consistency'' of our thermodynamics of BEH under the minimal set of assumptions. 
Since the concrete form of $X_b$ is not specified, the following discussions and calculations are very formal. 
However we can imply that the thermodynamic consistency is satisfied. 
Moreover, it is also verified that $X_b$ is an extensive variable and proportional to $r_w^2$.

Following the assumption~SdS-3, the internal energy $E_b$ of BEH can be defined by the argument of statistical mechanics, 
\eqb
\label{eq:limit.Fb.Eb}
 E_b \defeq - \left.\pd{\ln Z_{cl}}{(1/T_b)}\right|_{A, X_b = \mbox{const.}}
 = \pd{(F_b/T_b)}{(1/T_b)} = F_b + T_b\,S_b \, ,
\eqe
where Eq.\eref{eq:pre.F} is used in the second equality, and the definition of $S_b$ in Eq.\eref{eq:limit.Sb.def} is used in the third equality. 
Under the length size scaling~\eref{eq:limit.scaling}, this $E_b$ satisfies the scaling law of thermodynamic functions, $E_b \to \lambda\,E_b$. 
The third equality in Eq.\eref{eq:limit.Fb.Eb}, $E_b = F_b + T_b\,S_b$, is regarded as the Legendre transformation between $F_b$ and $E_b$, which determines $E_b$ to be a function of $(S_b,A,X_b)$. 
On the other hand, in thermodynamic argument, the internal energy is the thermodynamic function which is regarded as a function of only extensive state variables. 
Hence, it is verified that $X_b$ must be an extensive state variable. 
(The proportionality of $X_b$ to $r_w^2$ will also be shown mathematically at the end of this subsection.)

Next, in order to see the first law, we need the intensive state variables which are thermodynamically conjugate to $A$ and $X_b$. 
By the assumption~SdS-3, the conjugate state variable is defined by the appropriate differential of a thermodynamic function. 
In an analogy with the ordinary pressure of ordinary gases, the \emph{surface pressure} (at the heat wall) $\sigma_b$ is defined formally as
\seqb
\eqab
\label{eq:limit.sigmab.def}
 \sigma_b && \defeq - \pd{F_b(T_b,A,X_b)}{A} \\
&&
\label{eq:limit.sigmab}
\begin{aligned}
 &= -\frac{1}{8 \pi r_w}\,
      \frac{1}{(\partial_M T_b)\,(\partial_H X_b) - (\partial_M X_b)\,(\partial_H T_b)} \\
 & \quad \times
    \Bigl[\, \left\{\, (\partial_H F_b)\,(\partial_{r_w} T_b)
                     - (\partial_{r_w} F_b)\,(\partial_H T_b)\, \right\}\,(\partial_M X_b)\\
 & \qquad\quad + \left\{\, (\partial_{r_w} F_b)\,(\partial_M T_b)
                     - (\partial_M F_b)\,(\partial_{r_w} T_b)\, \right\}\,(\partial_H X_b)\\
 & \qquad\quad + \left\{\, (\partial_M F_b)\,(\partial_H T_b)
                     - (\partial_H F_b)\,(\partial_M T_b)\, \right\}\,(\partial_{r_w} X_b)
    \,\,\, \Bigr] \, ,
\end{aligned}
\eqae
\seqe
where Eq.\eref{eq:pre.3parameters} is used in the second equality. 
Under the length size scaling~\eref{eq:limit.scaling}, this $\sigma_b$ satisfies the intensive scaling law, $\sigma_b \to \lambda^{-1} \sigma_b$. 
The state variable given by the same definition with $\sigma_b$ appears also in single-horizon thermodynamics~\cite{ref:euclidean.sch,ref:euclidean.kn,ref:euclidean.ds} to ensure the thermodynamic consistency. 
(See Appendix~B of paper~\cite{ref:euclidean.ds} for thermodynamic meanings of $A$ and $\sigma_b$.)

The intensive state variable $Y_b$ conjugate to $X_b$ is defined formally as
\eqb
\label{eq:limit.Yb}
 Y_b \defeq \pd{ F_b(T_b , A , X_b)}{X_b}
 =
 \frac{(\partial_M F_b)\,(\partial_H T_b) - (\partial_H F_b)\,(\partial_M T_b)}
      {(\partial_M X_b)\,(\partial_H T_b) - (\partial_H X_b)\,(\partial_M T_b)} \, ,
\eqe
where Eq.\eref{eq:pre.2parameters} is used in the second equality. 
Under the length size scaling~\eref{eq:limit.scaling}, when $X_b$ is scaled as an extensive variable $X_b \to \lambda^2\,X_b$, then this $Y_b$ satisfies the intensive scaling law, $Y_b \to \lambda^{-1}\,Y_b$.

Then by definitions of $S_b$, $\sigma_b$ and $Y_b$, we get
\eqb
 dF_b(T_b,A,X_b) = - S_b \, dT_b - \sigma_b \, dA + Y_b \, dX_b \, .
\eqe
The first law follows this relation via the Legendre transformation in Eq.\eref{eq:limit.Fb.Eb},
\eqb
 dE_b(S_b,A,X_b) = T_b \, dS_b - \sigma_b \, dA + Y_b \, dX_b \, .
\eqe

Concerning the internal energy, the Euler relation is interesting from the point of view of thermodynamics, because it gives a restriction on the form of state variables. 
By the scaling laws of extensive variable and thermodynamic function, we get
\seqb
\eqb
 \lambda\, E_b(S_b \,,\, A \,,\, X_b) = E_b(\lambda^2 S_b \,,\, \lambda^2 A \,,\, \lambda^2 X_b) \, .
\label{eq:limit.euler.1.beh}
\eqe
This denotes that $E_b(S_b,A,X_b)$ is the homogeneous expression of degree $1/2$. 
Operating the differential $\partial_{\lambda}$ on Eq.\eref{eq:limit.euler.1.beh}, we get
\eqb
 \frac{1}{2}\, E_b(S_b \,,\, A \,,\, X_b) = T_b\,S_b - \sigma_b\,A + Y_b\,X_b \, .
\label{eq:limit.euler.2.beh}
\eqe
\seqe
This relation~\eref{eq:limit.euler.2.beh} is obtained from the scaling behavior~\eref{eq:limit.euler.1.beh}. 
Furthermore by the well-known \emph{Euler's theorem on the homogeneous expression}, the scaling behavior~\eref{eq:limit.euler.1.beh} is also obtained from the relation~\eref{eq:limit.euler.2.beh} (which is proven by the vanishing differential $\partial_{\lambda}[\,\lambda^{-1} E_b(\lambda^2 S_b,\lambda^2 A,\lambda^2 X_b)\,] = 0$). 
Hence Eqs.\eref{eq:limit.euler.1.beh} and~\eref{eq:limit.euler.2.beh} are equivalent. 
As shown below, we find the Euler relation~\eref{eq:limit.euler.2.beh} is consistent with the assumption~SdS-4:

By the Legendre transformation in Eq.\eref{eq:limit.Fb.Eb} and the Euler relation~\eref{eq:limit.euler.2.beh}, we get a relation, $F_b = T_b\,S_b - 2\,\sigma_b\,A + 2\,Y_b\,X_b$. 
Then substituting Eqs.\eref{eq:limit.Sb}, \eref{eq:limit.sigmab} and~\eref{eq:limit.Yb} into this relation, we obtain
\eqb
 k_1\,\pd{X_b}{M} + k_2\,\pd{X_b}{H} + r_w\,k_3\,\pd{X_b}{\,r_w}
 = 2\,k_3\,X_b \, ,
\label{eq:limit.pde.k}
\eqe
where
\seqb
\eqab
\label{eq:limit.k1}
 k_1 &\defeq& - F_b\,\left( \partial_H T_b \right) - T_b\,\left( \partial_H F_b \right)\\
 &&\quad - r_w\,\left[\, \left( \partial_H F_b \right)\,\left( \partial_{r_w} T_b \right)
                     - \left( \partial_{r_w} F_b \right)\,\left( \partial_H T_b \right)
                \,\right]
\nonumber \\
\label{eq:limit.k2}
 k_2 &\defeq& F_b\,\left( \partial_M T_b \right) + T_b\,\left( \partial_M F_b \right) \\
 &&\quad - r_w\,\left[\, \left( \partial_{r_w} F_b \right)\,\left( \partial_M T_b \right)
                     - \left( \partial_M F_b \right)\,\left( \partial_{r_w} T_b \right)
                \,\right]
\nonumber \\
\label{eq:limit.k3}
 k_3 &\defeq& - \left( \partial_M F_b \right)\,\left( \partial_H T_b \right)
         + \left( \partial_H F_b \right)\,\left( \partial_M T_b \right) \, .
\eqae
\seqe
The concrete forms of $k_i$ ($i = 1$, $2$, $3$) are obtained from the differentials of $T_b$ and $F_b$ shown in Eq.\eref{eq:limit.dTb/dM}~$\sim$~\eref{eq:limit.dFb/dH.dTb/dH}, and result in relations, 
\eqb
\label{eq:limit.k.rel}
 k_1 = M\,k_3 \quad,\quad k_2 = -H\,k_3 \, .
\eqe
Then Eq.\eref{eq:limit.pde.k} reduces to
\eqb
 M\,\pd{X_b}{M} - H\,\pd{X_b}{H} + r_w\,\pd{X_b}{\,r_w} = 2\,X_b \, .
\label{eq:limit.pde.beh}
\eqe
This partial differential equation (PDE) is equivalent to the relation~\eref{eq:limit.euler.2.beh} which is also equivalent to the relation~\eref{eq:limit.euler.1.beh}. 
Therefore, if a solution $X_b$ of our PDE~\eref{eq:limit.pde.beh} exists, then the $X_b$ satisfies the extensive scaling behavior $X_b \to \lambda^2\,X_b$ under the length size scaling~\eref{eq:limit.scaling}. 
Indeed, the general solution of PDE~\eref{eq:limit.pde.beh} is expressed as
\eqb
 X_b(M,H,r_w) = r_w^2 \,\tilde{\psi}_b( M/r_w , H r_w ) \, ,
\label{eq:limit.Xb.sol}
\eqe
where $\tilde{\psi}_b(x,y)$ is an arbitrary function of two arguments. 
Obviously this $X_b$ is proportional to $r_w^2$, and satisfies the extensive scaling law under the length size scaling~\eref{eq:limit.scaling}. 
It is also obvious that the arbitrary functions $\Psi_b(\kappa_c r_w)$ in Eq.\eref{eq:limit.X.1} and $\Psi_b(H r_w)$ in Eq\eref{eq:limit.X.2} are consistent with $\tilde{\psi}_b(M/r_w , H r_w)$ in Eq.\eref{eq:limit.Xb.sol}, since $\kappa_c$ in Eq.\eref{eq:limit.kappa} is expressed as a function of $M/r_w$ and $H r_w$. 
Hence we find that the Euler relation~\eref{eq:limit.euler.2.beh} is consistent with the assumption~SdS-4, which implies that the internal energy $E_b$ and also the free energy $F_b$ are defined well in our thermodynamics of BEH. 
The well-defined free energy guarantees the thermodynamic consistency. 
Now it has been checked that the minimal set of assumptions introduced in Sec.\ref{sec:limit.assumption} constructs the ``consistent'' thermodynamics for BEH.

\subsubsection{Physical expectation of the breakdown of entropy-area law for BEH}

Some comments which suggest the breakdown of entropy-area law for BEH may be possible. 
Let us try to give two comments: 
For the first, recall the meaning of state variable $X_b$, which expresses the thermodynamic effect on BEH due to the external gravitational field produced by CEH. 
Furthermore it is worth pointing out that the CEH temperature $T_c$ depends on $\kappa_c$ which has dependence on $(M\,,\,H)$, not on $H$ solely. 
Then, in the assumption~SdS-4, it may be natural that the quantity $\kappa_c r_w$, not $H r_w$, is the characteristic variable of CEH's gravity and the extensive variable $X_b$ is expressed by $X_b = r_w^2 \Psi_b(\kappa_c r_w)$ as required in Eq.\eref{eq:limit.X.1}. 
If this is true, then the breakdown of entropy-area law for BEH is concluded as explained at Eq.\eref{eq:limit.Sb.2}.

Next, recall that, in Sec.\ref{sec:limit.assumption}, our thermal equilibrium systems for BEH and CEH are compared qualitatively with the magnetized gas. 
By the differential of free energy $F_{\rm gas}(T_{\rm gas} , V_{\rm gas} , \vec{H}_{\rm gad})$, the entropy $S_{\rm gas}$ and pressure $P_{\rm gas}$ of the gas are defined by (see for example \S52, 59 and~60 in Landau and Lifshitz~\cite{ref:ll}),
\eqb
 S_{\rm gas} \defeq
  -\pd{F_{\rm gas}(T_{\rm gas} , V_{\rm gas} , \vec{H}_{\rm ex})}{T_{\rm gas}} \quad,\quad
 P_{\rm gas} \defeq
  -\pd{F_{\rm gas}(T_{\rm gas} , V_{\rm gas} , \vec{H}_{\rm ex})}{V_{\rm gas}} \, .
\eqe
This implies that all state variables of the gas depend on the external field $\vec{H}_{\rm ex}$. 
Therefore, the entropy of the gas under the influence of external magnetic field deviates from the entropy without external magnetic field. 
Hence, for our thermal equilibrium system for BEH, it may be naturally expected that BEH's entropy under the influence of external gravitational field of CEH does not satisfy the entropy-area law which holds for BEH in single-horizon spacetimes.

The above two comments have no rigorous mathematical verification, but seem to be physically reasonable. 
Furthermore, one additional comment which support the breakdown of entropy-area law for BEH will be given in Sec.\ref{sec:conc}.

\subsection{Breakdown of Entropy-Area Law for the Cosmological Event Horizon}
\label{sec:limit.ceh}

We examine whether the entropy-area law holds for ``consistent'' thermodynamics of our thermal equilibrium system for CEH.
Discussion in this section goes parallel to Sec.\ref{sec:limit.beh}. 
However the integration constant in Euclidean action, which is determined with referring to de~Sitter canonical ensemble, enables us to find a reasonable evidence of the breakdown of entropy-area law for CEH.

\subsubsection{Temperature and free energy of CEH}

By the assumption~SdS-3, the temperature $T_c$ of CEH is defined by Eq.\eref{eq:pre.T}, which relates $T_c$ to the proper length in the imaginary time direction at the boundary $\partial D_c$,
\eqb
 T_c \defeq \left[\, \int_0^{\beta_c}\,\sqrt{f_w}\,d\tau \,\right]^{-1}
     = \frac{\kappa_c}{2\,\pi\,\sqrt{f_w}} \, ,
\label{eq:limit.Tc}
\eqe
where $\beta_c$ is the imaginary time period~\eref{eq:limit.period.ceh} and $f_w$ is in Eq.\eref{eq:limit.fw}. 
This $T_c$ coincides with the Hawking temperature obtained by Gibbons and Hawking~\cite{ref:temperature}, and the factor $\sqrt{f_w}$ is the Tolman factor ~\cite{ref:tolman} which expresses the gravitational redshift affecting the Hawking radiation propagating from CEH to observer at $r_w$. 
Under the length size scaling~\eref{eq:limit.scaling}, this temperature satisfies the extensive scaling law, $T_c \to \lambda^{-1}\,T_c$.

As defined in assumption~SdS-2, the extensive state variable of system size for our thermal equilibrium system is the surface area of heat wall,
\eqb
 A \defeq 4\,\pi\,r_w^2 \, .
\eqe

By the assumption~SdS-3, the free energy $F_c$ of CEH in Eq.\eref{eq:limit.Fc_abst} is defined by Eq.\eref{eq:pre.F},
\eqb
 F_c(T_c,A,X_c) \defeq - T_c\,I_{Ec} = r_c\,\sqrt{f_w} + \frac{3\,M - r_c}{2\,\sqrt{f_w}} \, .
\label{eq:limit.Fc}
\eqe
Under the length size scaling~\eref{eq:limit.scaling}, this free energy satisfies the scaling law of thermodynamic functions, $F_c \to \lambda\,F_c$. 
Furthermore, by the same discussion given in Sec.\ref{sec:limit.beh}, it is also mathematically verified that the free energy $F_c$ must be a function of three independent variables, which verifies the working hypothesis~SdS-1.

Let us show some differentials for later use:
\seqb
\eqab
 \pd{T_c}{M} &=&
  \frac{1}{2 \pi r_c^2 f_w^{3/2}}
  \left[ \frac{r_c}{2 r_w} \left( 3 H^2 r_c^2 - 1 \right)
       - \frac{3 H^2 r_c^2 + 1}{3 H^2 r_c^2 - 1}\, f_w \right]
\label{eq-limit.dTc/dM} \, , \\
 \pd{T_c}{H} &=&
  \frac{1}{4 \pi H r_c f_w^{3/2}}
  \left[ \left( 1 - \frac{2 M}{r_w} \right) \left( 3 H^2 r_c^2 -1 \right)
       - \frac{3 H^2 r_c^2 + 1}{3 H^2 r_c^2 - 1}\,\, \frac{2 M f_w}{r_c} \right]
\label{eq-limit.dTc/dH} ,
\eqae
and
\eqab
\label{eq:limit.dFc/dM}
 && \pd{F_c}{M} =
  \pd{\left[\,(r_c - r_w)\,\sqrt{f_w}\,\right]}{M}
  + \frac{1}{2 f_w^{3/2}}
    \left[ - \frac{r_c - 3 M}{r_w}
           + \frac{3 H^2 r_c^2 + 1}{3 H^2 r_c^2 - 1}\, f_w \right] \, , \\
\label{eq:limit.dFc/dH}
 && \begin{aligned}
 \pd{F_c}{H} = \,\,& \pd{\left[\,(r_c - r_w)\,\sqrt{f_w}\,\right]}{H} \\
  &+ \frac{1}{2 f_w^{3/2}}
    \left[- \left( 2 r_w + r_c - 7 M - 2 H^2 r_w^3 \right) H^2 r_w^2
          + \frac{2 H r_c^3 f_w}{3 H^2 r_c^2 - 1} \right] \, ,
    \end{aligned}
\eqae
\seqe
where the differentials in Eqs.\eref{eq:limit.diff.rc} and~\eref{eq:limit.diff.kappac} are used. 
Then we get
\seqb
\label{eq:limit.dFc.dTc}
\eqab
 \pd{F_c}{M} &=& \pd{\left[\,(r_c - r_w)\,\sqrt{f_w}\,\right]}{M} - \pi\,r_c^2\,\pd{T_c}{M}
\label{eq:limit.dFc/dM.dTc/dM} \, , \\
 \pd{F_c}{H} &\not\propto& \pi\,r_c^2\,\pd{T_c}{H}
\label{eq:limit.dFc/dH.dTc/dH} \, ,
\eqae
\seqe
where the definition of $r_c$, $f(r_c) = 0$, is used in the first relation. 
These relations are important to get a reasonable evidence of the breakdown of entropy-area law for CEH in next subsection.

Here one might naively expect that Eqs.\eref{eq:limit.dFc/dM} and~\eref{eq:limit.dFc/dH} would be obtained by replacing $r_b$ with $r_c$ in Eqs.\eref{eq:limit.dFb/dM} and~\eref{eq:limit.dFb/dH}, and also a relation $\partial_M F_c = - \pi r_c^2\, \partial_M T_c$ would be expected. 
However the first terms in the right-hand sides in Eqs.\eref{eq:limit.dFc/dM}, ~\eref{eq:limit.dFc/dH} and~\eref{eq:limit.dFc/dM.dTc/dM} appear, because of the difference of integration constant in Euclidean action as seen in Eqs.\eref{eq:limit.Isub.beh} and~\eref{eq:limit.Isub.ceh}. 
The integration constant of $I_{Ec}$ can not be obtained by replacing $r_b$ with $r_c$ in that of $I_{Eb}$.

\subsubsection{Entropy of CEH}

By the assumption~SdS-3, the entropy $S_c$ of CEH is defined as the thermodynamic conjugate variable to $T_c$,
\eqb
 S_c \defeq - \pd{F_c(T_c,A,X_c)}{T_c}
 = -\frac{(\partial_M F_c)\,(\partial_H X_c) - (\partial_H F_c)\,(\partial_M X_c)}
       {(\partial_M T_c)\,(\partial_H X_c) - (\partial_H T_c)\,(\partial_M X_c)} \, ,
\label{eq:limit.Sc}
\eqe
where Eq.\eref{eq:pre.2parameters} is used in the second equality. 
Under the length size scaling~\eref{eq:limit.scaling}, this entropy satisfies the extensive scaling law, $S_c \to \lambda^2\,S_c$. 
From this definition and the assumption~SdS-4 together with Eqs.\eref{eq:limit.dFc.dTc}, we obtain an evidence of the breakdown of entropy-area law for CEH by the reductive absurdity as follows:

Assume that the entropy-area law holds for CEH, $S_c = \pi\,r_c^2$. 
Then Eq.\eref{eq:limit.Sc} and $S_c = \pi\,r_c^2$ reduce to a PDE of $X_c$,
\seqb
\eqb
\label{eq:limit.pde.1}
 J_M\,\pd{X_c}{M} + J_H\,\pd{X_c}{H} = 0 \, ,
\eqe
where  Eq.\eref{eq:limit.dFc/dM.dTc/dM} is used, and
\eqb
 J_M \defeq \pd{F_c}{H} + \pi\,r_c^2\,\pd{T_c}{H} \quad,\quad
 J_H \defeq - \pd{\left[\,(r_c - r_w)\,\sqrt{f_w}\,\right]}{M} \, .
\eqe
\seqe
We find $J_M \not\equiv 0$ due to Eq.\eref{eq:limit.dFc/dH.dTc/dH}, and $J_H \not\equiv 0$ due to $\partial_M r_c \not\equiv 0$ and $\partial_M f_w = -2/r_w$.

For the case of $X_c = r_w^2 \Psi_c(\kappa_b r_w)$ given in Eq.\eref{eq:limit.X.1} of assumption~SdS-4, the PDE~\eref{eq:limit.pde.1} results in a contradiction as follows: 
Note that, because the surface gravities $\kappa_b$ and $\kappa_c$ are independent as functions of two variables $(M,H)$ due to non-zero \emph{Wronskian} $(\partial_M \kappa_b)\,(\partial_H \kappa_c) - (\partial_H \kappa_b)\,(\partial_M \kappa_c) \not\equiv 0$, the three quantities $(\kappa_b,\kappa_c,r_w)$ can be regarded as independent variables instead of $(M,H,r_w)$. 
Here, the transformation of independent variables between two pairs $(M,H,r_w)$ and $(\kappa_b,\kappa_c,r_w)$ is interpreted as the coordinate transformation in the state space of thermal equilibrium states of CEH. 
Then we find $\partial_{\kappa_c} X_c \equiv 0$ for $X_c = r_w^2 \Psi_c(\kappa_b r_w)$, and the PDE~\eref{eq:limit.pde.1} reduces to 
\eqb
 \left( J_M\,\pd{\kappa_b}{M} + J_H\,\pd{\kappa_b}{H} \right)\,\pd{X_c}{\kappa_b} = 0 \, ,
\eqe
which gives $\partial_{\kappa_b} X_c \equiv 0$ due to $J_M\,\partial_M \kappa_b + J_H\,\partial_H \kappa_b \not\equiv 0$. 
On the other hand, the form of $X_c = r_w^2 \Psi_c(\kappa_b r_w)$ means $\partial_{\kappa_b} X_c \not\equiv 0$, since $\Psi_c$ is not constant as explained in assumption~SdS-4. 
Hence we find the PDE~\eref{eq:limit.pde.1}, which is equivalent to the entropy-area law, contradicts Eq.\eref{eq:limit.X.1} of assumption~SdS-4.

Next, for the case of $X_c = r_w^2 \Psi_c(M/r_w)$ given in Eq.\eref{eq:limit.X.2}, the PDE~\eref{eq:limit.pde.1} results in a contradiction as follows: 
With regarding the three quantities $(M,H,r_w)$ as independent variables, Eq.\eref{eq:limit.X.2} means $\partial_H X_c \equiv 0$ and the PDE~\eref{eq:limit.pde.1} gives $\partial_M X_c \equiv 0$ due to $J_M \not\equiv 0$. 
On the other hand, the form of $X_c = r_w^2 \Psi_c(M/r_w)$ means $\partial_M X_c \not\equiv 0$, since $\Psi_c$ is not constant. 
Hence we find the PDE~\eref{eq:limit.pde.1}, which is equivalent to the entropy-area law, contradicts Eq.\eref{eq:limit.X.2} of assumption~SdS-4.

\emph{The above discussions imply the breakdown of entropy-area law by the reductive absurdity under the minimal set of assumptions introduced in Sec.\ref{sec:limit.assumption}. 
Now it is concluded that we find a ``reasonable'' evidence of the breakdown of entropy-area law for CEH in SdS spacetime, where the ``reasonableness'' means that our discussion retains the ``thermodynamic consistency'' as shown in next subsection.}

\subsubsection{Thermodynamic consistency of CEH}

The remaining part of this subsection is for the ``thermodynamic consistency'' of our thermodynamics of CEH under the minimal set of assumptions. 
It is also verified that $X_c$ is an extensive variable and proportional to $r_w^2$. 
The discussion for the confirmation of thermodynamic consistency of BEH given in Sec.\ref{sec:limit.beh} is applied to CEH.

By the assumption~SdS-3, the internal energy $E_c(S_c,A,X_c)$ of CEH, the surface pressure $\sigma_c$ at heat wall and the intensive variable $Y_c$ conjugate to $X_c$ are defined by
\eqab
\label{eq:limit.Ec}
 E_c
 &\defeq& - \left.\pd{\ln Z_{cl}}{(1/T_c)}\right|_{A, X_c = \mbox{const.}}
  = \pd{(F_c/T_c)}{(1/T_c)}
  =  F_c + T_c\,S_c \, , \\
 \sigma_c &\defeq& - \pd{F_b(T_c,A,X_c)}{A} \nonumber \\
\label{eq:limit.sigmac}
 &=& \mbox{Eq.\eref{eq:limit.sigmab} with replacing $(F_b , X_b)$ with $(F_c , X_c)$} \\
 Y_c
 &\defeq& \pd{ F_c(T_c , A , X_c)}{X_c} \nonumber \\
\label{eq:limit.Yc}
 &=& \mbox{Eq.\eref{eq:limit.Yb} with replacing $(F_b , X_b)$ with $(F_c , X_c)$} \,,
\eqae
where the relation in Eq.\eref{eq:limit.Ec}, $E_c(S_c,A,Y_c) = F_b(T_c,A,Y_c) + T_c\,S_c$, is regarded as the Legendre transformation, and Eqs.\eref{eq:pre.3parameters} and~\eref{eq:pre.2parameters} are used in the second equalities in $\sigma_c$ and $Y_c$. 
Under the length size scaling~\eref{eq:limit.scaling}, $E_c$ satisfies the scaling law of thermodynamic functions $E_c \to \lambda\,E_c$, and $\sigma_c$ and $Y_c$ satisfy the intensive scaling law $\sigma_c \to \lambda^{-1}\,\sigma_c$ and $Y_c \to \lambda^{-1}\,Y_c$. 
We find that, since the internal energy is a function of only extensive state variables, the state variable $X_c$ of BEH's gravitational effect on CEH should be an extensive variable. 
(The proportionality of $X_c$ to $r_w^2$ will also be shown mathematically at the end of this subsection.)

Then by these definitions of $E_c$, $\sigma_c$ and $Y_c$ together with the definition of $S_c$, we get the first law for CEH,
\eqb
 dE_c(S_c,A,Y_c) = T_c\,dS_c - \sigma_c\,dA + Y_c\,dX_c \, .
\eqe
Furthermore, the scaling behavior required in assumption~SdS-2 results in the same Euler relations as in Eqs.\eref{eq:limit.euler.1.beh} and~\eref{eq:limit.euler.2.beh},
\seqb
\eqab
\label{eq:limit.euler.1.ceh}
 \lambda\, E_c(S_c \,,\, A \,,\, X_c) 
   &=& E_c(\lambda^2 S_c \,,\, \lambda^2 A \,,\, \lambda^2 X_c) \, , \\
\label{eq:limit.euler.2.ceh}
 \frac{1}{2}\, E_b(T_c \,,\, A \,,\, X_c)
   &=& T_c\,S_c - \sigma_c\,A + Y_c\,X_c \, .
\eqae
\seqe
These two relations are mathematically equivalent by the Euler's theorem on the homogeneous expression.

By the Legendre transformation in Eq.\eref{eq:limit.Ec} and the Euler relation~\eref{eq:limit.euler.2.ceh}, we get a relation, $F_c = T_c\,S_c - 2\,\sigma_c\,A + 2\,Y_c\,X_c$. 
Then substituting $S_c$, $\sigma_c$ and $Y_c$ into this relation, we obtain a PDE of $X_c$,
\eqb
 l_1\,\pd{X_c}{M} + l_2\,\pd{X_c}{H} + r_w\,l_3\,\pd{X_c}{\,r_w}
 = 2\,l_3\,X_c \, ,
\label{eq:limit.pde.l}
\eqe
where $l_i$ ($i = 1$, $2$, $3$) are defined formally by the same definitions of $k_i$ in Eqs.\eref{eq:limit.k1}, \eref{eq:limit.k2} and~\eref{eq:limit.k3} by replacing $(F_b\,,\,T_b)$ with $(F_c\,,\,T_c)$.
Using the differentials of $T_c$ and $F_c$ shown at Eq.\eref{eq-limit.dTc/dM}~$\sim$~\eref{eq:limit.dFc/dH.dTc/dH}, we find $l_1 = M\,l_3$ and $l_2 = -H\,l_3$ which is the same with Eq.\eref{eq:limit.k.rel}.
Then our PDE~\eref{eq:limit.pde.l} reduces to the same PDE in Eq.\eref{eq:limit.pde.beh}, and its general solution is
\eqb
 X_c(M,H,r_w) = r_w^2 \,\tilde{\psi}_c( M/r_w , H r_w ) \, ,
\label{eq:limit.Xc.sol}
\eqe
where $\tilde{\psi}_c(x,y)$ is an arbitrary function of two arguments. 
Obviously this $X_c$ is proportional to $r_w^2$, and satisfies the extensive scaling law under the length size scaling~\eref{eq:limit.scaling}. 
It is also obvious that the arbitrary functions $\Psi_c(\kappa_b r_w)$ in Eq.\eref{eq:limit.X.1} and $\Psi_c(M/r_w)$ in Eq\eref{eq:limit.X.2} are consistent with $\tilde{\psi}_c(M/r_w , H r_w)$ in Eq.\eref{eq:limit.Xc.sol}, since $\kappa_b$ in Eq.\eref{eq:limit.kappa} is expressed as a function of $M/r_w$ and $H r_w$. 
Hence we find that the Euler relation~\eref{eq:limit.euler.2.ceh} is consistent with the assumption~SdS-4, which implies that the internal energy $E_c$ and also the free energy $F_c$ are defined well in our thermodynamics of CEH. 
The well-defined free energy guarantees the thermodynamic consistency. 
Now it has been checked that the minimal set of assumptions introduced in Sec.\ref{sec:limit.assumption} constructs the ``consistent'' thermodynamics for CEH. 
Hence, our conclusion that the entropy-area law breaks down for CEH is reasonable.

\subsection{Supplement: Near Nariai Case}
\label{sec:limit.supplement}

From the above, we give some reasonable evidence of the breakdown of entropy-area law for BEH and CEH in SdS spacetime. 
Our analysis is exact for parameter range, $0 < \sqrt{27}\,M H <1$ and $r_b < r_w < r_c$, where the first inequality ensures that the BEH and CEH is non-degenerate, $r_b < r_c$. 
It is obvious that our discussion is true of the near Nariai case (near extremal case of SdS spacetime), $r_b \simeq r_c$ ($\,\Leftrightarrow\,\sqrt{27}\,M H \simeq 1$)~\footnote{The metric of extremal SdS spacetime was found by Nariai~\cite{ref:nariai}, independently of the non-extreme SdS metric by Kottler~\cite{ref:kottler}.}. 
However, note that the temperatures of horizons are equal at the exact Nariai case, $T_b = T_c$ at $r_b = r_c$. 
Then, finally in this section, we analyze the near Nariai case (near extremal case of SdS spacetime) of our two thermal equilibrium systems of BEH and CEH.

Note that, in the exact Nariai case, the two horizons degenerate and our two thermal equilibrium systems of horizons disappear. 
Hence we consider the near Nariai case as a perturbation of the exact Nariai case. 
We introduce two independent small parameters, $\delta_w$ and $\delta_{bc}$, defined by 
\eqb
 r_w \defeqr 3M + \delta_w \quad,\quad
 r_c \defeqr r_b + \delta_{bc} \,,
\eqe
where we require $\delta_{bc} \ll M$ which means the near Nariai case.
The parameter $\delta_w$ controls the position of the heat wall, and satisfies, $-(3 M - r_b) < \delta_w < r_c - 3 M$.

In the following, we expand the temperatures and free energies of our two thermal equilibrium systems by the small parameters $\delta_w$ and $\delta_{bc}$, in which the 0-th order values are of the Nariai limit $\delta_{bc} \to 0$ and $\delta_w \to 0$. 
We measure the size of the exact Nariai spacetime by the mass parameter $M$. 
Then, it is useful to introduce the following supplemental small parameters, $\delta_H$ and $\delta_\alpha$, defined by
\eqb
 H \defeqr \frac{1}{\sqrt{27}\,M} - \delta_H \quad,\quad
 \alpha \defeqr \frac{\pi}{2} - \delta_\alpha \,,
\eqe
where $\alpha$ is given by $\sin\alpha = \sqrt{27} M H$. 
One of three parameters $(\,\delta_{bc}\,,\,\delta_H\,,\,\delta_\alpha\,)$ is independent. 
By the definition of $\alpha$, $\sin\alpha = \sqrt{27} M H$, and Eq.\eref{eq:limit.r}, we obtain
\eqb
 2 \sqrt{27}\,M\,\delta_H = \delta_\alpha^2 + O(\delta_\alpha^4) \quad,\quad
 \delta_{bc} = 2 \sqrt{3}\,M\,\delta_\alpha\,
                 \left[\, 1 + \frac{1}{3}\,\delta_\alpha^2 + O(\delta_\alpha^4) \,\right] \,.
\eqe
Furthermore, from Eq.\eref{eq:limit.fw}, we obtain
\eqb
 f_w = \frac{\delta_{bc}^2}{36 M^2}\,
       \left[\, 1 - 4 \left(\frac{\delta_w}{\delta_{bc}}\right)^2 \,\right]
       + O(\delta_{bc}^4) 
       + O(\delta_{bc}^2\,\delta_w) \,.
\eqe
The requirement $f_w > 0$  means $\bigl|\delta_w/\delta_{bc}\bigr| < 1/2$.

From the above, we obtain the near Nariai value of temperatures~\eref{eq:limit.Tb} and~\eref{eq:limit.Tc},
\seqb
\eqb
 T_b = T_N\,\left[\, 1 + \frac{\delta_{bc}}{9 M} + O(\delta_{bc}^2) \,\right] \quad,\quad
 T_c = T_N\,\left[\, 1 - \frac{\delta_{bc}}{9 M} + O(\delta_{bc}^2) \,\right] \,,
\eqe
where
\eqb
 T_N = \frac{1}{6 \pi M}\,
       \left[\, 1 - 4 \left(\frac{\delta_w}{\delta_{bc}}\right)^2 \,\right]^{-1/2}\,
       \left[\, 1 + O(\delta_w) + O(\delta_{bc}^2) + O(\delta_w\,\delta_{bc}^2) \,\right] \,.
\eqe
\seqe
This means that, in the near Nariai case, the temperatures of our two thermal equilibrium systems are equal up to the leading term. 
Hence, at the leading term approximation, the total system composed of two horizons is in a thermal equilibrium state in the near Nariai case. 
Then, as discussed above, we expect that the entropy-area law holds at the leading term approximation in the near Nariai case. 
To see it, let us show the free energies~\eref{eq:limit.Fb} and~\eref{eq:limit.Fc} in the near Nariai case,
\seqb
\eqab
 F_b &=& 3 M - \frac{3 M}{2}
             \left[\, 1 - 4 \left(\frac{\delta_w}{\delta_{bc}}\right)^2 \,\right]^{-1/2}
      + O(\delta_{bc}) + O(\delta_w) + O(\delta_{bc}\,\delta_w) \\
 F_c &=& F_b - 3 M \,.
\eqae
\seqe
Here one may think that the difference $3M = F_b - F_c$ results in the difference between entropies of BEH and CEH. 
But the definition of entropy, $S \defeq -\partial F(T,A,X)/\partial T$, is important. 
Up to the leading term, the system size is $A = 4 \pi (3 M)^2$ and $M$ is fixed in calculating the entropy. 
Therefore the difference $3M$ does not mean the difference between horizon entropies. 
The entropy of BEH is equal to that of CEH at the leading term approximation in the near Nariai case. 
However, unfortunately, we can not check if the entropy-area law recovers at the leading term approximation, because the state variables $X_b$ and $X_c$ are not specified and the partial derivative $-\partial F/\partial A$ can not be calculated. 
At present, we simply expect that the entropy-area law holds at the leading term approximation in the near Nariai limit.

\section{Conclusion}
\label{sec:conc}

\subsection{Necessary and Sufficient Condition for the Entropy-Area Law}
\label{sec:conc.condition}

The main part of this article was Sec.\ref{sec:limit}. 
In that section, in order to research whether the thermal equilibrium is the necessary and sufficient condition to ensure the entropy-area law, we have carefully constructed \emph{two thermal equilibrium systems} individually for BEH and CEH in SdS spacetime, and the ``consistent thermodynamics'' have been obtained for BEH and CEH under the minimal set of assumptions. 
The need of those assumptions was discussed in Sec.\ref{sec:pre.partition}. 
In the construction of the two thermal equilibrium systems, the role of cosmological constant in the consistent thermodynamics has also been pointed out in the working hypothesis~SdS-1, which has also been recognized in de~Sitter thermodynamics in Sec.\ref{sec:ds.lambda}. 
In our analysis, Euclidean action method was used with referring to Schwarzschild and de~Sitter canonical ensembles to determine the integration constants (subtraction terms). 
As a result, we have found a reasonable evidence for the breakdown of entropy-area law for CEH in Sec.\ref{sec:limit.ceh}, while the validity of the law for BEH could not be judged but the key issue on BEH's entropy has been clarified in Sec.\ref{sec:limit.beh}. 
If the breakdown of the law for BEH is verified, then it means:
\begin{itemize}
\item
\emph{
Thermal equilibrium of individual horizon in multi-horizon spacetime is just a necessary condition of entropy-area low.
}
\item 
\emph{
The necessary and sufficient condition of entropy-area law is the thermal equilibrium of the total system composed of several horizons in which the net energy flow among horizons disappears.
}
\end{itemize}

Concerning BEH, we have already suggested two physical discussions which support the breakdown of entropy-area law for BEH. 
Furthermore, here we suggest an additional discussion to support the breakdown: 
Note that, while the CEH temperature $T_c$~\eref{eq:limit.Tc} is obtained from BEH temperature $T_b$~\eref{eq:limit.Tb} by the simple replacement of $(r_b\,,\,\kappa_b)$ with $(r_c\,,\,\kappa_c)$, the CEH free energy $F_c$~\eref{eq:limit.Fc} can not be obtained from BEH free energy $F_b$~\eref{eq:limit.Fb} by such a simple replacement. 
This ``asymmetry'' of $F_b$ and $F_c$ is due to the asymmetry of integration constant of BEH's Euclidean action~\eref{eq:limit.Isub.beh} and that of CEH's one~\eref{eq:limit.Isub.ceh}. 
Then, it is naively expected that the coefficients $l_i$ of PDE~\eref{eq:limit.pde.l} do not satisfy the same relation~\eref{eq:limit.k.rel} as $k_i$. 
However, we find at Eq.\eref{eq:limit.pde.l} that the relation~\eref{eq:limit.k.rel} holds for both coefficients $k_i$ and $l_i$, and the same expression of general solutions of $X_b$ and $X_c$ are obtained as shown in Eqs.\eref{eq:limit.Xb.sol} and~\eref{eq:limit.Xc.sol}. 
This may imply that the same consistent structure of thermodynamics holds for BEH and CEH even though the forms of free energies are asymmetric. 
If this implication is true, then, since the entropy-area law breaks down for CEH, the law for BEH may also break down.

\subsection{Special Role of Cosmological Constant}
\label{sec:conc.constant}

Usually the cosmological constant $\Lambda$ is not regarded as variable in the framework of horizon thermodynamics. 
Indeed, in the micro-canonical ensemble of de~Sitter horizon (see references~\cite{ref:sds.special.2,ref:micro} and Sec.\ref{sec:ds.micro}), it is not necessary to regard $\Lambda$ as variable. 
Here note that, in general thermodynamical and statistical mechanical arguments, both micro-canonical and canonical ensembles can produce the same thermodynamic formalism of the system under consideration. 
Hence, the existence of micro-canonical ensemble of de~Sitter thermodynamics indicates the existence of canonical ensemble of de~Sitter thermodynamics. 
Then, as shown in Sec.\ref{sec:ds.lambda}, we can explicitly recognize in the framework of canonical ensemble of de~Sitter thermodynamics that $\Lambda$ should be regarded as a working variable. 
Furthermore, this is also true of SdS thermodynamics as discussed in Sec.\ref{sec:limit.assumption}.
The following role of $\Lambda$ is worth emphasizing;
\begin{itemize}
\item \emph{The canonical ensemble of CEH in both de~Sitter and SdS spacetime constructs the ``generalized'' thermodynamics in which $\Lambda$ behaves as a working variable, and the physical process is described by the constant $\Lambda$ process.}
\end{itemize}

\subsection{Two Discussions}
\label{sec:cocn.discussion}

Finally in this article, we make two discussions. 
One of them is on the quantum statistics of underlying quantum gravity, and the other is on SdS black hole evaporation as a non-equilibrium process. 
These are independent of each other.

\subsubsection{Discussion about quantum statistics of gravity}

There are some existing discussions on quantum nature of gravity under the existence of cosmological constant, e.g. in papers by Parikh and et al~\cite{ref:parikh}. 
Those papers seems to be interested in some holographic principle. 
However, let us give a discussion from different point of view, which is rather interested in a ``quantum statistical'' property of gravity.

The analysis in the main text of this article is based on the Euclidean action method. 
This is equivalent to assume that the basic principle of statistical mechanics of ordinary laboratory systems works well in calculating the partition function of the canonical ensemble for our two thermal equilibrium systems. 
Here let us emphasize that the basic principle of statistical mechanics of ordinary laboratory systems is the \emph{principle of equal a priori probabilities}~\cite{ref:sm,ref:ll}~\footnote{
When this principle is applied to the micro-canonical ensemble, the Boltzmann's relation $S = \ln W$ is obtained, where $S$ and $W$ are respectively the entropy and the number of states. And when this principle is applied to the canonical ensemble, the free energy is obtained by the relation in Eq.\eref{eq:pre.F}.
}.
Hence, if the analysis in this article and the comments in previous subsection are true (to imply the breakdown of entropy-area law for BEH), then it suggests that the principle of equal a priori probabilities results in the breakdown of entropy-area law for multi-horizon spacetimes in which horizon temperatures are not equal and a net energy flow among horizons exits. 
In other words, if the statistics of micro-states of quantum gravity obeys the principle of equal a priori probabilities, then the entropy-area law breaks down for the multi-horizon spacetimes.

Then what will be suggested if we adopt the other point of view? 
Let us dare to give priority to the entropy-area law, and assume that the entropy-area law holds for the two thermal equilibrium systems constructed in the assumption~1. 
Under this assumption, the discussion in previous paragraph implies that the principle of equal a priori probabilities and the Euclidean action method is not suitable to the quantum statistics of gravity in the multi-horizon spacetimes. 
In this case, the underlying quantum gravity should be formulated to yield the special statistic property of micro-states of gravity, which comes to obey the principle of equal a priori probability in the case of single-horizon limit~\cite{ref:euclidean.sch,ref:euclidean.kn,ref:euclidean.sads,ref:euclidean.ds}.

\subsubsection{Discussion on SdS black hole evaporation}

Turn our discussion to the second one, which is completely separated from the above discussion of quantum statistics. 
The second discussion is on SdS black hole evaporation process: 
Hereafter the heat wall introduced in the assumption~1 is \emph{removed}. 
Let us note the inequality $T_b > T_c$ due to Eq.\eref{eq:limit.kappa.rel}, which means the existence of a net energy flow from BEH to CEH due to the exchange of Hawking radiation emitted by two horizons. 
This means that, as mentioned in Sec.\ref{sec:intro}, the region~I in SdS spacetime is in a \emph{non-equilibrium state}, and the SdS spacetime evolves in time due to the energy flow. 
This time evolution is the SdS black hole evaporation process. 
Here note that Hawking temperature is usually much lower than the energies $E_b$ and $E_c$ when the horizon is not quantum but classical size~\cite{ref:hr}. 
Then the evolution of BEH and CEH during the SdS black hole evaporation can be described by the so-called \emph{quasi-static process}, in which thermodynamic states of BEH and CEH at each instant of the evolution can be approximated well by thermal equilibrium states. 
(Thermodynamic state of BEH evolves on a path in the state space of thermal equilibrium states. 
Also the thermodynamic state of CEH do the same, but the path in state space on which CEH evolves is different from that of BEH.) 
This implies that the matter field of Hawking radiation is responsible for the non-equilibrium nature of SdS spacetime. 
Because the matter field is in a non-equilibrium state, the total system composed of SdS spacetime and the matter field of Hawking radiation is in a non-equilibrium state, even when the horizons are individually in equilibrium states~\footnote{
The matter field of Hawking radiation is neglected in the main context of this article, because its energy scale is negligible for classical size horizons~\cite{ref:hr}. 
However, when we proceed to the research on the non-equilibrium nature of SdS spacetime and its time evolution, it is necessary to consider the matter field of Hawking radiation which is responsible for the non-equilibrium nature of SdS evaporation process.
}. 
If we can formulate a general non-equilibrium thermodynamics for arbitrary matter fields which are enclosed by two thermal bodies of different temperatures, then a non-equilibrium SdS thermodynamics may be obtained by applying the non-equilibrium thermodynamics to matter fields of Hawking radiation.

For non-self-interacting matters, a two-temperature non-equilibrium thermodynamics has already been constructed~\cite{ref:noneq.rad}. 
Therefore, under the assumption that the matter field of Hawking radiation is non-self-interacting (e.g. a minimal coupling massless scaler field~$\phi$ satisfying $\Box \phi = 0$), the non-equilibrium evolution process of SdS spacetime may be described by using the non-equilibrium thermodynamics of non-self-interacting matters~\cite{ref:noneq.rad}. 
Indeed, the non-equilibrium thermodynamics of non-self-interacting matters has already been applied to the evaporation process of Schwarzschild black hole~\cite{ref:evapo}, and has revealed the detail of evaporation process as a non-equilibrium process and verified the so-called generalized second law for the evaporation process~\cite{ref:gsl}. 
However the non-equilibrium thermodynamics of non-self-interacting matters requires to know \emph{a priori} the equilibrium state variables of thermal bodies among which the non-equilibrium matter field is enclosed. 
Therefore, before applying the non-equilibrium thermodynamics~\cite{ref:noneq.rad} to SdS spacetime, we have to specify the state variables $X_b$ and $X_c$.

Finally for self-interacting matters, under the condition that its non-equilibrium nature is not so strong, some non-equilibrium thermodynamics have already been constructed~\cite{ref:noneq}. 
Therefore, under the assumption that the strength of non-equilibrium nature is not so strong (e.g. not so large temperature difference), the non-equilibrium evolution process of SdS spacetime may be described by using an appropriate non-equilibrium thermodynamics~\cite{ref:noneq}. 
Concerning self-interacting matters, as far as the author knows, no one has been applied the theories~\cite{ref:noneq} even to Schwarzschild black hole evaporation process.


\label{lastpage-01} 
\end{document}